\documentclass[authoryear]{article}
\usepackage[english]{babel}
\usepackage[utf8]{inputenc}
\usepackage{johd}

\usepackage{amsmath}
\usepackage{amssymb}
\usepackage{ifthen}
\usepackage{array}
\usepackage{theorem}

\usepackage{dcolumn}
\usepackage{booktabs}				% for output of xtable
\usepackage[export]{adjustbox}			% http://ctan.org/pkg/adjustbox		to top-align figures
\usepackage{caption}
\usepackage{subcaption}
\usepackage{enumerate}				% For enact enumerations and changing spacing for lists
\usepackage{paralist}				% For inparaenum

\usepackage{comment}
\excludecomment{mycomment}
\newcommand{\Vast}{\bBigg@{5}}
\newcommand{\R}{\ensuremath{\mathbb{R}}}
\newtheorem{lemma}{Lemma}
\newtheorem{proposition}{Proposition}

\title{Competition and Collaboration in Crowdsourcing Communities: What happens when peers evaluate each other?}

\author{\\\normalsize{\textit{Forthcoming in Organization Science, 2024}}\\\\
Christoph Riedl$^{a}$$^{*}$, Tom Grad$^{b}$, Christopher Lettl$^{c}$ \\
       \small $^{a}$D'Amore-McKim School of Business, Northeastern University, Boston, MA \\
       \small $^{b}$Department of Strategy and Innovation, Copenhagen Business School, 2000 Frederiksberg, Denmark \\
       \small $^{v}$Department of Strategy and Innovation, Vienna University of Economics and Business, 1020 Vienna, Austria \\\\
       \small $^{*}$Corresponding author: \tt{c.riedl@northeastern.edu} \\
}
\date{\today}
\begin{document}
\maketitle

\begin{abstract} 
\noindent Crowdsourcing has evolved as an organizational approach to distributed problem solving and innovation. As contests are embedded in online communities and evaluation rights are assigned to the crowd, community members face a tension: they find themselves exposed to both competitive motives to win the contest prize and collaborative participation motives in the community. The competitive motive suggests they may evaluate rivals strategically according to their self-interest, the collaborative motive suggests they may evaluate their peers truthfully according to mutual interest. Using field data from Threadless on 38 million peer evaluations of more than 150,000 submissions across 75,000 individuals over10 years and two natural experiments to rule out alternative explanations, we answer the question of how community members resolve this tension. We show that as their skill level increases, they become increasingly competitive and shift from using self-promotion to sabotaging their closest competitors. However, we also find signs of collaborative behavior when high-skilled members show leniency toward those community members who do not directly threaten their chance of winning. We explain how the individual-level use of strategic evaluations translates into important organizational-level outcomes by affecting the community structure through individuals’ long-term participation. While low-skill targets of sabotage are less likely to participate in future contests, high-skill targets are more likely. This suggests a feedback loop between competitive evaluation behavior and future participation. These findings have important implications for the literature on crowdsourcing design, and the evolution and sustainability of crowdsourcing communities.
\end{abstract}

\noindent\\

% #####################################################################
% 						INTRODUCTION
% #####################################################################
\section{Introduction}

Crowdsourcing as an organizational approach to distributed problem solving and innovation \citep{afuah2023reflections} has come a long way. Originally conceived as an approach through which firms solicit contributions from independent individuals the past decade has seen a significant evolution in crowdsourcing design. Two aspects stand out. First, crowdsourcing contests are now often embedded in online communities \citep[e.g.,][]{jeppesen2010marginality,riedl2018learning,grad2022rivalry,zaggl2023governing,riedl2024scraplab}. Second, many of these crowdsourcing communities have adopted open participation architectures that assign evaluation rights to the crowd \citep{majchrzak2013towards,dahlander2019organizations,blohm2016rate}. 

When community members are asked to evaluate their rivals in contest-based communities, they face a tension as they find themselves exposed to both competitive motives to win the contest prize and collaborative participation motives in the community \citep{deutsch1949theory,fiske1992four}. Community members may evaluate their rivals fairly according to the collaborative ideal of the community \citep[][calls this \textit{equality matching}]{fiske1992four} or they may act in self-interest and evaluate them strategically in order to maximize their own chances of winning the contest (\textit{market pricing}). 
% \hl{\[Option 1\] After all, self-interested behavior refers to actions of an individual or entity being conducted for the sole purpose of achieving personal benefits \citep{smith2010theory,cropanzano2005self}. In contrast, collaborative behavior refers to actions of individuals or entities being conducted in pursuit of a common goal and thus for the purpose of achieving mutual benefits \citep{deutsch1949theory, fjeldstad2012architecture}. Examples of collaborative behavior include knowledge sharing and contributing to joint problem solving, providing fair peer evaluations, encouraging new members, or the maintenance of shared resources \citep{li2007fairness,chambers2020robust,chiu2015understanding}.}
That is, the incentives designed to encourage effort during idea generation (i.e., the contest prize) introduce a competitive element in the otherwise collaborative idea-evaluation process. This may thus tempt idea generators to violate collaborative community norms and evaluate ideas in strategic self-interest. Embedding contests in online communities appears to have created a mismatch between organizational context and incentives \citep{gallus2022relational}: the organizational context is collaborative yet the incentives are competitive. Such incongruent incentive schemes can backfire with unintended consequences and lower participation. 
% Given the incongruent incentive scheme and the resulting presence of both competitive and collaborative motives, it is not clear how individuals react to this tension and how they evaluate their peers. 
It is not clear how individuals evaluate their peers as they face this tension between competitive and collaborative motives; and in case they act in self-interest and evaluate strategically: who do they target and with what consequences? 

Collaborative aspects of mutual interest, such as helping behavior, reciprocity, and a desire for fairness \citep{wasko2000one,gebauer2013dark,franke2003communities,Bauer2016} are key characteristics of communities which bring individuals with shared interests in member or social welfare together \citep{faraj2011network,majchrzak2013towards}. Shared values and norms play a crucial role in sustaining good community citizenship and mutually interested collaborative behavior \citep{chiu2015understanding,ivaturi2019framing}. 
Individuals may consequently choose to evaluate fairly and truthfully along the meritocratic ideal. 
On the other hand, when contests are embedded in such communities, the self-interested motivation to win the contest prize is in conflict with the mutually interested participation motivation of the community. The economics literature on strategic behavior in contests \citep[see][for a summary]{konrad2009strategy} provides precise predictions for the self-interested evaluation behavior we may expect to see in crowdsourcing contests. It predicts that competitors engage in sabotage---defined as the effort of one individual that reduces the performance of another---and self-promotion---a form of unproductive effort that makes one’s own contributions appear better without increasing the quality of the contributions itself \citep{lazear1981rank,magee2008social,konrad2009strategy}.\footnote{In terms of rating behavior this corresponds to rate oneself up (self-promotion) and to rate competitors down (sabotage). There are other forms of strategic behavior such as strategic entry decisions \citep{liu2014crowdsourcing}, cheap talk \citep{archak2010money}, and reciprocal rating \citep{hutter2011communitition}.}
However, it is not at all clear if these predictions of self-interested behavior from contest theory bear out given the community setting. Indeed, research has shown that community members sometimes act collaboratively even when they are direct competitors \citep{franke2003communities,harhoff2003profiting} and that they are sometimes overly positive in their peer evaluations \citep{aadland2019friends,klapper2023strategic}. 
% How community members resolve this tension is not clear. 
Investigating sabotage and self-promotion together is crucial. While sabotage decreases the quality signal, \mbox{(self-)promotion} increases it and the two may cancel each other out. Sabotage also comes with an externality: beyond the saboteur it also increases the chances of winning for all other competitors \citep{konrad2009strategy}. As a result, the presence of one form of strategic behavior affects the effectiveness of the other. Community members will therefore consider the two forms of strategic behavior jointly as they attempt to resolve the tension between competition and cooperation.

Past work has repeatedly identified such conflicting competitive and collaborative motivations underlying behavior both in online communities and social situations more broadly  \citep{deutsch1949theory,lewis2000exploring}. Self-interested behavior refers to actions of an individual or entity being conducted for the sole purpose of achieving personal benefits \citep{smith2010theory,cropanzano2005self}. In contrast, collaborative behavior refers to actions of individuals or entities being conducted in pursuit of a common goal and thus for the purpose of achieving mutual benefits \citep{deutsch1949theory, fjeldstad2012architecture}. Examples of collaborative behavior include knowledge sharing and contributing to joint problem solving, providing fair peer evaluations, encouraging new members, or the maintenance of shared resources \citep{li2007fairness,chambers2020robust,chiu2015understanding}. These self-interested (competitive) and mutual-interested (collaborative) motivations are often simultaneously present and they interact with each other in complex ways \citep[e.g.,][]{adler2011combining,franke2003communities, Bauer2016,bullinger2010community,hutter2011communitition,nambisan2010different,roberts2006understanding, majchrzak2013towards,chambers2020robust}. 

However, existing studies do not offer precise predictions of how individuals resolve tensions between competitive and collaborative participation motives when asked to evaluate peers in crowdsourcing contests in which they are competing. 
%While recent research on crowdsourcing communities has started to acknowledge the existence of strategic behavior 
Recent research on crowdsourcing communities has started to acknowledge the existence of strategic behavior \citep[e.g.,][]{hofstetter2018should,archak2009optimal,liu2014crowdsourcing,hutter2011communitition,chen2020conan,deodhar2022influence}.
Notably, \citet{klapper2023strategic} highlight that peer evaluations, when transparent, offer the possibility for strategic behavior as they can provide individuals with a platform to shape their own reputation. However, their work is focussed on non-competitive environments with transparent evaluations which is only the case in some peer evaluation settings. 

Therefore we do not fully understand how and why strategic behaviors arise, its dynamics, nor how it affects the structure of the crowdsourcing community via long-term participation \citep[cf.][]{balietti2021incentives} when evaluations are anonymous and competitive. Thus, we complement existing research by offering insights for these manifestations of peer evaluations. The covert nature combined with competition is likely to spur an additional strategic behavior beyond the self-promotion found in prior studies \citep{edelman2015social,klapper2023strategic}, namely sabotage of competitors. Thus, we provide a more comprehensive view on strategic behaviors in crowdsourcing communities that differentiates between self-promotion and sabotage and investigates their complementary effect.

Our paper addresses two research questions: (1) \textit{How do participants in crowdsourcing communities resolve the tension between the competitive and collaborative participation motive when asked to evaluate their peers?}  (2) \textit{How does the way participants resolve the tension change the composition of the community when some individuals are more motivated by the competitive aspects of the contest?} A core moderator for self-interested behavior suggested by contest theory is skill\footnote{Skill in terms of generating high-quality contest entries not in terms of acting strategically, although the two may be correlated.} because it affects the chance to win the contest and thus affects the gain an individual can expect from making strategic peer evaluations \citep{boudreau2016performance}. This suggests that the strength of competitive and collaborative motives may depend on skill and that the tension is greatest for highly skilled participants who are most likely to win the contest, while it is less acute for those of lower skill who know they are unlikely to win. 

We investigate these questions using longitudinal panel data from a leading crowdsourcing community (Threadless), analyzing more than 38 million peer evaluations of 150,000 ideas by 75,000 individuals over 10 years. Using self-interested behavior predicted by contest theory as a framework, we look at self-promotion, sabotage, and skill as a moderator to identify both the culprits as well as the targets of strategic behavior. We then draw on the rich online community literature to explain collaborative behavior that is currently not well explained by contest theory. We also explain how the individual-level use of strategic evaluations translates into important organizational-level outcomes by affecting the community structure through individuals’ long-term participation.

We have two main findings. First, we find behavior that is both consistent with self interest and mutual interest in a rich and nuanced way. Consistent with self-interested behavior predicted by contest theory, we find lower-skilled individuals do not sabotage but self-promote. As their skill level increases, they increasingly adopt more competitive strategies and shift from using self-promotion to sabotaging their closest competitors (other high-skill individuals). On the other hand, we also find signs of collaborative behavior when they show leniency toward those community members who do not threaten their chance of winning. Using insights from two natural experiments---an evaluation rule change and a change in incentives---we rule out alternative explanations and establish that the observed behavior is indeed strategically motivated. Second, we find that the future participation of the targets of sabotage is potentially affected, depending on their skill level. Low-skill targets of sabotage are \textit{less} likely to participate in future contests, while high-skill targets are \textit{more} likely. This suggests a feedback loop between competitive evaluation behavior and future participation. Individuals increasingly act strategically as their skill level increases, and they find the resulting fierce competition so engaging that they increase their future participation even though it makes them the targets of sabotage. 

Our paper makes three key contributions. First, we extend prior research on the competitive and collaborative nature in crowdsourcing \citep{hutter2011communitition,franke2003communities, Bauer2016,nambisan2010different}. Our theorizing explains why community members sometimes wear the competitive hat rather than the collaborative one when they evaluate their peers. Specifically, we theorize that leniency is a crucial collaborative element that allows individuals to better justify self-interested strategic evaluations through a form of moral licensing \citep{blanken2015meta}. Second, we challenge the assumption in much community research that the tension between competitive and collaborative participation motives is a tension across individuals due to stable attributes \citep[some people are competitive while others are collaborative;][]{lakhani2003hackers, reuben2015taste, erat2012white, belenzon2015motivation} by showing that the tension is instead context specific and can be described coherently based on the competitiveness of the situation (which depends on the skill of the individuals involved). Third, our work contributes to our understanding of important organizational-level outcomes by showing how ostensible negative strategic behavior can have positive long-term effects. We theorize how the self-reinforcing dynamic of strategic behavior affects the social structure of communities \citep{kim2018external,faraj2011network,huang2014crowdsourcing,piezunka2019idea,hofstetter2018successive}.

% #####################################################################
%          				     Background
% #####################################################################
% !TEX root = Strategic Behavior V36_OrgSci_Round2_v4.tex

%#############################################
%                  Theoretical Model
% Catalini-OrgSci-Slack.pdf is great example for presentation of formal models with proofs in appendix for OrgSci/Management audience
%#############################################
\section{Contest-Theoretic Intuition} \label{sec:model}

How would evaluation behavior look like in a world without community in which evaluators act only according to self interest?
In this section, we develop a simple theoretical framework to formalize predictions of self-interested evaluation behavior for heterogeneous participants in crowdsourcing contests with an open participation architecture. Like much formal modeling work, our model is a simplified version of what we might expect going on in real contests but which can help build credible behavioral foundations about mechanisms that may explain observed behavior \citep{knudsen2022formal}. %makadok2022formal
%Our model is designed to help us better understand a key aspect of skill heterogeneity: who uses which form of strategic behavior (sabotage and self-promotion) and who are the likely victims of sabotage? The model will then guide our empirical investigation of the mechanism behind the use of strategic behavior when idea selection is delegated to idea generators. 
%It is important to note that we set up the model within the peer-voting in crowdsourcing context to better integrate theory and empirics. The basic premises and predictions of the model are, however, more general and apply to other contests with sabotage and self-promotion. The two key assumptions--(a) incentives from idea generation spill over to idea selection and (b) contest winners are determined through a peer-based mechanism--are, by no means, specific to the crowdsourcing setting. % These assumptions apply to a range of settings, including scientists acting as peer reviewers on journal submissions \citep[e.g.,][]{teplitskiy2018sociology} and grant proposals \citep[e.g.,][]{li2017expertise}, or employees submitting ideas at a given deadline and then have the opportunity to affect the selection outcome in some way \citep[e.g.,][]{reitzig2013biases}.\fxnote{what to do about this last part.} \hl{could cut this part. Interestingly the klapper, piezunka Dahlander paper on strategic behav. has a table these aswell}

% ########################################################
\subsection{Background}
% ########################################################
Sabotage and self-promotion are conceptually similar in that they affect the relative ranking within contests: since the contest winner is determined based on the \textit{relative} rank-order of contestants, any action that increases the likelihood to win for one contestant by necessity implies that the likelihood of other contestants is reduced. The economics literature points to one important difference between self-promotion and sabotage with regard to their effect on the relative ranking \citep{konrad2009strategy}. Self-promotion affects the likelihood of the culprit to win relative to everyone else (the relative likelihood to win among all others remains the same) while sabotage has an important externality. The action of the culprit affects the relative likelihood of the target and \textit{all other contestants} to win: everyone’s likelihood to win is affected, not only the likelihood of the culprit committing the sabotage. As we will show in our economic model, this distinction has important implications on the cost-benefit calculation to determine which form of strategic behavior to employ. In the Appendix (Section \ref{sec:AppendixEconLitRev}) we provide a brief overview of the economics literature on strategic behavior. There we show that while sabotage is well understood theoretically, empirical insights are scarce (especially with regard to the effect of skill) and that so far prior work has not modeled sabotage and self-promotion together. It is thus unclear how the two affect each other.

% Whereas existing work on the behavior and structure of online communities has often used social network analysis methods, we rely on an econometric approach based on a game-theoretic model. We contend that this approach has three advantages. First, contest theory seems to explain many phenomena in crowdsourcing contests well—such as the amount of effort participants will exert \citep{boudreau2011incentives,boudreau2016performance,korpeouglu2018incentives,hu2021joint}. Second, it allows us to disentangle strategic evaluations from truthful evaluations that would be in line with the meritocratic community ideal \citep{felin2017firms,johnson2014emergence}. Third, it allows us to explain the dynamics in crowdsourcing communities with their multiple layers of activities and motivational drivers that a static structural perspective may not reveal \citep{faraj2016special,kilduff2003social}. This is crucial, since skill is thought of as a dynamic construct \citep{stewart2005social,dahlander2011progressing}.

\subsection{Basic Model Setup}
To keep the modeling tractable and the exposition simple, we build on previous models of sabotage in a single contest (so call one-shot contest), actors that are differentiated by skill, and a single winner prize \citep[see, e.g.,][]{harbring2011sabotage, Konrad2000}. We term the participants in the contest ``agents''. The contest consists of three types of agents: high type agents, low type agents, and neutral outsiders. The contest proceeds as follows. First, agents will make contest submissions (every agent makes exactly one submission). Second, the value of each submission is determined through peer evaluation. Third, the submission with the highest value (determined through peer evaluation) wins the single contest prize ($M$). %The contest prize can be monetary but may also include status markers such as symbolic awards.
 
First we look at contest submissions. Low and high types produce a contest submission (i.e., an idea) of low ($b_l$) and high quality ($b_h$), respectively. Without loss of generality, we normalize the evaluation scale to range from 0 and 1. Furthermore, by construction the quality of a low submission is lower than that of a high submission ($0 < b_{l} < b_{h} < 1$). Neutral outsiders are agents who participate only in the evaluation phase of the contest but do not enter their own submission into the contest. As they are not competing for the winner prize, they have no incentive to act strategically and evaluate sincerely, thus they help us establish the ``true quality'' of a submission. 

Second, to determine the value of a submission, we make the simplifying assumption that every agent (each outsider, low, and high type) evaluates every submission. Since every submission thus receives the same amount of evaluations, we can determine the value of a submission by summing up all the evaluations each submission receives. While evaluations can be sincere (i.e., $b_{l}$ or $b_{h}$, respectively), an agent can also sabotage any other agent, evaluating their submission with $0$ or promote them by evaluating their submission with $1$.
As an example, in the case of sincere evaluation by everyone (no promotion and no sabotage), the value of a high type would be $b_h$ times the number of high types ($h$), low types ($l$), and outsiders ($n$; i.e., $v_h = b_h(n+l+h)$). The only other assumption required for our model is that there are fewer high types than low types, and even more outsiders than low types (i.e., $n > l > h$), which should be realistic in most cases.

Third, in order to determine how to act during peer evaluation (i.e., evaluate sincerely, sabotage, or promote) agents consider the cost and benefits of their actions. The benefit for an agent derives from that agents likelihood of winning the contest multiplied by the prize of the contest ($M$)---that is, the incentive of the idea generation phase of the contest. To calculate an agent's likelihood of winning, we rely on established contest literature and for simplicity model this as a Tullock contest \citep{Tullock1980}. In a Tullock contest, the probability of winning is proportional to the value of an agent's submission (i.e., the sum of the evaluations the agent receives on her submission) in relation to the total contest output (i.e., aggregate evaluations of all contest submissions). As far as the costs are concerned, we assume that while evaluating sincerely is free of costs, sabotaging and promoting incurs costs ($c_s$ and $c_p$, respectively). Costs associated with promotion and sabotage might arise from the costs of identifying suitable targets \citep{harbring2007sabotage,Munster2007}, the moral costs associated with lying \citep{abeler2018preferences, gneezy2018lying}, or violating social norms \citep{elster1989social}. In our empirical setting, evaluating is anonymous but this is not an assumption of our model. If evaluation is anonymous, this simply means that the costs associated with promotion and sabotage may be relatively small as there are no reputation costs associated with it (but moral costs may still be exist). 

From this setup, we can already derive four important insights. \textit{Insight 1.} The damage done by sabotage to a high type (evaluating $0$ instead of $b_h$) is larger than the damage done to a low type (evaluating $0$ instead of $b_l$ because $b_l < b_h$ by construction). Since the cost of a single act of sabotage is the same whether a high type or low type are targeted. This suggests that targeting high types is more attractive. \textit{Insight 2.} Following a parallel argument, the benefit gained by promoting a low type is larger than the benefit gained by a high-type (i.e., $1 - b_l > 1 - b_h$). \textit{Insight 3.} Promoting anyone other than oneself is not rational: one would incur the cost of promoting, but has no benefit (in fact, one reduces one's own chance of winning by promoting others). With this setup, it is immediately clear that sabotaging oneself and promoting any agent other than oneself are not helpful in winning the contest prize and no rational agent would engage in such behavior. In the following we thus speak of only ``self-promotion'' and simplify the notation to consider at most one promotion act per contest entry. Together with the previous insight, this suggests that self-promotion will be more attractive to low types. \textit{Insight 4.} All neutral outsiders evaluate others sincerely as they have nothing to gain from strategic behavior (since they did not submit to the contest their chance of winning the prize is zero) yet would incur cost of evaluating strategically. 

% The last point implies that the number of outsiders in the contest affects the effectiveness of any type of strategic behavior performed by the high and low type agents: the more outsiders there are, the lower the effect that any individual strategic evaluation has. 
Regarding the last point, note that neutral outsiders are not necessarily agents who never compete---they simply do not compete in the current contest. We will use this feature to identify strategic behavior in our empirical analysis where agents do not enter every contest: they evaluate as neutral outsiders in some weeks and evaluate as competitors with stakes from idea generation---and incentives to evaluate strategically---in the contest in other weeks.

% ###############################
% ##           Self-Promotion                   ##
% ###############################
\subsection{Self-Promotion}
Self-promotion increases an agent's chance of winning. Since the expected gain from self-promotion is higher for a low type than a high type, there is a cost boundary at which all low type agents decide to self-promote, but none of the high type agents self-promote. If the cost of self-promotion are low enough, all high types will of course self-promote in addition to all low types who will continue to self-promote. Even low types will not self-promote if the cost exceeds their expected gain (see Appendix \ref{sec:AppendixModel} for a formal proof).
% Lastly, the weight that any self-promotion act carries, decreases as the size of the contest---number of all agents who submit---increases. As a result, we would expect fewer agents to self-promote in larger contests. 

% ###############################
% ##                 Sabotage                      ##
% ###############################
\subsection{Sabotage}
The decision to sabotage---and who to sabotage---is more complicated. If an agent sabotages another agent, she decreases that agent's output and thus decreases total contest output, which increases her probability of winning the contest (in the utility function the numerator of the agent's output remains the same while the total contest output in the denominator decreases). However, this decrease in the denominator also benefits all other agents and thus sabotage has---contrary to self-promotion---an important negative externality \citep{Konrad2000}. In the Appendix we first show a proof that the marginal gain from sabotaging one more agent of a given type is increasing in the number of agents of that type that are already being sabotaged. This means that an agent will either sabotage \textit{all} agents of a given type, or \textit{none} of that type. 
Second, we show that high types have more to gain from sabotaging other agents of a given type (all low types or all high types), than low types have to gain from sabotaging those the same agents (see proof in appendix). This result follows from the externality associated with sabotage: the negative externality of sabotage that improves everyone else's chance of winning has a relatively larger impact for low types than high types. Hence, the relative gain from sabotaging other agents is larger for high types than low types. 
Third, we show that both high and low types have more to gain from sabotaging high types than low types, which follows from the fact that the damage done by sabotage to a high type is higher than the damage done to a low type.

A key question now is whether low types or high types find it more attractive to sabotage low types, or low types sabotage high types. That is, at what cost of sabotage will one group start to sabotage the other group (while also considering everyone else's behavior). In the appendix, we go through the exercise of calculating all bounds to establish the precise order according to which groups of agents will decide to sabotage other groups of agents. The bounds reveal that low types have more to gain from sabotaging high types, than high types have to gain from sabotaging low types. The intuition behind this result rests on the fact that damage done by sabotage to high types is higher than the damage done to low types.

% ###############################
% ##                 Predictions                   ##
% ###############################
\subsection{Predictions for Empirical Analysis}
Our model allows us to make several predictions of the self-interested behavior of crowdsourcing participants. The first is that self-promotion is prevalent: most agents, including low type agents, self-promote. The second is that high types are the most likely to sabotage. The third is that high types target each other with their sabotage. In addition to guiding our empirical investigation---in particular with regard to the crucial role of skill---these predictions also serve as a baseline expectation for self-interested strategic behavior which we can use to explore contrasting collaborative behavior. 

\setlength{\parindent}{15pt}

% #####################################################################
\section{Empirical Context}
% #####################################################################

Our empirical setting is Threadless, a prototypical crowdsourcing community \citep{majchrzak2016trajectories}. Threadless is a crowdsourcing and e-commerce platform that hosts weekly T-shirt design contests \citep{Nickell2010}. Since 2001, the site has developed into a leading crowdsourcing platform, pioneering a community-based business model. The company involves its community of 1.5 million members in nearly all aspects of the innovation process and does not employ any in-house designers \citep{Lakhani2008}. The platform draws on the creative talent of designers from across the world and has a distinct focus on building and nurturing an online community of creative individuals \citep{riedl2018learning}. 

% [2023-08-15] OrgSci Round 4: Wow - i forgot we wrote this -- good think we take this out and replace it with "competition vs. cooperation
% Threadless has three organizational features associated with future organizations and work. (1) Open boundaries and sustained crowdsourced value creation \citep{malhotra2021future}: Threadless provides an online platform on which the crowd is empowered to generate as well as pre-select ideas and on which a social network evolves. (2) Lack of formal hierarchy and facilitation of peer feedback \citep{malhotra2021future,riedl2018learning}: Threadless has designed its organization with flat structures, enabling agile knowledge flows and democratized learning through features like peer evaluation and forums in which contributors can provide and receive peer feedback. However, an element of formal hierarchy is that Threadless retains residual control over the final selection of contest winners. (3) Temporal, locational, and affiliative autonomy \citep{malhotra2021future}: Threadless resembles a globally distributed collective of contributors who work both when and where they desire. They also self-select into the Threadless online community and innovation contests; therefore contributors are not employees in the conventional sense. 

% [2023-08-15] OrgSci Round 4: new part
The tension between competition and collaboration is apparent in our study context, the Threadless T-shirt design community. Threadless describes itself as being an ``inspiring design community’’ \citep[][title page]{Nickell2010}. Like in many other crowdsourcing communities \citep{majchrzak2013towards,majchrzak2016trajectories}, Threadless users freely reveal their work by posting draft designs, ask for and provide feedback on designs in the forum, inspire each other with their work, read interviews, and learn from each other \citep{Nickell2010,riedl2018learning}. According to the founder, Threadless is about ``real community, friendships, and working on fun, cool projects together” \citep[][inside cover flap]{Nickell2010}. Community members value participation as an experience in itself with constructive and developmental spirit \citep{brabham2010threadless}, making the community ethos front and center. The community has also developed strong social norms supportive of collaborative behavior \citep{Bauer2016}. Threadless as a company is seen as a member in the community in which it participates rather than being its owner \citep[][p. 45]{Nickell2010}. The designs printed on T-shirts and sold on the community’s e-commerce page are chosen by the community and most of the proceeds from sales are distributed to the community \citep{Nickell2010}. And yet, winning a design contest is coveted. The name of the designer of a winning design is printed on each T-shirt thus giving credit to the individual. Winners are inducted into an elite alumni club, are invited to events, appear in interviews on the community’s blog, and are profiled in books and teaching cases \citep{Lakhani2008,Nickell2010}. Other communities even post public leader boards \citep{grad2022rivalry}. Cash prizes for winning a contest have increased over the years \citep{Nickell2010}. Designers have also been honored in additional \textit{Designer of the Year} and \textit{Most Printed Design} awards \citep{Nickell2010}. Ultimately, being a successful contest winner can be a launch pad for design careers beyond Threadless \citep{Nickell2010}. 

Contests are divided into two distinct and temporally separated stages: (1) entering a contest by submitting a design, and (2) evaluating the submitted designs. To enter a contest, Threadless provides a designer kit and template, and submissions from any standard software program or digitized drawings are accepted, thus creating low barriers to entry. The submission section of the site is static and identical every week. Through a simple form, Threadless enables designers to upload their design, provide a thumbnail, title, and short description. The submission site provides no information as to how many designs have been submitted, or by who, or what their quality might be. At this stage, submissions undergo an editorial review by Threadless staff before they go up for a seven-day rating period.\footnote{Threadless lists several reasons for declining a submission at this stage: designs using copyrighted material, duplication of prior work, inappropriate content, technical errors with the image file such as low resolution, text only designs, designs that use too many colors, are too large or cannot be printed for other reasons. The rating period is seven days with the exception of the change in rules about dropping design earlier which we exploit in the natural experiment on self-promotion in Section \ref{Section:NESelfPromotion}.}

Threadless posts designs publicly on their website for peer rating, using a scale from 0 to 5 stars. The “score submissions” page shows a grid of all submissions in the running with thumbnails, title (possibly trimmed to fit the grid), and username. To navigate designs, participants can toggle between “currently in the running'' and “archive'', the design they submitted themselves, and a single “filter by keyword'' search. Clicking on the thumbnail or title leads to the scoring page of a specific design which shows the full title and description, a thread of comments, and the days remaining of the scoring period. The page also shows a counter of how many others have already scored the design, but no average rating is shown to avoid social influence and herding. The average rating is shown once the scoring period is over and the submission has entered the “archive'', but individual ratings are never revealed. That is, ratings are completely anonymous to the community (but not to us as the researchers).

Among the submissions that received the highest average rating, Threadless typically selects three to six designs per week as contest winners. We show the rating percentile of contest winners in the Appendix (Figure \ref{fig:appendix:ratingDistribution}). On average, contest winners scored in the 95$^{th}$ percentile of all designs submitted that week. The median rank of printed designs is in the 98$^{th}$ percentile. The designs of contest winners are then printed and subsequently sold through the Threadless e-commerce site. Designers receive a cash prize and additional store credit if their design was selected for printing. Threadless does not publish a global ranking with respect to designers’ past design evaluations as other crowdsourcing platforms do (e.g., Topcoder). Threadless generated \$30 million in revenue in 2012 and provided over \$775,000 in prizes over the observation period. We focus on the regular weekly contests from 2001 to 2011, excluding special and themed contests that Threadless also hosts occasionally from our analysis . The data was provided to us by Threadless directly (i.e., it was not scraped from the website) under an NDA. The authors declare that they have no relevant financial or non-financial competing interests to report.

% #####################################################################
\subsection{Identification Strategy} 
% #####################################################################
To empirically identify strategic contest behavior, we leverage the fact that the competition-collaboration tension is most acute for designers who participate in the evaluation when also being active submitters, and less acute when they only participate in the evaluation process and are not themselves competing: 
a designer may evaluate the submissions of others without having her own design in the running in one week, and then enter her own design in the contest \textit{and} evaluate in the next week.\footnote{See Section \ref{sec:ExtendedDescriptives} for more details on the empirical patterns.}
We model the effects of participating in a contest on evaluation behavior using a difference-in-difference approach. That is, we model within individual differences between evaluations cast on competitors versus non-competitors (thus controlling for unobserved individual differences such as, e.g., being a harsh critic in general) and within submission differences between evaluations cast by those who submitted to the contest and those that did not (thus controlling for submission quality). Evaluations cast by an individual in a contest that the individual did not compete in serve as control for those contests in which they did. At the same time, evaluations of the same submission from individuals who are not themselves competing in the contest serve as controls for individuals who are competing in the contest. In terms of our model, a contestant who did not submit a contest entry is a neutral outsider, and a (high/low type) contestant otherwise.

Note that ``neutral outsiders'' are other individuals who submit to contests in general, just not to the one of the current week. Therefore, outsiders are still designers who are in a good position to judge the quality of submissions and not consumers who never submit. Why are outsiders participating in the rating? These outsiders appear to be predominantly motivated to participate in the rating process in order to engage in the community and learn vicariously from the work of others \citep{riedl2018learning}. We make the assumption that these outsiders are ``neutral'' in the sense that they have no direct incentive to evaluate strategically since they do not have anything to gain from doing so while facing moral costs of lying if they evaluated strategically rather than truthfully. This assumption seems particularly plausible given the community focus of Threadless \citep{Lakhani2008}. This assumption not withstanding, it is possible that rivalries among a small set of competitors may cause them to evalute strategically even when not competing \citep{kilduff2010psychology}. %If this is the case, then this should bias our estimates of strategic behavior downward. This is because the within individual counterfactual of ``neutral'' ratings would also be strategic. Our results could then be interpreted as the \textit{additional} strategic behavior when competing compared to some baseline strategic behavior based on rivalries. 
We explore this in the Appendix \ref{sec:dyadFE}.\footnote{Note that fake accounts are not of great concern to our analysis. While fake accounts may exist on Threadless in principle, our analysis includes only ratings cast by designers---users who also make design submission. To qualify as “designer” a user needs to make a design submission that passes Threadless’ basic muster of being a valid submission (a real t-shirt design on the Threadless template, cannot be blank, cannot be an obvious copy of other work, etc.), and have their design submission actually put up for voting.% Outsiders who never make a valid design submission are not included in our data.
}

We model sabotage as the change in probability that individual $i$ submits a 0-star rating---the lowest possible rating---on submission $j$ when having submitted to the same contest relative to not having submitted to the same contest. Conversely, we model self-promotion as the change in probability of rating one's own submission with 5-stars---the highest possible rating---when rating one's own submission versus rating someone else 5-stars.\footnote{By using the two extreme outcomes, 0-stars and 5-stars respectively, we use a conservative measure of sabotage and self-promotion respectively, because (a) we only include the likely report of maximal lies \citep{gneezy2018lying}, and (b) we make sure not to conflate the two behaviors. An alternative specification for sabotage would be, e.g., to use the probability that a design gets assigned a 0 or 1-star rating. However, this would not just include sabotage (all designs that are of higher quality) but also self-promotion (all 0-star designs that get assigned a 1-star rating) and therefore conflate the two behaviors. A similar argument can be made for self-promotion.} The two key explanatory variables are (a) a dummy variable indicating whether an individual is a participant in the contest (i.e., whether $i$ has submitted to the same contest $c$ and is in the running for the prize); and (b) a dummy variable indicating if the individual is rating his or her own submission (this is only possible if the other dummy variable is also 1). That is, we estimate the following equations:

\begin{align}
	%Pr(rating_{ijc}=0) &=  logit^{-1}(\beta_{21}*\text{Submitted to same contest}_{ic} + \alpha_i + \alpha_c + \epsilon_{ijc})	 \label{eq:regression:sabotage} \\
	\text{0-Star Rating}_{ij} &=  \beta_{11} \text{Submitted to same contest}_{ij} + \beta_{12} \text{Rate own submission}_{ij} + \alpha_i + \alpha_s + \epsilon_{ij}	 \label{eq:regression:sabotage} \\
	\text{5-Star Rating}_{ij} &=  \beta_{21} \text{Submitted to same contest}_{ij} + \beta_{22} \text{Rate own submission}_{ij} + \alpha_i + \alpha_s + \epsilon_{ij} 		 \label{eq:regression:selfPromotion} 
	%Pr(rating_{ijc}=5) &=  logit^{-1}(\beta_{11}*\text{Rate own submission}_{ij} + \alpha_i + \alpha_c + \epsilon_{ijc}) 		 \label{eq:regression:selfPromotion} 
\end{align}

\noindent where 0-Star Rating$_{ij}$ is an indicator that is $1$ if the rating submitted by individual $i$ on the submission $j$ is a 0-star rating, 5-Star Rating$_{ij}$ is an indicator that is $1$ if the rating submitted by individual $i$ on the submission $j$ is a 5-star rating, $\alpha_i$ are individual-level and $\alpha_s$ submission-level fixed effects, and $\epsilon_{ij}$ are error terms.

%To estimate heterogeneous effects by skill level, we include measures for source skill (agent casting the rating) and target skill (agent entering the submisison being rated), 
We use this approach to separately identify the two types of strategic behavior outlined in the theoretical model above:
  \begin{itemize}
	\item \textit{Self-promotion:} If a 5-star vote is assigned to one's own submission, this indicates self-promotion ($\beta_{22}$ in Eq.~\ref{eq:regression:sabotage}). We provide robustness tests using a natural experiment to address concerns that this would not reflect strategic behavior but rather the result of overconfidence or an increased preference fit in Section \ref{Section:NESelfPromotion}
	\item \textit{Sabotage:} Sabotage is indicated if 0-star votes are assigned more liberally when competing in the same contest versus being an outsider who has no stakes in the contest (controlling for how likely the submission is to receive 0-stars overall; $\beta_{11}$ in Eq.~\ref{eq:regression:sabotage}). We provide results from a second natural experiment to provide evidence that low ratings are strategic rather than a result of endogenous entry into contests in Section \ref{Section:NESabotage}
	% \item \textbf{Withholding of top-scores:} In addition to the sabotage and self-promotion effects discussed in our model, we also investigate a weaker form of sabotage which is indicated when individuals withhold a top-score when (coefficient of interest: $\beta_{11}$) when an individual is competing in the same contest. This allows us explore additional richness in the empirical analysis while keeping the theoretical model tractable. 
\end{itemize} 

We estimate our models as linear probability models for ease of interpretation \citep{Angrist2001,Greene2012}. 
%Linear probability models are widely used as useful approximations of approaches that model limited dependent variables explicitly \citep{Heckman1997}. 
A discussion of the usefulness of this approach can be found in \cite{Angrist2001} and \cite{Moffitt2001}. Furthermore, notwithstanding the limitations of this approach, efficient estimation approaches for OLS exist for large datasets that support demeaning of multiple fixed effects. We estimate linear probability models using the lfe package for R \citep{Gaure2013}, which supports demeaning of multiple fixed effects. This algorithmic approach is mandatory as our dataset is extremely large with more than 38 million observations, 74,525 individual fixed effects, and 154,086 submission fixed effects (or 511 contest fixed effects).%\footnote{The naive approach of estimating the model with dummy variables for the two fixed effects is computationally intractable even using advanced high-performance computing.}

To investigate how sabotage and self-promotion vary across skill levels we compute skill as the time lagged average submission quality.\footnote{We use the quality of the best previous submission as a robustness check which is consistent with our results, see Section \ref{appendix:Het} in the Appendix.} We compute skill for both the individual casting the rating (we refer to this as the \textit{source}) and the individual who made the contest entry being evaluated (we refer to this as the \textit{target}). 
We then estimate versions of the models in Equation \ref{eq:regression:selfPromotion} and \ref{eq:regression:sabotage} in which we include source and target status and their interactions with the \textit{Submitted to same contest} and \textit{Rate own submission} dummy variables. Since target skill is time-invariant at the submission level, the main coefficient drops out. % We compute \textit{Contest size} as the number of submissions to a contest (z-scored). 

% #####################################################################
\subsection{Data} 
% #####################################################################
Our data consists of an unbalanced cross-panel of 74,525 individuals participating in 511 contests, with over 38 million ratings on 154,086 contest entries (see Table \ref{tab:countVotes}). The skill distribution is right skewed (Appendix figure \ref{fig:ContestSize}a) and thus, our empirical setting matches our model, which assumes fewer high types than low types. The average contest size is 407 submissions (Appendix figure \ref{fig:ContestSize}b shows the entire distribution).%\footnote{The bimodal distribution of contest sizes is an artifact of doubling contest sizes in the early years (2001-2005) and a smaller increase in the later years.}  
We show descriptive statistics and correlations of main variables in Table \ref{descriptivesEconStyle}.

\textbf{% latex table generated in R 3.2.2 by xtable 1.7-4 package
% Thu Nov  2 17:58:59 2017
\begin{table}[ht]
\centering
{\scriptsize
\begin{tabular}{rrr}
  \hline
  & \multicolumn{2}{c}{Rate own  Submission} \\
\cline{2-3} \\[-3px]
Submitted to same contest & No & Yes \\ 
  \hline
No & 30,246,070 (13.58\%) &   \\ 
  Yes & 7,772,772 (10.73\%) & 114,914 (74.58\%) \\ 
   \hline
\end{tabular}
}
\caption{Summary of rating behavior. Parentheses show probability of rating conditional on submitting.}
\label{tab:countVotes}
\end{table}
 }

%\fxnote{Think about showing rating distribution for three subgroups here (even when not very telling because differences become only apparent when considering skill). Alternatively we could show rating distribution across different skills?}

% latex table generated in R 3.6.2 by xtable 1.8-4 package
% Sat Feb  5 13:40:36 2022
\begin{table}[ht]
\centering
\begingroup\scriptsize
\begin{tabular}{rrrrrllllll}
  \hline
 & Mean & SD & Min & Max & (1) & (2) & (3) & (4) & (5) & (6) \\ 
  \hline
\textit{Rating} (1) & 1.93 & 1.53 & 0.00 & 5.00 &  &  &  &  &  &  \\ 
  \textit{Submitted to same contest} (2) & 0.21 & 0.41 & 0.00 & 1.00 &  0.03  &  &  &  &  &  \\ 
  \textit{Rate own submission} (3) & 0.00 & 0.05 & 0.00 & 1.00 &  0.11  &  0.11  &  &  &  &  \\ 
  \textit{Contest size (4)} & 406.73 & 131.88 & 2.00 & 848.00 &  0.07  &  0.11  &  0.00  &  &  &  \\ 
  \textit{Average score in contest} (5) & 1.93 & 0.23 & 1.26 & 2.76 &  0.15  &  0.04  &  0.00  &  0.50  &  &  \\ 
 \textit{Source skill} (6) & 2.26 & 0.61 & 0.00 & 5.00 &  0.02  &  0.05  &  0.00  &  0.17  &  0.29  &  \\ 
  \textit{Target skill} (7) & 2.28 & 0.57 & 0.00 & 5.00 &  0.27  & -0.01  & -0.01  &  0.13  &  0.29  &  0.08  \\ 
   \hline
\end{tabular}
\endgroup
\caption{Descriptive statistics and correlations of main variables. All correlations are significant at $p < 0.001$.}
\label{descriptivesEconStyle}
\end{table}

\section{Results}

%##############################
% \subsection{Strategic Behavior}
%##############################
We start with baseline estimates of a difference-in-difference model estimated (Table \ref{table:mainRegression}).\footnote{We cluster standard errors on the submission level to account for possible dependence of ratings of the same submissions and in the same contest. One may alternatively be concerned about capturing auto-correlation at the rater level. We examine the robustness of our specification in \ref{tab:HetEffectMeanIndivLevel}. The clustered standard errors are virtually identical to robust standard errors. Furthermore, $p$-values are generally much smaller than $0.001$ (i.e., $1e-9$) so that our substantial conclusions are not affected by the level of clustering of standard errors.}
With a continuous measure of ratings as dependent variable (Model 1), we find no signs of sabotage but rather slight leniency toward ones' competitors ($\beta = 0.024$; $p < 0.001$) and strong signs for self-promotion ($\beta=2.820$; $p < 0.001$). Moving to a linear probability model (Model 2) we find also find overall leniency: a slightly lower likelihood to rate 0-stars when an individual submitted to the same contest ($\beta = -0.008$; $p < 0.001$). That is, community members do not appear to act strategically against their competitors but instead show leniency. 
% However, the likelihood to sabotage increases significantly in larger contests (Model 3: $\beta = 0.008$; $p < 0.001$). 
Note that our model with individual and submission fixed effects controls for the possibility that some contests include more low-quality submissions that deserve 0-stars. Next, we estimate a linear probability model on casting a 5-star rating. We find a significant level of self-promotion (Model 3: $\beta = 0.855$; $p < 0.001$). The effect of self-promotion is extremely strong. Seventy-five percent of individuals rate their own submissions and the overwhelming majority of them does so with the highest possible rating: 97\% of self-votes are 5-stars. Given that the average rating for non-self-ratings is only 1.80 stars, this indicates that contestants are self-promoting with dramatically inflated ratings.

\begin{table}[h!]
\begin{center}
\begin{scriptsize}
\begin{tabular}{@{\extracolsep{5pt}}l D{.}{.}{4.4} D{.}{.}{4.4} D{.}{.}{4.4} }
\toprule
 & \multicolumn{1}{c}{OLS} & \multicolumn{2}{c}{Linear Probability} \\
 \cline{2-2} \cline{3-4} \\[-5pt]
 Dependent Variable: & \multicolumn{1}{c}{\bf Rating} 	& \multicolumn{1}{c}{\bf Sabotage} 	 & \multicolumn{1}{c}{\bf Self-Promotion} \\
				 & 							 & \multicolumn{1}{c}{0-Star Rating} & \multicolumn{1}{c}{5-Star Rating} \\
 \cline{2-2} \cline{3-3} \cline{4-4} \\[-5pt]
 & \multicolumn{1}{c}{(1)} & \multicolumn{1}{c}{(2)} & \multicolumn{1}{c}{(3)} \\
\midrule
Submitted to same contest: Yes                       	& 0.024^{***} & -0.008^{***} & -0.002^{***}  \\
                                                     			& (0.001)     & (0.000)          & (0.000)          \\
Rate own submission: Yes                             	& 2.820^{***} & -0.195^{***}  & 0.855^{***}   \\
                                                     			& (0.003)     & (0.001)          & (0.001)        \\
\textit{Individual} 		& \multicolumn{1}{c}{\textit{Fixed}} & \multicolumn{1}{c}{\textit{Fixed}}& \multicolumn{1}{c}{\textit{Fixed}}  \\
\textit{Submission}		& \multicolumn{1}{c}{\textit{Fixed}} & \multicolumn{1}{c}{\textit{Fixed}}& \multicolumn{1}{c}{\textit{Fixed}}	 \\
\midrule
Adj.~R$^2$                              & 0.382       & 0.372       & 0.210            \\
Num.~obs.                                            & \multicolumn{3}{c}{38,102,880}  \\
\bottomrule
\multicolumn{4}{l}{\tiny{$^{***}p<0.001$, $^{**}p<0.01$, $^*p<0.05$}}
\end{tabular}
\end{scriptsize}
\caption{Estimates for strategic behavior in contest. Standard errors are in parentheses, clustered at the submission level.}
\label{table:mainRegression}
\end{center}
\end{table}
	% SE are virtually identical to robust SE

% #####################################
% ##                   Heterogeneity by Skill            ##
% #####################################
% \subsection{Heterogeneous contestant skill: High-skilled target other high-skilled}

Next, we investigate if there is heterogeneity in the use of strategic behavior among those of different skill. We estimate a variation of Equation \ref{eq:regression:selfPromotion} and include measures of idea generation skill for both the source and the target of the rating, an interaction between source and target skill, and interaction terms between the \textit{Submitted to same contest} dummy and source and target skill, respectively (Table \ref{tab:HetEffect}).\footnote{This analysis relies on a reduced sample as we need to exclude (a) all ratings that an individual casts before making their first submission (because for those individuals we cannot compute idea generation skill), and (b) all ratings on every contestant's first submission (again, because idea generation skill is still unknown for those individuals). We check whether the use of subsamples drives our findings in Section \ref{appendix:Het} in the Appendix but find no evidence of this.} We find significant heterogeneity in sabotaging behavior. We find that higher-skill individuals are more likely to be the source of sabotage (Model 1: $\beta = 0.012$; $p < 0.001$) and are more likely to be the target of sabotage ($\beta = 0.019$; $p < 0.001$). This pattern mirrors the predictions of our theoretical model. 
% Controlling for skill, past winners are also more likely to commit sabotage (Model 2: $\beta = 0.012$; $p < 0.001$) and more likely to be targets of sabotage ($\beta = 0.013$; $p < 0.001$).
Turning to self-promotion, we estimate a similar model using the \textit{Rate own submission} dummy variable but include only the interaction with source skill, since source and target are the same in the case where a contestant is voting on themselves. We find that higher-skill individuals are significantly less likely to self-promote (Model 2: $\beta = -0.055$; $p < 0.001$) compared to lower-skill contestants. This matches our model prediction that low-skill agents are more likely to engage in self-promotion. 
% We also find that past winners are less likely to self-promote (Model 4: $\beta = -0.068$; $p < 0.001$).

\begin{table}[h!]
\begin{center}
\begin{scriptsize}
\begin{tabular}{@{\extracolsep{5pt}}l D{.}{.}{3.3} D{.}{.}{3.3}   }
\toprule
 & \multicolumn{2}{c}{Linear Probability} \\
 \cline{2-3}\\[-5pt]
 Dependent Variable: & \multicolumn{1}{c}{\bf Sabotage} 	 & \multicolumn{1}{c}{\bf Self-Promotion} \\
				 & \multicolumn{1}{c}{0-Star Rating}   & \multicolumn{1}{c}{5-Star Rating} \\
				\cline{2-2} \cline{3-3} \\[-5pt]
				& \multicolumn{1}{c}{(1)} & \multicolumn{1}{c}{(2)} \\
\midrule
Submitted to same contest: Yes            & -0.071^{***} & 0.001^{***}   \\
                                                              & (0.001)            & (0.000)        \\
Rate own submission: Yes                    & -0.194^{***} & 0.957^{***}    \\
                                                              & (0.001)      & (0.004)          \\
Source skill                                            & -0.028^{***} & -0.007^{***} \\
                                                              & (0.001)           & (0.001)          \\
Source skill $\times$ Target skill            & -0.005^{***} & 0.008^{***}    \\
                                                              & (0.000)        & (0.000)      \\
% Source is past winner                            &              & -0.031^{***} &              & 0.016^{***}  \\
%                                                               &              & (0.000)      &              & (0.000)      \\
% SourceIsAlumniFALSE:TargetIsAlumniTRUE                        &              & 0.015^{***}  &              &              \\
%                                                               &              & (0.004)      &              &              \\
% SourceIsAlumniTRUE:TargetIsAlumniTRUE                         &              & 0.012^{**}   &              &              \\
%                                                               &              & (0.004)      &              &              \\
Submitted to same contest: Yes \\
\quad $\times$ Source skill    		& 0.012^{***}             &              \\
                                                           & (0.000)                &              \\
\quad $\times$ Target skill        		& 0.019^{***}             &              \\
                                                       	& (0.000)                &              \\
% \quad $\times$ Source is past winner &              & 0.012^{***}  &              &              \\
%                                                               &              & (0.000)      &              &              \\
% \quad $\times$  Target is past winner &              & 0.013^{***}  &              &              \\
%                                                              &              & (0.000)      &              &              \\
Rate own submission: Yes \\
\quad $\times$ Source skill              &                           & -0.055^{***}  \\
                                                              &                           & (0.002)          \\
% \quad $\times$ Source is past winner       &              &              &              & -0.068^{***} \\
%                                                               &              &              &              & (0.002)      \\
\textit{Individual} 		& \multicolumn{1}{c}{\textit{Fixed}} & \multicolumn{1}{c}{\textit{Fixed}}  \\
\textit{Submission}		& \multicolumn{1}{c}{\textit{Fixed}} & \multicolumn{1}{c}{\textit{Fixed}}  \\
\midrule
% Num. obs.                                                     & 27188751     & 27188751     & 27188751     & 27188751     \\
% R$^2$ (full model)                                            & 0.395        & 0.396        & 0.249        & 0.249        \\
% R$^2$ (proj model)                                            & 0.002        & 0.002        & 0.051        & 0.051        \\
Adj. R$^2$                                        & 0.391        	    & 0.243       	      \\
% Adj. R$^2$ (proj model)                                       & -0.006       & -0.005       & 0.044        & 0.044        \\
% Num. groups: UserID                                           & 50709        & 50709        & 50709        & 50709        \\
% Num. groups: SubmissionID                                     & 154077       & 154077       & 154077       & 154077       \\
Num.\ obs.                                                    & \multicolumn{2}{c}{18,787,584} \\
\bottomrule
\multicolumn{3}{l}{\tiny{$^{***}p<0.001$; $^{**}p<0.01$; $^{*}p<0.05$}}
\end{tabular}
\end{scriptsize}
\caption{Estimates for strategic behavior with skill heterogeneity. Standard errors are in parentheses, clustered at the submission level.}
\label{tab:HetEffect}
\end{center}
\end{table}

\begin{figure}[t!] % htbp
	\begin{center}
		\includegraphics[width=0.49\linewidth,valign=c]{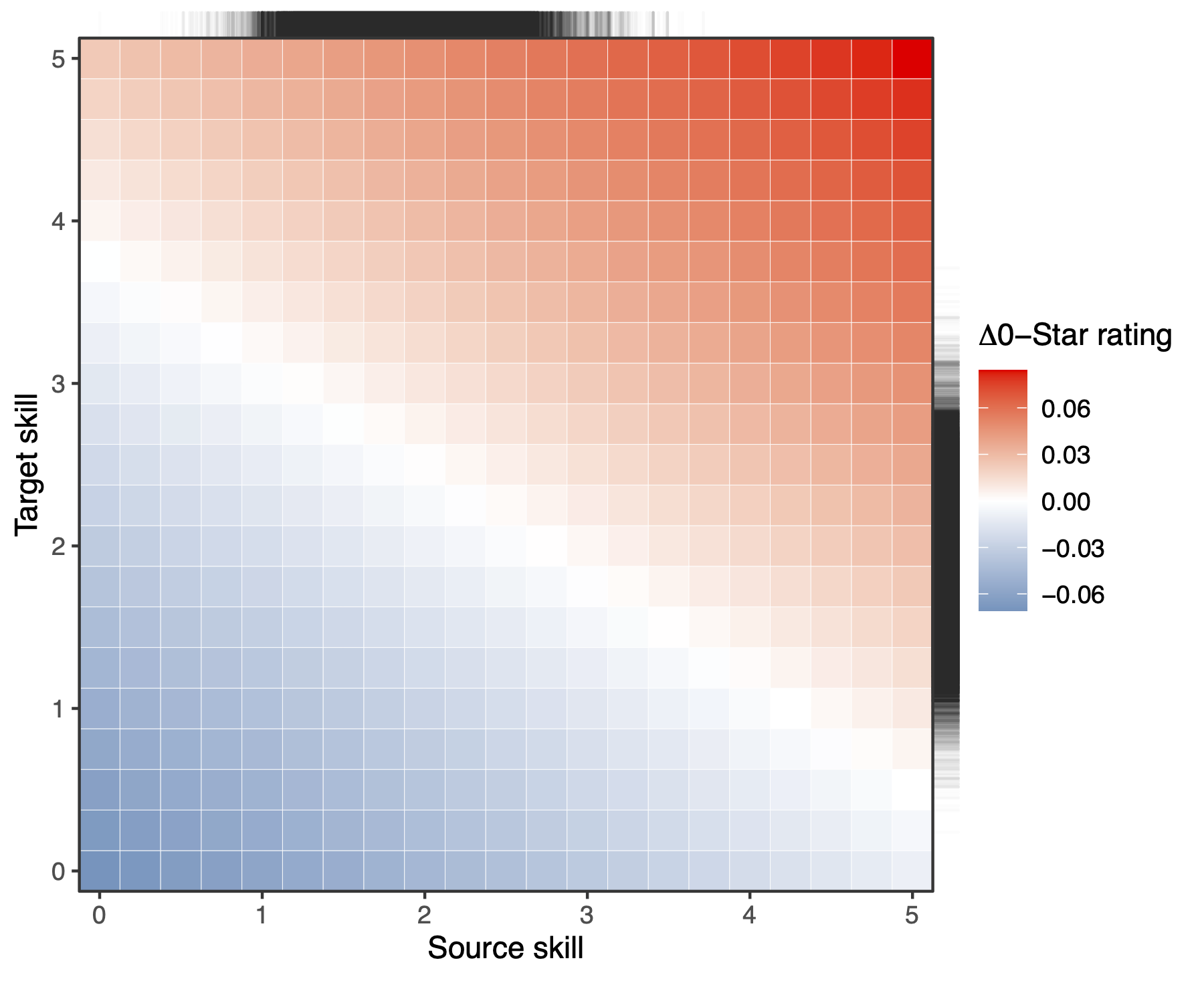}
		\includegraphics[width=0.40\linewidth,valign=c]{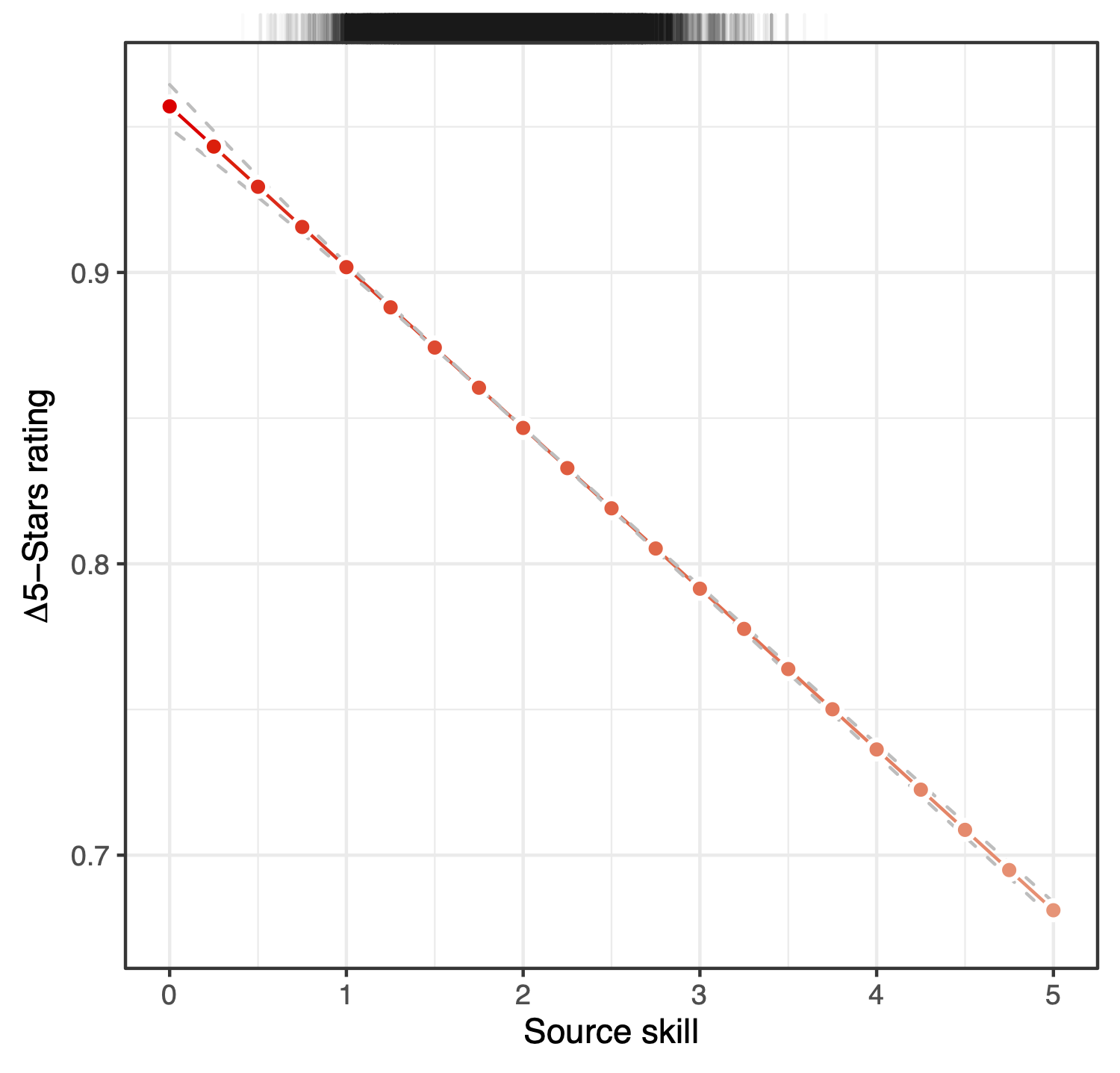}
		\caption{Strategic behavior by competitors of heterogeneous skill levels.}
		\begin{flushleft}
		\begingroup
		\leftskip4em
		\rightskip\leftskip
		\footnotesize{\textit{Note: }For these figures the mean submission quality is used as measure of skill. 
			\textbf{Panel A (Sabotage).} Relative change in probability of rating 0-stars when competing compared to not competing across skill levels. Outer rugs show distribution of data. There are  4,512 (1,091) observations for source (target) skill $\geq 3.5$.
			\textbf{Panel B (Self-promotion).} Relative change in probability of rating 5-stars when rating own submission compared to submissions by others of same skill (error band is 95\% confidence interval).} 	% CONFIRMED: Check with code, it is 95% CI not SE
		\par
		\endgroup
		\end{flushleft}
		\label{fig:HetMean}
	\end{center}
\end{figure}

To better interpret the heterogeneity across skill levels, we compute predicted values for the expected \textit{change} in the likelihood to rate 0-stars when competing versus not competing (Figure \ref{fig:HetMean}). The heat map in Panel A shows that high-skill contestants target other high-skill contestants with an up to 6\% increased likelihood of assigning a 0-star rating when competing compared to their baseline when not competing. The null-line marking the start of sabotaging behavior is not perfectly symmetric: high-skill contestants are sabotaged by contestants of all other skill levels. This matches our theoretical predictions that low types sabotaging high types is the second most lucrative form of sabotage (after high types sabotaging high types). We similarly compute the relative \textit{change} in probability of self-promotion (since source and target skill are the same in the case of self-rating, this analysis is effectively the diagonal of the heatmap in Panel A). Panel B in Figure \ref{fig:HetMean} shows that the likelihood that contestants engage in self-promotion decreases with their own skill. This finding is in line with the predictions from our theoretical model that self-promotion is less effective among high types.\footnote{We perform three important robustness tests. 
First, we repeat the same analysis using the quality of the \textit{best} design an individual submitted in the past (instead of the average of all past submissions) and find almost identical coefficients (Appendix Table \ref{tab:HetEffectMax} and Figure \ref{fig:HetMax}). 
Second, we explore alternative measures of skill: tenure, experience, and being a past winner (Appendix Table \ref{tab:Mechanisms}). We find no significant effect of tenure. We find mixed results for experience . We find a lower likelihood to sabotage for those with more evaluation experience but higher likelihood to sabotage for those with more submission experience. This is consistent with the interpretation that competitive motivation is the driver behind strategic behavior (i.e., those who are competitively motivated and make many submissions, sabotage more while those who may be more socially motivated and submit many evaluations). We find that past winners act significantly more strategically. They are both much more likely to sabotage, and other past winners are the targets.
Being a past winner could be seen as socially validated skill. 
Third, we explore whether the reported effect of strategic behavior could be explained by dyadic rivalries instead (\citet{kilduff2010psychology}; Section \ref{sec:dyadFE}). We find that sabotage happens outside, and on top of, any pre-existing dyadic rivalries. Even within the same dyad, we still find a difference in individuals' rating behavior between competing for the same prize vs.~not competing which indicates strategic behavior. In summary, dyadic rivalry patterns are not sufficient to explain strategically motivated sabotage.}

As further evidence for the use leniency as mutual-interested compensation for self-interested strategic evaluation, the Appendix (Table \ref{tab:Leniency}) shows regressions on the contest-level using standard deviation of the ratings that individuals cast as dependent variable. Those regressions show that when individuals have submitted to the same contest, they make better use of the full rating spectrum and submit ratings with a higher standard deviation. The effect is amplified by skill so that higher-skilled individuals submit ratings with an even higher standard deviation. This suggests that leniency is a phenomenon within a single contest: individuals accumulate credits on an invisible scorecard by promoting some community members (giving them higher ratings than they deserve) and then spend these credits on self-interested strategic evaluations (giving them lower ratings than they deserve).

% #####################################
% ##            Mechanisms of Sabotage             ##
% #####################################
\section{Mechanisms of Strategic Behavior}
Why do some community members behave strategically? Is it really driven by incentives of the idea generation phase spilling over to idea evaluation or are there other alternative explanations? Exploring our data further, we show that 
\begin{inparaenum}[(a)]
	\item sabotage is strategically motivated as a result of idea generation incentives spilling over to the idea evaluation,
	\item self-promotion is strategically motivated and not just a result of overconfidence,
	\item those who act strategically mostly use sabotage and self-promotion together rather than substituting one with the other, and finally
	\item we briefly explore the effect that strategic behavior has on the selection of contest winners.
\end{inparaenum}

% ######################################
% ##         Incentives spill over nat.exp.             ##
% ######################################
\subsection{Sabotage is a result of idea generation incentives}\label{Section:NESabotage}
If peer evaluation is strategically motivated by idea generation incentives spilling over to idea evaluation, we would expect an increase in the probability to assign 0-stars in contests with higher incentives \citep[or more precisely, the prize spread between the contest winner and the contest loser(s); e.g.,][]{Lazear1989,harbring2011sabotage}. Conversely, if idea evaluation is not strategically motivated, we would expect no change or possibly even a decrease in the probability to rate 0-stars after the incentive change because submission quality can be expected to increase as a result of increased effort due to the incentive effect \citep{jeppesen2010marginality}. We leverage a platform change as a natural experiment---a change in prize money awarded to contest winners. The prize for winning the weekly contest doubled from \$500 to \$1,000 in 2005 (see Figure \ref{fig:BlogPrize} in the Appendix). This doubling of the contest prize was announced on the company's blog on the day it took effect and therefore, community members had no prior knowledge of it and were not able to withhold submissions or otherwise alter their behavior in anticipation of it.%
\footnote{Note that this natural experiment also helps us address another potential alternative explanation. An alternative explanation for the observed negative rating behavior could be seen in the endogenous decision to enter contests \citep[see, e.g.,][]{Bockstedt2016}. Contestants may submit entries to contests where they expect that competition is weak and, therefore, deserve lower ratings. Following the argument of endogenous entry we would expect either no change in 0-star ratings as contestants should not be able to time their contest entry around the incentive change. 
}

We use a difference-in-difference design to identify the causal effects of the incentive changes. We use a 6-months time window before and after the incentive change. We estimate the same linear probability model for 0-Star ratings with individual- and submission-level fixed effects as before (Table \ref{tab:Natural}). We find a significant 2.2\% increase in the probability to rate 0-stars when competing after the incentive change (Model 1: $\beta = 0.022$; $p < 0.001$). The effect is stronger (2.3\%) in the first quarter after the incentive change and slightly weaker (2.0\%) in the second quarter (Model 2: $\beta = 0.023, \beta = 0.020$; both $p < 0.001$). Furthermore, we find that submission quality (as determined by ratings of neutral outsiders who have no incentive to rate strategically) is significantly higher after the incentive change ($\mu_{before} = 1.18$, $\mu_{after} = 1.39$; t-test $p < 0.001$). This indicates that the incentive change, as intended, indeed increases the average quality in the idea generation phase but also increases sabotage in the idea evaluation phase. A placebo test in which we selected a fake date four weeks before the actual incentive change provides further evidence for the causal impact of the incentive change on the rating behavior (Model 3: $\beta = -0.009$; $p < 0.001$) as it rather indicates a negative time trend.

\begin{table}[h!]
\begin{center}
\begin{scriptsize}
\begin{tabular}{@{\extracolsep{5pt}}l D{.}{.}{4.3} D{.}{.}{4.3} D{.}{.}{4.3} }
\toprule
 & \multicolumn{3}{c}{Linear Probability} \\
 \cline{2-4}\\[-5pt]
 Dependent Variable: & \multicolumn{3}{c}{\textbf{Sabotage:} 0-Star Rating} \\
 \cline{2-4} \\[-5pt]
 & \multicolumn{1}{c}{Base} & \multicolumn{1}{c}{Persistence} & \multicolumn{1}{c}{Placebo Test}  \\
 \cline{2-2}  \cline{3-3} \cline{4-4}  \\[-5pt]
 & \multicolumn{1}{c}{(1)} & \multicolumn{1}{c}{(2)} & \multicolumn{1}{c}{(3)}  \\
\midrule
Submitted to same contest: Yes                         & -0.037^{***} & -0.037^{***} & -0.035^{***} \\
                                                          			& (0.000)      & (0.000)      & (0.000)      \\
Rate own submission: Yes                                 	& -0.179^{***} & -0.179^{***} & -0.179^{***} \\
                                                          			& (0.000)      & (0.000)      & (0.000)      \\
Submitted to same contest: Yes\\
\quad $\times$ After	& 0.022^{***}  &              & 0.028^{***}  \\
                                                          			& (0.000)      &              & (0.000)      \\
\quad $\times$ 1st quarter after &              & 0.023^{***}  &              \\
                                                          &              & (0.000)      &              \\
\quad $\times$ 2nd quarter after &              & 0.020^{***}  &              \\
                                                          &              & (0.000)      &              \\
\quad $\times$ Fake after        &              &              & -0.009^{***} \\
                                                          &              &              & (0.000)      \\
\textit{Individual} 		& \multicolumn{1}{c}{\textit{Fixed}} & \multicolumn{1}{c}{\textit{Fixed}}& \multicolumn{1}{c}{\textit{Fixed}} \\
\textit{Submission}		& \multicolumn{1}{c}{\textit{Fixed}} & \multicolumn{1}{c}{\textit{Fixed}}& \multicolumn{1}{c}{\textit{Fixed}} \\
\midrule
%R$^2$                                                     & 0.405        & 0.405        & 0.405        \\
Adj.\ R$^2$                                               & 0.390        & 0.390        & 0.390        \\
Num.\ obs.                                                & \multicolumn{3}{c}{825,504}  \\
\bottomrule
\multicolumn{4}{l}{\tiny{$^{***}p<0.001$, $^{**}p<0.01$, $^*p<0.05$}}
\end{tabular}
\end{scriptsize}
\caption{Natural experiments of sabotage behavior after an incentive changes from \$500 to \$1,000 prize money using $\pm 6$ months an observation windows. Standard errors in parentheses, clustered at the submission level.}
\label{tab:Natural}
\end{center}
\end{table}

% ######################################
% ##        Self-promotion vs. Overconfidence    ##
% ######################################
\subsection{Self-promotion is strategically motivated and not only overconfidence}\label{Section:NESelfPromotion}
Why do community members evaluate their own submissions so highly? High evaluations for one's own submission could reflect overconfidence of contestants in their own abilities \citep{Benabou2002, Camerer1999b} or an increased preference fit of one's own creative work compared to the work of others \citep[e.g.,][]{berg2016balancing, Franke2010}. To explore the mechanism behind the high self-evaluation, we leverage another natural experiment around a change in the scoring mechanism. 

Before the change, every contest entry was posted on the website for the full 7-day rating period. After the change in 2005, contest entries with a rating below 1.5 stars were eliminated from the rating process before the 7-day rating period was complete (see Figure \ref{fig:BlogScoring} in the Appendix).\footnote{Specifically, contest entries that had received ratings from 100 different people but the average of those rating was less than 1.5 stars were eliminated. Note that the average rating still was not visible during rating process but only to the platform owner. This rule change was introduced to help the platform owner to focus on a smaller set of contest entries by removing low quality submissions from the contests earlier.} The change was announced and immediately implemented. If rating one's own submission is not intended strategically but only reflect biased perception and overconfidence, we should not see any change in rating behavior around this rule change. If, however, rating on one's own submission is intended strategically, such an evaluation should happen earlier to maximize the chance to progress past the 100 vote / 1.5 stars cutoff. 

\begin{figure}[htbp]
	\begin{center}
		\includegraphics[width=0.49\linewidth]{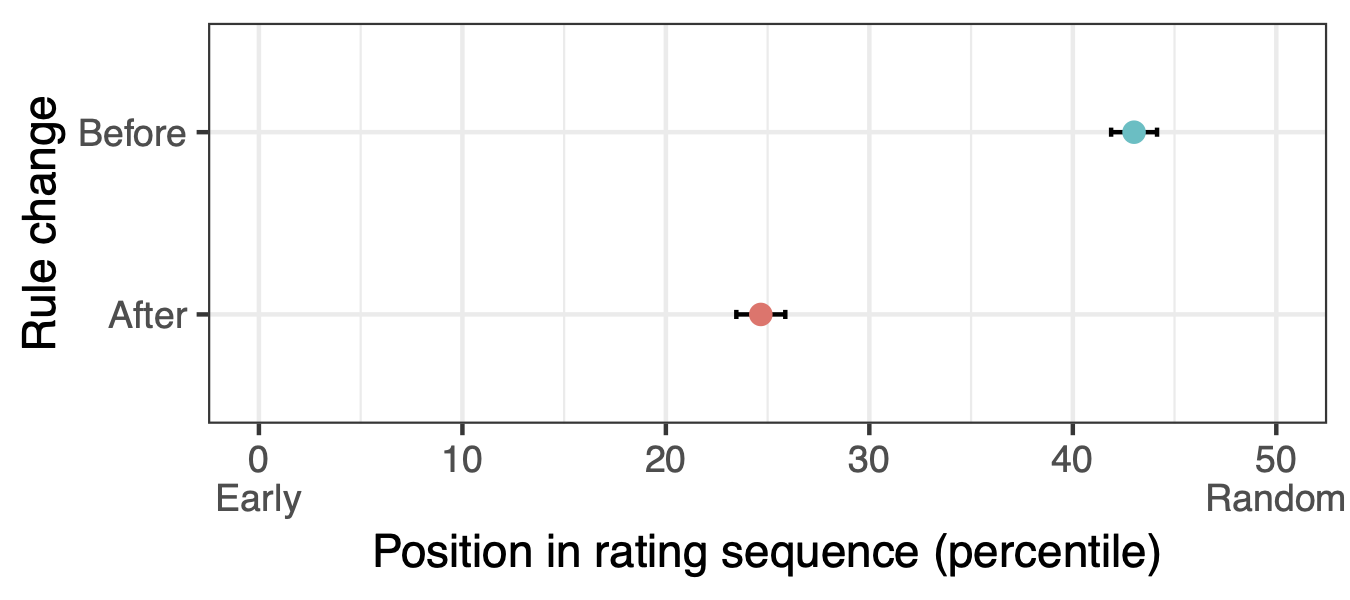}
		\caption{Analysis of Rating Sequence.}
		\vspace{12pt}
		\small{\textit{Note: }Self-rating happens significantly earlier after the rule change.}	
		\label{fig:votingRule}
	\end{center}
\end{figure}

We analyze the sequence of evaluations for each contest entry and test when in the sequence the self-promoting evaluation was cast (Figure \ref{fig:votingRule}). By focusing on the relative position in the evaluation sequence we control for any behavior change that might occur for community members evaluating the same contest entry. We focus our analysis on the 20-week period around the evaluation rule change ($\pm 10$ weeks before/after the rule change). We find that before the rule change, self-evaluation happens roughly in the middle of the sequence (43rd percentile). That is, self-evaluations were cast roughly on a random day during the evaluation period. After the rule change, self-evaluations are cast significantly earlier, falling in the 25th percentile of the evaluation sequence. Our analysis of this natural experiment suggests that evaluating one's own submission has a strategic intention (make it past the 1.5 stars cutoff) rather than simply a reflection of overconfidence or preference for one's own submission.

% #####################################
% ##                       Substitution                        ##
% #####################################
\subsection{Sabotage and self-promotion are mostly used together}\label{sec:SourcesOfSabotage}

Do individuals use sabotage and self-promotion together to maximize the impact of their strategic behavior or do they substitute one for the other, maybe to atone for moral cost? As noted before, investigating this question empirically is difficult as we can identify sabotage only probabilistically through the difference-in-difference approach. However, we can get some meaningful leverage to identify possible substitution by using self-promotion (or rather the lack thereof) as our starting point. Remember that self-promotion is very common: 86\% of individuals who had a chance to self-promote (i.e., submitted to a contest and then rated at least some of their competitors) chose to do so and 97\% of those self-ratings are 5-stars. Conversely, only 14\% of individuals who had a chance to self-promote did not do so.

Using the same modeling approach with individual- and submission-level fixed effects as in the main analysis, we find (Table \ref{tab:Substitution}) that the likelihood to rate the competition with 0-stars if one has also self-promoted in the same contest is significantly lower than for those who did not (Model 1, $\beta = -0.002; p < 0.01$; note that having self-promoted in a contest implies also having submitted to that contest). This suggests that some individuals do in fact sabotage somewhat less when they have self-promoted. Note, however, that this coefficient estimate captures the effect of marginal individuals: individuals who only sometimes self-promote but not always (as otherwise the individual fixed effect captures this time-invariant behavior). As such, the estimate may mask the more specific effect among that group of more strategic individuals who always self-promote and for which the model cannot pick up variation in behavior. To check this possibility, we estimate the same model, but without individual-level fixed effects (Model 2). Here we find a large positive effect: those who have self-promoted also have a much higher likelihood to sabotage ($\beta = 0.031; p < 0.001$). So while the small group of marginal raters may indeed commit less sabotage when they also self-promoted, this is not the case for large majority of individuals who tend to use sabotage and self-promotion together to maximize their likelihood of winning the contest.

% #####################################
% ##                       Org-Level Outcome -- Idea Selection                        ##
% #####################################
\subsection{Consequences of strategic behavior: Effect on selection of contest winners}\label{sec:IdeaSelection}
Does strategic behavior affect idea selection? To quantify the impact that strategic evaluations have on idea selection---compared to which ideas would be selected if all evaluations were truthful and followed the meritocratic ideal of the community---we perform two supplementary analyses. The first assesses the effect of self-promotion, the second of sabotage (see Appendix \ref{sec:LongTerm}). For self-promotion, keeping everything else constant, we find that the winner changes in 1.4\% of contests (7 out of 511) and in 5.5\% (28 of 511) there is a change in at least one of the top three ranks. For sabotage, we find that the winner changes in 12\% of the contests and in 48\% there is a change among the top three. The effect is especially pronounced in close contests where the contest winner would change in 25\% of cases, while 65\% would see a change in the top three.

\begin{mycomment}
% #####################################
% ##            Mechanisms of Sabotage             ##
% ##      (tenure, experience, past winner)       ##
% #####################################
\subsection{Sabotage is driven by relative skill differences}\label{sec:SourcesOfSabotage}
Lastly, we investigate why some competitors act more strategically than others. In our model we propose that the resulting differences in probability win for individuals of heterogeneous skill explain why higher-skilled contestants engage more in strategic behaviors. However, an alternative explanation could be that strategic behaviors are just learned by being part of the community over time. We leveraging the panel structure of our data to look at temporal dynamics (Table \ref{tab:Mech} in the Appendix). We introduce interaction terms between the \textit{Submitted to same contest} dummy with the time-varying individual attributes of \textit{tenure}, \textit{experience} (submission and rating), and being a \textit{past winner}. Note that all these analyses control for the heterogeneous effect of skill. We find evidence to support the conclusion that higher expectations about winning a contest (i.e. being a past winner) explain the sources and targets of sabotage (Model 3; $\beta = 0.021; p < 0.001$ and $\beta = 0.013; p < 0.001$, respectively) while socialization in the community (tenure) has no effect (Model 1) and experience have either no or mixed effect on sabotage (Model 2). Thus our results indicate that once the ambiguity about their own skill, or the skill of their competitors is resolved through past success signals, competitors engage in more sabotage (against other competitors who are past winners). That is, past success explains not only the sources but also the targets of sabotage.  

%\hl{@Tom: as discussed let us not frame this as status concerns. Rather past winner with controlled for skill indicates that the socially approved confirmation of being a top designer (=past winner) creates a desire/pressure to win again and thus induces such competitors to use sabotage as a means to increase their chances of winning again}

%Why do some competitors engage in strategic behaviors more than others? \citet{edelman2015social} showed how status concerns from negative comparisons can spur deceptive self-promotion. We explore if status concerns could also be an explanation for engaging in sabotage. We leveraging the panel structure of our data to look at temporal dynamics (Table \ref{tab:Mechanisms} in the Appendix where we also explore several other possible mechanisms). We introduce interaction terms between the \textit{Submitted to same contest} dummy with different time-varying individual attributes. Note that these analyses control for the heterogeneous effect of skill. We find evidence to support the conclusion that \hl{high status individuals---those who have won a contest in the past---sabotage significantly more} (Model 3; $\beta = 0.009; p < 0.01$) while socialization in the community (tenure) has no effect (Model 1) and experience have either no or mixed effect on sabotage (Model 2).

%\hl{@Tom flag -- please rewrite ``high status'' interpretation}

% #############################################
% ##              Alumni Badge Experiment                 ##
% #############################################
\subsection{Victims of Sabotage}\label{sec:VictimsOfSabotage}
We now turn to the questions of how individuals decide \emph{who} to sabotage. Since rating is anonymous, reputational and other social concerns that are usually a major deterrent of sabotage are unclear in this setting.\footnote{In fact, the increase of sabotage committed by past winners shown in Model 3 above suggest that concerns about losing reputation from engaging strategic behaviors are not a major concern: The most reputable competitors in the community arguably are past winners, but they sabotage more, not less.}
Conversely, competitors do not simply sabotage everyone else either, suggesting that there is a driving mechanism and that there is a cost associated with sabotaging others just as our theoretical model assumes ($c_{s}$). We argue that these costs come at least partially from search cost of identifying targets of sabotage ---which may be particularly so given the contests large size which would make it difficult to sabotage all others (which would be 406 others on average and up to 847 in the largest contest). We observe exogenous variation in search cost over the study period and can exploit it to investigate the role of search cost as underlying mechanism that directs sabotage. 

On July 16, 2007, Threadless introduced the ``Alumni Club'': All competitors who won at least one contest would become members of the club and be awarded the ``Alumni Badge''. The website was then changed to show this badge as a small icon next to the user name. That is, after the system change, one could easily identify contest submissions by high-skill idea generators by simply scanning for the Alumni Badge next to the user name associated with a submission. We exploit this variation in search cost which would affect the amount of sabotage received by \textit{past winners} in the period \textit{after} the system change. If search cost affect sabotage behavior, sabotage of past winners should increase after the system change, as these high-skill idea generators are now easy to identify and distinguish from low-skill idea generators. 

% Intersting idea: we could run a super sharp version of this analysis: look at the sabotage a designer receives in the week they make their first winning submission. that is: they are already high-skill, but the system doesnt show them with the alumni badge yet
We again use a difference-in-difference design to identify the causal effect of search cost on sabotage through the variation caused by the introduction of the Alumni Club. Specifically, we add an interaction term between a dummy indicating the \textit{After} period and a dummy indicating whether the submission being rated on was made by a contestant who---at the moment of having made the submission---was a member of the alumni club and hence would have been shown with the Alumni Badge next to his/her user name (and, of course, the \textit{Submitted to same contest} dummy). Using a time window of $\pm6$ months around the platform change, we first see that, overall, there is less sabotage after the change (Model 1, $\beta = -0.004; p < 0.001$). Despite the time trend, however, sabotage of past winners is significantly higher after the system change (Model 2, $\beta = 0.004; p < 0.001$). Using the variation in search cost before and after the system change we are able to show that high-skill idea generators who have won in the past are more likely to be sabotaged after the system change which made identifying them much easier.

\input{../../01-Analysis-Writeup-LaTeX/table_NATURAL_Alumni24weeks_MANUAL}
\end{mycomment}

% #############################################
% ##              Sabotage - Targets of Sabotage                 ##
% #############################################
% \section{Consequences of Sabotage}

% #####################################################################
%          				     Long-Term
% #####################################################################

% #############################################
% ##              Sabotage - Targets of Sabotage                 ##
% #############################################
\section{Consequences of Sabotage on Long-Term Participation}\label{sec:LongTerm}
\begin{table}[h!]
\begin{center}
\begin{scriptsize}
\begin{tabular}{@{\extracolsep{5pt}}l D{.}{.}{5.5} D{.}{.}{5.5} D{.}{.}{5.5} D{.}{.}{5.5}}
\toprule
 & \multicolumn{4}{c}{Hazard: Participate in Next Round} \\
 \cline{2-5}\\[-5pt]
 & \multicolumn{2}{c}{Sabotage} & \multicolumn{2}{c}{Leniency} \\
 \cline{2-3} \cline{4-5} \\[-5pt]
 & \multicolumn{1}{c}{Skill} & \multicolumn{1}{c}{Skill}  & \multicolumn{1}{c}{Skill}  & \multicolumn{1}{c}{Skill}  \\
 & \multicolumn{1}{c}{Continuous} & \multicolumn{1}{c}{Quintiles}  & \multicolumn{1}{c}{Continuous}  & \multicolumn{1}{c}{Quintiles}  \\
 \cline{2-2} \cline{3-3} \cline{4-4} \cline{5-5} \\[-5pt]
 & \multicolumn{1}{c}{(1)} & \multicolumn{1}{c}{(2)} & \multicolumn{1}{c}{(3)} & \multicolumn{1}{c}{(4)} \\
\midrule
Sabotage received                   & -0.017^{\dagger} & -0.017       &                  &              \\
                                    & (0.010)          & (0.019)      &                  &              \\
Sabotage received \\
\quad $\times$ Skill    & 0.066^{***}      &              &                  &              \\
                                    & (0.009)          &              &                  &              \\
\quad $\times$ Skill Q1 &                  & -0.158^{***} &                  &              \\
                                    &                  & (0.027)      &                  &              \\
\quad $\times$ Skill Q2 &                  & -0.076^{**}  &                  &              \\
                                    &                  & (0.027)      &                  &              \\
\quad $\times$ Skill Q4 &                  & 0.075^{**}   &                  &              \\
                                    &                  & (0.026)      &                  &              \\
\quad $\times$ Skill Q5 &                  & 0.119^{***}  &                  &              \\
                                    &                  & (0.026)      &                  &              \\
Leniency received                   &                  &              & -0.110^{***}     & -0.044       \\
                                    &                  &              & (0.018)          & (0.028)      \\
Leniency received \\
\quad $\times$ Skill    &                  &              & -0.123^{***}     &              \\
                                    &                  &              & (0.011)          &              \\
\quad $\times$ Skill Q1 &                  &              &                  & 0.166^{***}  \\
                                    &                  &              &                  & (0.028)      \\
\quad $\times$ Skill Q2 &                  &              &                  & 0.117^{***}  \\
                                    &                  &              &                  & (0.029)      \\
\quad $\times$ Skill Q4 &                  &              &                  & -0.208^{***} \\
                                    &                  &              &                  & (0.034)      \\
\quad $\times$ Skill Q5 &                  &              &                  & -0.525^{***} \\
                                    &                  &              &                  & (0.046)      \\
\underline{Controls}\\
\quad Skill                               & 0.165^{***}      &              & 0.170^{***}      &              \\
                                    & (0.013)          &              & (0.013)          &              \\
\quad Skill Q1                            &                  & -0.341^{***} &                  & -0.443^{***} \\
                                    &                  & (0.035)      &                  & (0.035)      \\
\quad Skill Q2                            &                  & -0.084^{*}   &                  & -0.114^{***} \\
                                    &                  & (0.032)      &                  & (0.033)      \\
\quad Skill Q4                            &                  & 0.122^{***}  &                  & 0.096^{**}   \\
                                    &                  & (0.035)      &                  & (0.035)      \\
\quad Skill Q5                            &                  & 0.225^{***}  &                  & 0.036        \\
                                    &                  & (0.044)      &                  & (0.043)      \\
\quad Submission quality      & 0.976^{***}      & 0.982^{***}  & 0.821^{***}      & 0.797^{***}  \\
                                    & (0.027)          & (0.028)      & (0.034)          & (0.036)      \\
\quad Submission popularity   & 0.067^{***}      & 0.066^{***}  & 0.071^{***}      & 0.070^{***}  \\
                                    & (0.008)          & (0.008)      & (0.008)          & (0.008)      \\
\quad Competition size        & -0.000           & -0.000       & -0.000^{\dagger} & -0.000^{*}   \\
                                    & (0.000)          & (0.000)      & (0.000)          & (0.000)      \\
\quad Average contest rating  & 0.953^{***}      & 0.933^{***}  & 0.845^{***}      & 0.779^{***}  \\
                                    & (0.053)          & (0.053)      & (0.056)          & (0.056)      \\
\midrule
AIC                                 & 912233.777       & 911871.273   & 922869.780       & 921779.500   \\
R$^2$                               & 0.317            & 0.321        & 0.327            & 0.337        \\
% Max. R$^2$                          & 1.000            & 1.000        & 1.000            & 1.000        \\
Num. obs.                           & 72162            & 72162        & 73018            & 73018        \\
Num. events                         & 48052            & 48052        & 48639            & 48639        \\
% Missings                            & 43046            & 43046        & 42190            & 42190        \\
PH test                             & 0.000            & 0.000        & 0.000            & 0.000        \\
\bottomrule
\bottomrule
\multicolumn{4}{l}{\scriptsize{$^{***}p<0.001$; $^{**}p<0.01$; $^{*}p<0.05$; $^{\dagger}p<0.1$}}
\end{tabular}
\caption{Analysis of long-term participation. Multiple event hazard model of participation in the next week.}
\label{table:longTerm}
\end{scriptsize}
\end{center}
\end{table}

This section investigates a key organizational-level outcome resulting from the emergence of strategic behavior: how does being the target of sabotage and leniency affect long-term participation and hence community structure? Retaining existing members is critical to sustaining the communities in which contests are embedded \citep{faraj2011knowledge,ransbotham2011membership}. While existing research has investigated the motivation to join communities \citep{lakhani2003hackers}, much less is known about individuals’ progression and long-term participation in communities \citep{smirnova2022building}. There is a rich body of research on fair processes which would suggest that the targets of unfair behavior (such as sabotage), or those who experience random shocks to their performance evaluation, adjust their future effort and engagement \citep[e.g.,][]{gomez1992determinants,balietti2021incentives}. Several studies that investigate sabotage in lab experiments find that being the target of sabotage increases the likelihood of dropout \citep{balietti2021incentives,Charness2014, chen2003sabotage}. 
% On the other hand, high-skilled individuals may be able to understand and anticipate the competitive environment, and hence not perceive sabotage as negative and random. High-skill individuals have likely built up an identification with the community. They may therefore perceive being the target of sabotage as a challenge and increase their future participation, while lower-skill individuals and those who have not yet received social validation for their contributions may perceive it as a threat and reduce their future participation \citep{to2020interpersonal}. Leniency, on the other hand, may act as a motivator for lower-skilled individuals, contributing to their challenge mindset, and thus increase their long-term participation. 
Notice that while we consider self-promotion of equal importance, this section focuses on the amount of sabotage and leniency received simply because the amount of self-promotion received is virtually identical across all individuals. 

% Such a pattern would lead to a stabilization of the community’s core of most valuable members and a renewal of those at the periphery. In any event, differences in the reaction to being the target of sabotage between low- and high-skill individuals inevitably affect the evolution of the community’s social structure.

% In sum, while prior work has suggested that skill and past success are likely to influence evaluation behavior in crowdsourcing, the extent to which cooperative or competitive motivations prevail remains unclear, as are their effects on long-term participation. By examining whether skill and past success can explain peer evaluations and the reaction to being evaluated strategically, we can better inform contest design, but also understand how it affects the crowdsourcing communities’ structure in terms of long-term participation.
% How does being targeted with sabotage affect long-term participation of individuals of different skill and those who have experienced past success? 

The analysis proceeds as follows. We analyze the likelihood to participate in next week's contest, based on the amount of sabotage and leniency an individual experienced on their current week's submission. For this analysis, it is important to recall that sabotage (leniency) can be identified only probabilistically. For example, a low-quality submission may deserve a 0-star rating even if it comes from a competitor. Conversely, certain individuals may simply be harsh critics who rate most submissions with 0-stars without any intention to sabotage them. To measure the amount of sabotage (leniency) that a submission received, we rely on our micro-level model of sabotage developed above. 

For each individual, we estimate the counterfactual likelihood to cast a 0-stars evaluation of neutral outsider (using Model 1 from Table \ref{tab:HetEffect}). We then compute the ``residual'' between the evaluation that was actually cast and a counterfactual without strategic behavior. That is, our micro-level model of sabotage allows us to estimate the degree to which any of the 0-star ratings are likely due to sabotage (accounting for submission fixed effects, evaluator fixed effects, and time varying covariates like source and target skill). We also compute the amount of leniency a submission received as ``residual'' between a non-0-stars rating that should have been a 0-star rating. We then aggregate the evaluation-level residual to the submission level by taking the average across all ratings. This then captures the level of sabotage (leniency) that a submission received and gives us a panel dataset in which we have an estimate for the degree of sabotage (leniency) received by an individual for each contest.
We then estimate a multiple event hazard model \citep{survival-book} to predict the likelihood of long-term participation.\footnote{We estimate an interval Cox proportional hazard model for multiple-events, controlling for individual hazard rate of participation, exploiting within-individual variation in received sabotage; robust standard errors clustered at the individual level. We include the key observables that a competitor observes after a contest: the rating their submission received, the number of ratings received, the level of competition (number of competitors and average quality of all submissions in the contest), as well as our focal variable of interest, the amount of sabotage received.} 

% https://sphweb.bumc.bu.edu/otlt/mph-modules/bs/bs704_survival/BS704_Survival6.html
We find strong positive main effects of skill (Table \ref{table:longTerm} Model 1: $\beta = 0.165; p < 0.001$): higher-skilled individuals are more likely to participate in future contests. There is a small marginally significant main effect of sabotage ($\beta = -0.017; p < 0.10$).
We find a significant positive interaction between the level of sabotage received and skill ($\beta = 0.066; p < 0.001$). That is, high-skilled community members who received one standard deviation more sabotage, are about 6.6\% more likely to participate in the next contest compared to those who received only the average amount of sabotage. 
To better interpret the effect across skill levels, we convert skill to quintiles, using the middle quintile as our reference category (Model 2). Looking at the interaction between the level of sabotage received and skill, we find that individuals in first and second quintile react negatively to being targets of sabotage: their likelihood of participation in the next round drop significantly by 16\% and 8\%, respectively ($\beta = -0.158; p < 0.001$ and $\beta = -0.076; p = 0.003$). For high-skilled competitors, we find the opposite: they react positively to sabotage and their likelihood to participate increases. There is an 8\% increase for individuals in the 4th skill quintile ($\beta = 0.075; p = 0.003$) and a 12\% increase ($\beta = 0.119; p < 0.001$) for individuals in the highest skill quintile in the expected hazard to participate in the next round for a one unit change in received sabotage (one standard deviation).\footnote{We can also interpret the coefficients of our control variables. Having done well in a contest (high rating, many ratings) makes future participation significantly more likely. Competition has mixed effects (positive effect of high average rating across all contest submissions, no effect of competition size).} 

The pattern is reversed for the amount of leniency that a submission receives. The lowest skilled individuals increase their future participation by 17\% (Model 4: $\beta = 0.166; p < 0.001$) when they receive an additional one standard deviation of leniency. The highest-skilled individuals on the other hand decrease their likelihood of future participation. For high-skilled individuals, we need to caution against placing too much weight on the coefficient for leniency: it is very rare that high-skilled individuals would submit a design of such low quality that it deserves a 0-star rating and hence there is virtually no room to show leniency toward high-skilled individuals.

% #####################################################################
% 					     DISCUSSION
% #####################################################################
\section{Discussion}
Crowdsourcing has evolved as an organizational approach to distributed problem solving and innovation. As contests are embedded in online communities and evaluation rights are assigned to the crowd, community members face a tension between competitive and collaborative participation motives. This tension tempts idea generators to violate community norms and evaluate their peers strategically. Using large-scale digital trace data from Threadless, a prototypical crowdsourcing community \citep{majchrzak2016trajectories}, we answer the questions of how community members balance the competitive and collaborative motives as they evaluate their peers, and how the individual-level decisions they make change the structure of the community. We show that as their skill level increases, the competitive motive increasingly wins out: community members shift from using self-promotion to sabotaging their closest competitors (other high-skill individuals). However, community members also act in the collaborative spirit of the community and show leniency toward those who do not directly threaten their own chance of winning. 

In addition to these immediate short-term effects, we also find surprising long-term effects of the fierce competition that ensues. While low-skill targets of sabotage are less likely to participate in future contests, high-skill targets are more likely. Furthermore, low-skilled beneficiaries of leniency are encouraged and are more likely to participate in the future. This suggests a feedback loop between competitive evaluation behavior and future participation. 

% ########################################################
\subsection{Theoretical Implications}
% ########################################################
Our findings have three important implications for the literature on the interplay between competitive and collaborative motives in crowdsourcing design, as well as the evolution and sustainability of crowdsourcing communities.

First, we extend prior research on the nature of competitive and collaborative behavior in crowdsourcing. The tension between competition and collaboration has been identified by several crowdsourcing researchers \citep{adler2011combining, franke2003communities, bullinger2010community,hutter2011communitition,majchrzak2013towards,boudreau2015open} and features prominently in the broader organizations literature \citep{deutsch1949theory,tsai2002social,lado1997competition,gallus2022relational}. This work has shown that the level of knowledge sharing and mutual support decreases in communities when members are also competing \citep{franke2003communities,harhoff2003profiting}. We complement this research with insights into evaluation behavior. We show that the competitive incentives designed to encourage effort during idea generation spill over into the collaborative community, creating a mismatch between the collaborative organizational context and competitive incentives \citep{gallus2022relational}. However, the resulting strategic behavior is not rampant and out of control as contest theory might predict. This suggests that embedding contests in communities drives up the cost of amoral strategic behavior even in a community without reputation costs (because peer ratings are anonymous) and low search costs (because sabotaging everyone would only cost a few mouse clicks). The tension between competitive and collaborative participation motives is felt most acutely by high-skilled individuals who are most likely to win (and thus have the most to gain). Hence, they are the most willing to accept the cost of violating community norms. 

However, community members not only sabotage, but they also show leniency. We theorize that showing leniency is a crucial collaborative element that allows individuals to better justify violating community norms with competitive evaluations through a form of moral licensing \citep{blanken2015meta}. Moral licensing describes behavior in which people who behave in a moral way can also display behaviors that are immoral, unethical, or otherwise problematic \citep[e.g.,][]{merritt2010moral}. Specifically, we theorize that individuals accumulate credits on an invisible scorecard by promoting some low-skilled community members (giving them higher ratings than they deserve) and then spend these credits on self-interested strategic evaluations (giving some rivals lower ratings and self-promoting their own work). Often, the more individuals are invested in the community, the stronger their perceived need to atone for the violation of community norms may be \citep{ashforth2001hat}. From the perspective of contest theory, leniency is unexpected collaborative behavior because promoting anyone other than oneself is counter to the self-interest of winning the contest. However, this leniency is not simply a collaborative act within the community. It has its own strategic nature as it is targeted toward those who least threaten one's own chance of winning. Recognizing that showing leniency toward low-skilled community members is strategically motivated has far-reaching implications for our understanding of interactions in communities \citep[e.g.,][]{wasko2005should,faraj2011network,ren2012building}. Future models of behavior in online communities need to consider not only interactions themselves but also how they are motivated. Thus, leniency appears to be a crucial collaborative element that allows individuals to better justify self-interested strategic evaluations. It alleviates the tension between the mismatched collaborative organization structure and the competitive incentive \citep{gallus2022relational}.

Second, our dyad-level approach offers complementary insights to previous studies that examined participation motivations at the individual level \citep[e.g.,][]{hippel2003open,wasko2005should}. Prior work focuses on competitive and collaborative participation motives as individual-level traits \citep[some individuals are competitive while others are collaborative;][]{ lakhani2003hackers, reuben2015taste, erat2012white, belenzon2015motivation} to explain why community members behave in certain ways. By contrast, we explain how the contextual factor of the competitive situation explains behavior. The same individual in the same contest may act competitively toward one individual yet show leniency toward another. 
Our theory and empirical results imply that within individual changes in skill and contextual factors (whether the individual is itself competing for the prize vs.~a neutral outsider, and who the target of the behavior is) make some interactions more (or less) competitive and explain the observed microlevel behavior. 
When community members are not themselves competing, they evaluate their peers fairly with equal treatment of everyone \citep[i.e., according to equality matching;][]{fiske1992four}. When they are competing, the evaluations depend on the competitiveness of the situation. The evaluation of those who do not threaten a community member's own chance of winning fall into a different category of relational behavior. Those evaluations are not governed by the competitive participation motivation \citep[i.e., they are not based on a rational cost-benefit calculation of market pricing;][]{fiske1992four} and are instead evaluated under a collaborative scheme of kindness and selfless generosity \citep[i.e., under communal sharing;][]{fiske1992four}.
As a result, skill (and social confirmation of skill from past success) is a key factor that shapes the strength and saliency of the self-interested motivation because it affects the gains individuals can expect from such behavior and can thus explain whether individuals wear a competitive or collaborative hat \citep{ashforth2001hat}. Differentiating behavior based on the target allows individuals to be competitive and cooperative at the same time, thus alleviating the tension \citep[cf.][]{waldman2019role}.
Individuals may have joined the community out of intrinsic motivation but once they have high skill (and have experienced success), the competitive motive takes over. This contributes to recent research which has started to acknowledge that there are changes and growth over individual ``careers'' in communities as they learn and improve their skill \citep{riedl2018learning,soda2021networks, smirnova2022building,dahlander2011progressing}. 

Our dyad-level investigation also adds to the nascent crowdsourcing literature that has started to acknowledge the existence of strategic behaviors \citep{hofstetter2018should,archak2009optimal,liu2014crowdsourcing,hutter2011communitition,chen2020conan,klapper2023strategic,deodhar2022influence} by providing well-identified empirical evidence for both sabotage and self-promotion in a prominent crowdsourcing platform. In particular, our findings complement prior research on peer evaluations by highlighting the importance of competition and covertness of evaluations: While \citet{klapper2023strategic} find that Wikipedia editors target negative evaluations at community members who are unlikely to retaliate, we show that when in direct competition---as it is the case in many crowdsourcing settings---negative evaluations are targeted towards the most prolific competitors. Furthermore, positive evaluations are more likely to be granted to those lower-skilled competitors who matter little for the overall outcome of the contest. Thus the dynamic reverses compared to the findings of \citet{klapper2023strategic}. This has important implications for crowdsourcing community organizers as it underscores the important roles of competition and transparency when designing the interactions of community members. While non-transparent peer evaluations limit the possibility to use peer evaluations as a means to portray oneself, it does reduce the costs to engage in sabotage behavior towards competitors. For settings where competition is prevalent it also highlights the usefulness of game-theoretic considerations to assess heterogeneous evaluation patterns.

On a more general note, we add important field evidence to the existing economics literature on sabotage that has mostly relied on lab experiments \citep{Lazear1989, Charness2014, harbring2011sabotage, Konrad2000} and found ambiguous evidence for the association between sabotage and skill \citep{Charness2014,harbring2007sabotage,chambers2020robust}.

Third, our work contributes to our understanding of the evolution and sustainability of online communities by showing how individual-level strategic behavior affects the organizational-level social structure of communities \citep{kim2018external,faraj2011network,huang2014crowdsourcing,piezunka2019idea,hofstetter2018successive}. We document a self-reinforcing dynamic in which increased member skill not only leads to more strategic behavior but also increased future participation. This complements past studies that have investigated the fluidity of communities focused on aspects of self-selection and open boundaries---i.e., dynamics \textit{across} individuals---with insights of dynamics \textit{within} individuals \citep{faraj2011knowledge,felin2017firms}. Contrary to the expectation that unfair behavior would reduce engagement \citep[e.g.,][]{gomez1992determinants,faullant2017fair, Franke2013,balietti2021incentives}, strategic behavior appears to facilitate future participation because it encourages low-skilled individuals with leniency and engages high-skilled community members in challenging competition through sabotage. Strategic behavior appears to play an important organizational role in stabilizing the core of a community by engaging members in intense competition. 		

Why does being a victim of sabotage make highly skilled community members more likely to participate in future contests? We theorize that the fierce competition among the high-skilled adds to their intrinsic motivation and promotion focus. Competition can be thrilling and exciting, even when outcomes are negative \citep{franken1995people}. Intrinsic motivation and corresponding theories may provide an explanation. First, according to self-determination theory \citep{ryan2000self} skilled individuals may experience a sense of competence and autonomy that can enhance their motivation. Having been a victim of sabotage could serve as a confirmation of being considered a true competitor, which could increase motivation and thus the likelihood of future participation. Furthermore, the fierce competition among the high-skilled adds to the status incentive of the competition and may be perceived as a challenge to overcome, rather than a threat \citep{to2020interpersonal}. Second, the challenge of competition is a trigger of a flow state \citep{csikszentmihalyi1990flow}. Higher-skilled individuals have a wider range in which flow can occur when the challenge-level rises compared to their lower-skilled counterparts. Being a victim of sabotage may thus increase the perceived challenge: while low-skilled individuals perceive sabotage as a threat, high-skilled individuals perceive it as a challenge \citep[promotion focus][]{to2020interpersonal}, making their participation in future contests more likely. 

Prior work has focused on collaborative behavior and explained why members in crowdsourcing act reciprocally toward others who have helped them in the past \citep{safadi2021contributes,perry2006social,jeppesen2010marginality,dahlander2016one}. By contrast, we explain why apparent reciprocity in terms of social network structure may also result from competitive behavior in which high-skilled members sabotage other high-skilled community members. We thus provide an additional explanation for the phenomenon of a consolidated community core of high-skill individuals that is a key characteristic of many crowdsourcing communities \citep{safadi2021contributes,perry2006social,jeppesen2010marginality,dahlander2016one}. Our contest theory model and empirical results imply that behaviors (and resulting network structures) which appear to be reciprocal and collaborative can in fact be deeply competitive. Together, this shows why strategic behavior must be incorporated in our theoretical understanding of crowdsourcing communities that have often been studied using structural social network approaches. The understanding of the implications of social structure in communities could therefore benefit from a more systematic integration of, and attention to, how actors’ behavioral motivation guides their behavior.

With the proliferation of social network studies focused on static structural aspects \citep{safadi2021contributes,perry2006social,jeppesen2010marginality,dahlander2016one}, the feedback loops between dynamic interactions and their effect on structure suggest important questions for future research \citep[e.g.,][]{soda2021networks,fulker2021spite,foley2021avoiding}. Our study is a first step in that direction. It provides an alternative explanation for why communities often have a core-periphery structure. Future work needs to consider not only the structure of community interaction itself but also the valence of interaction (e.g., are comments helpful or hurtful? Are ratings positive or negative?) and how the interactions are motivated. This may reveal more nuanced network structures in terms ``negative'' (competitive) ties \citep{labianca2006exploring}, adding to our understanding of social networks from structural analysis which has focused on ties based on social relationships (e.g., friendship), communication, and information flow \citep{wasserman1994social,borgatti2009network}.

% ########################################################
\subsection{Practical Implications}
% ########################################################
Our findings also have important practical implications. Counter to common sense, platform design that allows strategic behavior may not necessarily be bad. From an organizational perspective, strategic behavior is usually seen as destructive and thus undesirable. According to this view, organizations are well advised to limit its impact by designing incentives (and structures) that reduce strategic behavior. Our study reveals that this view is not always valid. The fierce competition among high-skilled community members which makes them targets of sabotage increases their future participation. It also leads to the encouragement of less skilled members through leniency. As a result, the competitive strategic behavior that ostensibly runs counter to the notion of a collaborative community does in fact have positive long-term effects on the community. Therefore, our study challenges conventional wisdom on strategic behavior: instead of eliminating strategic behavior entirely, the organizers of crowdsourcing communities may encourage it within controlled boundaries (e.g., anonymous voting).

% ########################################################
\subsection{Limitations, Generalizability, and Future Research}
% ########################################################
This paper is not without limitations. First, while Threadless is a prototypical crowdsourcing community and therefore highly suitable for generalization, there may still be some differences to other settings. For example, Threadless does not publish an overall ranking which may limit competitive motivation compared to other settings \citep[e.g., Topcoder publishes public rankings which can spur rivalry; see][]{grad2022rivalry}. Future research could therefore investigate the patterns of strategic behaviors in settings that differ from ours. The contest theory model can serve as a starting point for that research. The findings may apply to other contexts beyond crowdsourcing. As the theoretical reasoning and empirical setting of this study is concerned with individuals anonymously evaluating their rivals in an otherwise collaborative environment, the study’s results may generalize to contexts with similar characteristics such as peer evaluation in teams, academic publishing, grant awarding, or political races. Second, this paper focuses on two specific forms of strategic behavior on the individual level. The study of other forms of strategic behaviors that emerge between two or more participants over time, such as interpersonal rivalry and reciprocity, are exciting avenues for future research. Finally, our work opens new fields to investigate career dynamics in communities and the connection between individual behaviors and organizational-level outcomes like overall community structures.

\section*{Acknowledgements}
We gratefully acknowledge helpful comments by Jennifer Brown, Blair Davey, Imke Reimers, Ulrich Berger, Ben Greiner as well as participants at the NBER Summer Institute 2019, and participants of the Digital Innovation Workshop 2019 at Boston College. This research was funded in part by the National Science Foundation [Grant IIS-1514283] and benefitted from computing infrastructure available at the Network Science Institute funded by the US Office of Naval Research [N00014-17-1-2542].

\bibliographystyle{apa}
\bibliography{manual}

%######################################################################################################
%                                  Appendix
%######################################################################################################
\clearpage
\section*{Appendix}
% !TEX root = Strategic Behavior V36_OrgSci_Round2_v4.tex
% \renewcommand{\appendixpagename}{Appendix}

\appendix
% \appendixpage

%\renewcommand{\thesection}{\Alph{section}.} 

\renewcommand\thefigure{A.\Roman{figure}}    
\renewcommand\thetable{A.\Roman{table}}    
\setcounter{figure}{0}    
\setcounter{table}{0}    
\setcounter{page}{1}

% ########################################################
\section{A Brief Review of the Economics Literature on Strategic Behavior}
\label{sec:AppendixEconLitRev}
% ########################################################

The threat of sabotage in which actors engage in non-productive effort to reduce the performance of another actor has been at the forefront of economic literature and noted in the earliest work on rank-order tournaments \cite{Lazear1989,Dixit1987}. However, past empirical work on sabotage is scarce. Notable exceptions are the work on withholding helping behavior \citep{Drago1998,haas2010share}, which can be considered a weaker form of sabotage. The dearth of empirical insight into sabotage is (at least in part) attributable to the challenge of investigating sabotage empirically: actors try to hide their opportunistic behavior due to its immoral connotation so that “company data on sabotage is generally not available for research'' \citep[][p.~65]{dato2014gender}. 

Self-promotion is somewhat better understood empirically \citep{li2017expertise, keum2017influence,reitzig2013biases,edelman2015social, lerchenmueller2019gender}, but it is unclear how it interacts with sabotage and when individuals would choose one form of strategic behavior over the other (i.e., whether the two are complements or substitutes). Thus, there are limited empirical insights into strategic behavior beyond theoretical work \citep{Lazear1989,MilgromRoberts1988,Munster2007}, laboratory studies \citep{harbring2011sabotage, Charness2014}, or sports \citep{Balafoutas2012, Deutscher2013, Garicano2005}.\footnote{Of course, there are many other forms of strategic behavior besides sabotage and self-promotion in crowdsourcing and contest theory more generally. For example, \citet{liu2014crowdsourcing} look at strategic entry decisions, \citet{archak2010money} look at the use of “cheap talk” to deter others from entering, \citet{hutter2011communitition} look at instances in which individuals do not reciprocate cooperative efforts, and \citet{hofstetter2018should} look at \textit{socially} motivated reciprocal rating (as opposed to strategically motivated rating when actors are anonymous), \cite{chen2020conan} looks at collusion in micro-tasking where actors copy each others’ solutions, and \citet{scheiner2018participation} study unethical behavior using survey responses but not actual behavior.}

Prior work from the economics literature has suggested that strategic behavior is contingent on the skill of contestants \citep[e.g., ][]{schotter1992asymmetric, Dixit1987}. However, existing empirical evidence is ambiguous. Some studies find that low skill is associated with engagement in strategic behavior \citep{Carpenter2010, Charness2014} while others find evidence for higher skill \citep{harbring2011sabotage}. Yet others find no association between skill and strategic behavior at all \citep{dato2014gender}. Hence, the effect (or lack of it) remains unclear.

% ####################################
% ######## Theoretical Model #############
% ####################################
% NOTE: 2022-02-25 for OrgSci Round 2, change v_i to v_h and v_k to v_l, keeping v_i for a generic agent i
% This makes reading easier because it has the high / low in the name. however, its slightly confusing because $h$ refers to THE NUMBER of high types, not to a specific high type agent; and $l$ refers to THE NUMBER of low types
\section{Theoretical Framework} \label{sec:AppendixModel}

\subsection{Model Setup in More Formalized Terms}

We build on previous models of one-shot contest with sabotage \citep[see, e.g.,][]{harbring2011sabotage, Konrad2000}, skill heterogeneous agents, and a single winner prize. The contest consists of two types of agents: a set $H$ of $h$ high-skill agents, and a set $L$ of $l$ low-skill agents. We refer to these as high and low types, respectively. Furthermore, the contest involves a set $N$ of $n$ outsiders. 
We assume that $h < l < n$, which should be satisfied in most realistic contest settings. 
%Following \cite{Moldovanu2001} and \cite{Boudreau2016b}, we do not consider the effort choice of contestants directly but their choice of bid quality $b_i$ based on their skill level $a_i$.\fxnote{Do we need this sentence or is that just something econ heavy people would find interesting? Otherwise I find $a_i$ and $b_i$ more confusing than helpful. We could integrate $b_i$ in the next sentence} 
Low and high types produce a contest submission (i.e. idea) of low quality $b_l$ and high quality $b_h$, respectively, which they enter into the contest. Outsiders do not enter the contest and thus remain neutral. The submissions of all agents are rated by each outsider and by each high and low type agent on a bounded quality scale in $\R_{\ge 0}$ with $[r_{min}, r_{max}]$. Without loss of generality, we normalize the rating scale to the unit interval so that $r_{min}=0$ and $r_{max}=1$. We assume that the quality bid of low type agents' submissions is $b_{l}$, whereas the quality bid of high type agents' submissions is $b_{h}$ with $0 < b_{l} < b_{h} < 1$. 

While ratings can be sincere (i.e., $b_{l}$ or $b_{h}$, respectively), an agent can also sabotage any other agent, rating their submission at $r_{min}=0$; or promote them by rating their submission at $r_{max}=1$. Let $\Delta s_h = b_{h}$ be the damage done by sabotaging a high type and $\Delta s_l = b_{l}$ be the damage done by sabotaging a low type. Since $b_l < b_h$, it follows that the damage inflicted on a high type through sabotage is larger than on a low type (i.e., $\Delta s_l < \Delta s_h$).
Further, let $\Delta sp_h = 1 - b_{h}$ denote the benefit gained from promoting a high type and $\Delta sp_l = 1 - b_{l}$ denote the benefit gained from promoting a low type. Since $b_{l} < b_{h}$ it follows that low types have more to gain from promotion than high types (i.e., $\Delta sp_l > \Delta sp_h$). 

\begin{mycomment}
Figure \ref{fig:RatingInterval} illustrates the rating scheme.
\begin{figure}[tbh!]
	\begin{center}
		\includegraphics[width=0.40\linewidth]{FIGURES/RatingInterval.png}
	\end{center}
	\caption{Rating Interval.}
	\label{fig:RatingInterval}
\end{figure}
\end{mycomment}

We assume that while rating sincerely is free of costs, sabotaging another agent costs $c_s$ and promoting any agent (including oneself) costs $c_p$. Costs associated with promotion and sabotage might arise from the costs of identifying suitable targets \citep{harbring2007sabotage,Munster2007}, the moral costs associated with lying \citep{abeler2018preferences, gneezy2018lying} or violating social norms \citep{elster1989social}. 
As a consequence all $N$ outsiders rate contestants $h$ and $l$ sincerely. This implies that the size $n$ of the set of outsiders $N$ affects the effectiveness of any type of strategic behavior performed by the set of $H$ and $L$ agents: the more outsiders there are, the lower the effect that any individual strategic rating has. Effectively, $n$ can be considered a modifier in our model that governs how effective strategic behavior is. Note also that outsiders $n$ are not necessarily agents who never compete---they simply do not compete in the current contest. We will use this feature to identify strategic behavior in our empirical analysis where contestants do not enter every contest: they rate as neutral outsiders in some weeks and rate as competitors with stakes from idea generation---and incentives to rate strategically---in the contest in other weeks. 

% Value
The rating given by agent $i$ to agent $j$ is denoted by $v_{ij}$. The value of agent $j$'s submission is the sum of all the ratings that this focal agent receives: $v_j = \sum_{i \in N \cup L \cup H} v_{ij}$. Throughout the remainder, let $v_h$ denote the value of some arbitrary high type (she), and $v_l$ denote the value of some arbitrary low type (he). As an example, in the case of sincere rating by everyone (no promotion and no sabotage), the value of a high type is $v_h = b_h(n+l+h)$ and that of a low type is $v_l = b_l(n+l+h)$. Finally, we also assume that the difference between contest entries $b_l$ and $b_h$ is large enough so that the relative order between high and low types cannot be changed through strategic behavior alone. That is, $v_l < v_h$ even if all low types engage in self-promotion and all high-types are sabotaged by everyone. 
%\hl{It is important to note that we set up the model within the peer-rating context in crowdsourcing contests to better integrate theory and empirical analyses. The basic premises and predictions of the model are, however, more general and apply to other contests with promotion and sabotage. The key assumption---a fixed rating scale and resulting limits to potential gain from promotion and damage from sabotage---is, by no means, specific to contests in this setting.}\fxnote{We mention this earlier. I would cut it here and keep the one before. other way round works aswell but one can go. CR@TG}

\subsection{Utility Function}
The winner of the contest is determined based on the evaluations from $N$, $H$, and $L$. For simplicity, we model this as a Tullock contest \citep{Tullock1980} where the probability of winning is proportional to the aggregate rating $v_i$ plus an error term in relation to all other contest submissions. That is, each agent $i$'s probability of winning the contest is $p_i = \frac{v_i}{\sum_{j \in L \cup H} v_j} $. For simplicity, let $S = \sum_{j \in L \cup H} v_j$ denote the total value of all contestants and we write the Tullock contest success function as $p_i = \frac{v_i}{S}$. The winner of the contest receives a prize $M$, which is without loss of generality normalized to $1$. Given the equilibrium values for productive effort and the resulting quality bid in the ordinary Tullock contest, agents maximize their expected winning probability minus their total costs of sabotage and cost of promotion. The utility function for agents' strategic contest behavior can then be written as:

% ### Utility Function ###
\begin{align} 
	E \Pi_i = M p_i - c_s ( \sum_{j} sab_{ij}) - c_{p} sprom_{ii}), 
	\label{UtilityFunction}
\end{align}

where $M$ is the winner prize, $p_i$ is the probability of agent $i$ winning the prize, $sab_{ij}$ ($prom_{ij}$) is $1$ if agent $i$ sabotages (promotes) agent $j$ and $0$ otherwise, and $c_s$ and $c_p$ are the cost of sabotage and promotion, respectively.

Formally, a strategy of agent $i$ is a list $(r_{ij})$ of rating behavior towards all agents $j \in L \cup H$, where $r_{ij} \in \{\text{sincere}, \text{sabotage}, \text{promote}\}$. Since sabotaging oneself and promoting any agent other than oneself are clearly strictly dominated and payoffs are symmetric with respect to actions towards other agents of the same type, we only consider equivalence classes of strategies of the form $s_i = (hsab_i, lsab_i, sprom_i)$, where $hsab_i$ is the number of high type agents $i$ sabotages, $lsab_i$ the number of low type agents $i$ sabotages, and $sprom_i$ takes the value $1$ or $0$, indicating whether or not $i$ self-promotes. We now investigate the strategic behavior for self-promotion and sabotage.

% ####### Self-Promotion ###########
\subsection{Self-Promotion}
Self-promotion increases agent $i$'s value $v_i$ and total contest output $S$ by the same amount and hence increases agent $i$'s chance of winning. The size of the this increase of course depends on agent $i$'s type, the number of other high and low types in the contest, and the number of outsiders. 
% \fxnote{The size of the increase also depends on the behavior of the other contestants} true, but let's leave that equilibirum consideration out for now.

\begin{lemma}
	In contests with a sufficiently large performance gap $g$ between low and high types, the expected gain from self-promotion is higher for a low type than a high type.
	\label{lemma1}
\end{lemma}

% ######## Proof ###########
% Intuition of proof: to show that the term is greater than 0, show that each term is greater than zero. Denominator is clearly positive since $S$ is positive. Since $v_i > v_k$ it follows $4(v_i - v_k) > 0$. So the first $S ... $ term remains: given $S > v_i > v_k$ it follows that $3S + v_i > 4v_i > 4v_k$ and hence $4v_i - 4v_k > 0$. Together: since all other terms are $ > 0$, the whole equation is $>0$.
\paragraph{Proof.} The expected gain from self-promotion for a currently non-self-promoting high type agent $h$ is
	$\frac{v_h + \Delta sp_h}{S + \Delta sp_h} - \frac{v_h}{S} = \frac{\Delta sp_h (S-v_h)}{S(S+\Delta sp_h)}$, and 
	$\frac{v_l + \Delta sp_l}{S + \Delta sp_l} - \frac{v_l}{S} = \frac{\Delta sp_l (S-v_l)}{S(S+\Delta sp_l)}$ 
for a currently non-self-promoting low type agent $l$.
Considering the worst case value for a high skill agent gives us $v_h \geq b_{h}(n+1)$ (i.e., sincere rating by all $n$ outsiders and $h$ herself, all $l$ and $h-1$ sabotage and rate $0$).	% the n outsiders rate the high submission truthfully at b_high, even if others sabotage it can't be lower than this since we define the rating scale to be on the non-negative numbers range
Conversely, the best case value for a low type gives us $v_l \leq b_{l}(n + l + h)$ (i.e., sincere rating by everyone). 
% TODO: NOTE: Actually the best case value for a low types would also include her self-promotion vote

In a contest where the performance gap between high and low types is large with respect to the proportion of neutral outsiders $n$ (who vote truthfully) and the overall contest size (where $l$ and $h$ might potentially sabotage) we get $S > v_h > v_l$. 
Define $g$ as the performance gap between high and low types so that $v_h = g v_l$. Expressing $g$ as the inequality between the worst case value for a high type (receiving sabotage from all other high types and all low types) and the best case value for a low type (not being sabotaged at all)

\begin{align}
	g = \frac{b_h}{b_l} \frac{n+1}{n+l+h}.
\end{align}

This expression gives a lower bound on $g$ for a contest with a positive performance gap. A positive performance gap ensures that the high type values $v_h$ are higher than the low type values $v_l$ even if the high types are being sabotaged by all $H_{-i}$ and $L$ agents. This ensures that the $b_l < b_h$ relationship also holds for the values $v_l < v_h$. Note that this expresses the necessary size of the performance gap as a product of the difference in production skill and the proportion of neutral outsiders to overall contest size. Considering the smallest possible contest will have $n=3, l=2, h=1$ provides a lower bound for $g$ so that
\begin{align}
	g = \frac{b_h}{b_l}  \frac{n+1}{n+l+h} \ge \frac{b_h}{b_l}   \frac{2}{3}
\end{align}

Now consider a contest with a significant performance gap between high and low types so that $g \ge 1$ holds, then we can compare the expected gain from self-promotion between high and low types.

\begin{align}
	\frac{\Delta sp_l (S-v_l)}
	        {S(S+\Delta sp_l)} - \frac{\Delta sp_h (S-v_h)}
	                                                 {S(S+\Delta sp_h)} 
	=
	\frac{S((\Delta sp_l - \Delta sp_h)S + v_h - \Delta sp_l v_l) + \Delta sp_l(v_h - v_l)}
		{S(S+\Delta sp_h)(S+\Delta sp_l)}.
\end{align}

	This expression is $> 0$ as can be seen by substituting $g v_l$ for $v_h$ and $1-b_l$ for $\Delta sp_l$:

\begin{align}
 	S((\Delta sp_l - \Delta sp_h)S + v_h - \Delta sp_l v_l) \ge \\
	S[(\Delta sp_l - \Delta sp_h)S + g v_l - (1 - b_l) v_l] = 	\nonumber \\
	S[(\Delta sp_l - \Delta sp_h)S + (g - 1 + b_l) v_l] > 0 \nonumber
\end{align}
% ==> the other way of thinking about this is that if skill is a spectrum, there are ghigh types whose b_h is high enough that g >= 1 condition is true.

Thus, in contests where the performance gap between low and high types is large enough such that $g \ge 1$ is satisfied, the expected gain from self-promotion for a low type is larger than for a high type. \hfill $\blacksquare$
\vspace{0.5cm}

% Consider: n=3, l=2, h=1 with bl=2, bh=3 with {0, ..., 5}
% Worst case high type: vi = 3*3 + 0*2 + 0*1 = 9
% Best case low type: vk = 3*2 + 2*2 + 1*2 = 12
% (3*2 + 3*3) < S < (6*3 + 6*2)
% 15 < S < 30
% Then gain high: 2(S-9) / 9(9+2) = 2(S-9) / 99   [sub in S=15] ==> 0.12
% Gain low: 3(S-12) / 12(12+3)  = 3(S-12) / 180  [sub in S=15] ==> 0.05  ==> the gain for high types is higher than for low types

% ##############  everyone self promotes if it yields positive benefit ########
A high type will of course self-promote if the gain from self-promotion is higher than its costs. More specifically:

\begin{lemma}
	If, for a high type with value $v_h$ and a performance gap $g \ge 1$, $c_p < \frac{\Delta sp_h (S-v_h)}{S(S-\Delta sp_h)}$ in equilibrium, then all agents will engage in self-promotion.
	% If its small enough for the high types (who benefit less) to do it, the low types will already do it.
	\label{lemma2}
\end{lemma}

\paragraph{Proof.} If the high type did not engage in self-promotion, her winning probability would fall by $\frac{v_h}{S} - \frac{v_h - \Delta sp_h}{S-\Delta sp_h} = \frac{\Delta sp_h (S-v_h)}{S(S-\Delta sp_h)}$. By assumption, this is larger than $c_p$, so her net gain would be negative. She will therefore adhere to self-promotion. By Lemma \ref{lemma1}, this continues to hold for the low types if $g \ge 1$. Alternatively, low types will engage in self-promotion if $c_p < \frac{\Delta sp_l (S-v_l)}{S(S-\Delta sp_l)}$. \hfill $\blacksquare$

\vspace{0.5cm}

% NOTE: This is "wrong" it uses the ambiguous definition of S. Actually, what would happen for self-promotion of high-types is that low-types already sabotage

% ###########################
% ####### Sabotage ###########
% ###########################
\subsection{Sabotage}
% Note "marginal utility" already contrasts the cost of sabotaging one more unit with the resulting increase in winning probability from sabotaging one more unit. 
% E.g., c_s = .5, benefit from sabotaging one other is 1 => so you do it. Sabotaging a second one, your benefit (of sabotaging this additonal unit) is 1.1, while your cost is still .5, so you do that too, and so on.
% Your largest (maximal) marginal increases happens from sabotaging the last unit
% HOWEVER, the marginal increase from sabotaging all vs. none is larger than the marginal increase from only the last unit (because its the sum of all the marginal increases and includes the last unit)
% Question: Is the marginal benefit from sabotaging the first low type higher or lower than the first high type?

If an agent sabotages another agent, this decreases that agent's output $v_i$ and thus decreases $S$, which increases that agents probability of winning the contest. This decrease in $S$ also benefits all other agents and thus sabotage has---contrary to self-promotion---an important negative externality \citep{Konrad2000}. We first show that if an agent sabotages another agent, the agent sabotages all agents of that type.
% the marginal gain in probability is increasing in the number of other agents (of the same type) the focal agent sabotages.

% This is counter to standard theory which predicts that due to the externality it should be lower. Why?
% Maybe the word "marginal" is wrong.

\begin{lemma}
	The expected marginal gain in the probability of winning increases with the number of other agents (of the same type) being sabotaged. 
	\label{lemma:increasingGain}
\end{lemma}

\paragraph{Proof.} Consider a high type of value $v_h$ and sincere rating by everyone. Her probability of winning is $\frac{v_h}{S}$. If she sabotages one more high type, her probability of winning increases to $\frac{v_h}{S-b_h}$. The marginal gain in probability from this act of sabotage is 
	$\frac{v_h}{S-b_h} - \frac{v_h}{S} = \frac{b_h v_h}{S(S-b_h)}$. 
	% This is 100% correct; see fn2 in sabotage_maximal-marginal-increase.nb
We can then express the marginal gain of sabotage as a function of the number of other high types $x_h$ that the agent sabotages: $f(x_h) = \frac{x_h b_h v_h}{S(S-x_h b_h)}$. The first derivative of which is $f'(x_h) = \frac{b_h v_h}{(S-x_h b_h)^2}$ which is always positive for positive $x_h$ and given that $S > n b_h > h b_h > x_h b_h$ holds since by design $n > h$ and $h > x_h$ because the agent does not sabotage herself and at most sabotages $h-1$ high types. Thus, the marginal gain of sabotaging another high type increases with the number of high types already sabotaged. If the costs of sabotage are smaller than the (maximum) marginal gain from sabotaging the last other high type agent, a rational high type agent will therefore sabotage all other high type agents. An analogous argument shows that the marginal gain for a high type from sabotaging one more low type agent increases with the number of low type agents already sabotaged. The same arguments also hold for low types sabotaging other agents. \hfill $\blacksquare$
\vspace{0.5cm}

% ########### NEW SECTION on multiple thresholds
We can use Lemma \ref{lemma:increasingGain} to compute the bounds of sabotage costs $c_s$ when agents engage in strategic behavior. For example, consider a high type of value $v_h$ not currently sabotaging anyone and sincere voting by everyone else so that $v_h = b_h (n+l+h)$ and $v_l = b_l (n+l+h)$, and $S=h v_h + l v_l$. 
% Assuming sequential rationality 
% Applying the subgame perfect Nash equilibrium solution concept and backward induction
If she sabotages all other high types (remember she will not sabotage herself) her probability of winning is $\frac{v_h} {S-b_h(h-1)}$. If she were to sabotage only $h-2$ high types her probability of winning is $\frac{v_h} {S-b_h(h-2)}$.
Consequently, her maximal marginal increase in the probability of winning from sabotaging other high types results from sabotaging all other $h-1$ instead of $h-2$ high type agents and is of size 

\begin{align}
				& & \frac{v_h} {S-b_h(h-1)} - \frac{v_h} {S-b_h(h-2)} = \frac{v_h b_h}{(S-b_h(h-1))(S-b_h(h-2))} 
				\label{eq:boundHighSabHigh}
\end{align}

If the costs of sabotage $c_s$ are smaller than the maximal marginal gain in Eq.~\ref{eq:boundHighSabHigh}, a rational high type will therefore sabotage all other high types. In a similar fashion we can compute the bounds for all other acts of sabotage (high types sabotaging low types, low types sabotaging high types etc.). We show all bounds in the appendix. Inspecting the bounds, the following relationships between the bounds are apparent. 
% it is clear that due to the externality of sabotage, self-promotion is more effective and will happen before any type of sabotage. 
%Furthermore, high types have more to gain from sabotage than low types. Both high and low types have more to gain from sabotaging high types than low types.

\begin{lemma}
	Both high and low types have more to gain from self-promotion than from sabotaging any other types. 
	\label{selfOverSabotage}
\end{lemma}
\paragraph{Proof.} This follows directly from the externality associated with sabotage. \hfill $\blacksquare$
\vspace{0.5cm}

% ############## High Start Sabotaging High Types ##############
\subsection{High Types Start Sabotaging High Types}
\begin{lemma}
	High types have more to gain from sabotaging other agents of a given type (all low types or all high types), than low types have to gain from sabotaging those agents. 
	\label{highBenefitMore}
\end{lemma}

\paragraph{Proof.} Consider a high type of value $v_h$ not currently sabotaging anyone and sincere voting by everyone else so that $v_h = b_h (n+l+h)$ and $v_l = b_l (n+l+h)$, and $S=h v_h + l v_l$. 
High types will sabotage other high types if the cost of sabotage $c_s$ is lower than the maximum marginal increase of sabotaging the last unit (i.e., going from sabotaging $h-2$ to $h-1$).

\begin{align}
				& & \frac{v_h} {S-b_h(h-1)} - \frac{v_h} {S-b_h(h-2)} 	\\[+10pt]
	\Leftrightarrow  & & \frac{v_h(S-b_h(h-2)) -   v_h(S-b_h(h-1))}{(S-b_h(h-1))(S-b_h(h-2))} 	\nonumber\\[+10pt]
	\Leftrightarrow  & & \frac{v_h S - v_h b_h(h-2) - v_h S + v_h b_h(h-1)}{(S-b_h(h-1))(S-b_h(h-2))} 	\nonumber\\[+10pt]
	\Leftrightarrow  & & \frac{v_h b_h}{(S-b_h(h-1))(S-b_h(h-2))} \nonumber
\end{align}
\hfill $\blacksquare$
\vspace{0.5cm}

% ############## High Start Sabotaging Low Types ##############
\subsection{High Types Start Sabotaging Low Types}
There are two possible transition points into this equilibrium: High types may start sabotaging low types while high types already sabotage other high types, but no sabotage otherwise. Or high types may start sabotaging low types, after low types have already started to sabotage high types. That is, the exact bound for high types to sabotage low types, will depend on whether it falls before or after the bound of low types sabotaging high types (in terms of $c_s$) as this determines the behavior of low types, which in turn affects the costs of sabotaging behavior for high types. We compute both bounds and then compare them to establish this sequence. 

\begin{lemma}
	Both high and low types have more to gain from sabotaging high types than low types.
	\label{sabHighOverLow}
\end{lemma}

% ############### OPTION 1 ################
\subsubsection{Option H1: High types sabotage all other high types and low types do not sabotage anyone}
Consider a high type of value $v_h$ not currently sabotaging anyone and sincere voting by everyone else so that $v_h = b_h (n+l+h)$ and $v_l = b_l (n+l+h)$, and $S=h v_h + l v_l$. In this equilibrium, $h$ high types vote $-b_h$ on $h-1$ others. 

\begin{align}
				& & \frac{v_h} {S-h b_h(h-1) - b_l l} - \frac{v_h} {S-h b_h(h-1) - b_l(l-1)} 	\\[+10pt]
	\Leftrightarrow  & & \frac{v_h(S-h b_h(h-1) - b_l(l-1))    -   v_h(S-h b_h(h-1) - b_l l)}
					     {(S-h b_h(h-1)-b_l l)(S-h b_h(h-1) - b_l(l-1))} 	\nonumber\\[+10pt]
	\Leftrightarrow  & & \frac{v_h S - v_h h b_h(h-1) - v_h b_l(l-1)    - v_h S + v_h h b_h(h-1) + v_h b_l l}
					     {(S-h b_h(h-1)-b_l l)(S-h b_h(h-1) - b_l(l-1))} 	\nonumber\\[+10pt]
	\Leftrightarrow  & & \frac{v_h b_l}{(S-h b_h(h-1)-b_l l)(S-h b_h(h-1) - b_l(l-1))} \nonumber
\end{align}

% ############## OPTION 2 ##############
\subsubsection{Option H2: High and low types sabotage all high types}
Consider a high type of value $v_h$ not currently sabotaging anyone and sincere voting by everyone else so that $v_h = b_h (n+l+h)$ and $v_l = b_l (n+l+h)$, and $S=h v_h + l v_l$. In this equilibrium, $h$ high types vote $-b_h$ on $h-1$ others and $l$ low types vote $-b_h$ on $h$ high types. This turns out to be the relevant bound (see proof below).

\begin{align}
				& & \frac{v_h} {S-h b_h(h-1) - l b_h h - b_l l} - \frac{v_h} {S-h b_h(h-1) - l b_h h - b_l(l-1)} 	\\[+10pt]
	\Leftrightarrow  & & \frac{v_h (S-h b_h(h-1) - l b_h h - b_l(l-1)) - v_h (S-h b_h(h-1) - l b_h h - b_l l)}{(S-h b_h(h-1) - l b_h h - b_l l) (S-h b_h(h-1) - l b_h h - b_l l)}  	\nonumber\\[+10pt]
	\Leftrightarrow  & & \frac{v_h S  - v_h h b_h(h-1) - v_h l b_h h - v_h b_l(l-1) - v_h S + v_h h b_h(h-1) + v_h l b_h h + v_h b_l l}{(S-h b_h(h-1) - l b_h h - b_l l) (S-h b_h(h-1) - l b_h h - b_l l)}  	\nonumber\\[+10pt]
	\Leftrightarrow  & & \frac{v_h b_l}{(S-h b_h(h-1) - l b_h h - b_l l) (S-h b_h(h-1) - l b_h h - b_l l)}  	\nonumber\\[+10pt]
	\Leftrightarrow  & & \frac{v_h b_l}{(S-h b_h(h-l-1) - b_l l) (S - h b_h(h-l-1) - b_l l)} \nonumber
\end{align}
\hfill $\blacksquare$

% ############## Low Types Sabotaging High Types (while high already sabotage high) ##############
\subsection{Low Types Start Sabotaging High Types}
Similarly, there are two possible bounds when low types start to sabotage high types: one in the equilibrium where high types sabotage low types and one where they do not. 

% ############### OPTION 1 ################
\subsubsection{Option L1: High types sabotage only other high types, but no low types}
Consider a high type of value $v_h$ not currently sabotaging anyone and sincere voting by everyone else so that $v_h = b_h (n+l+h)$ and $v_l = b_l (n+l+h)$, and $S=h v_h + l v_l$. 
In this equilibrium, $h$ high types vote $-b_h$ on $h-1$ all others (high and low types) and low type start sabotaging $h$ high types with $-b_h$. This turns out to be the relevant bound (see proof below).

\begin{align}
				& & \frac{v_l} {S-h b_h(h-1) - b_h h} - \frac{v_l} {S-h b_h(h-1) - b_h(h-1)} 	\\[+10pt]
	\Leftrightarrow			& & \frac{v_l b_h} {(S-h b_h(h-1) - b_h h)(S-h b_h(h-1) - b_h(h-1))} 		\nonumber \\[+10pt]
	\Leftrightarrow			& & \frac{v_l b_h} {(S-h b_h h)(S - b_h h^2 + b_h)} \nonumber		
\end{align}

% ############### OPTION 2 ################
\subsubsection{Option L2: High types sabotage all other high types and low types}
Consider a high type of value $v_h$ not currently sabotaging anyone and sincere voting by everyone else so that $v_h = b_h (n+l+h)$ and $v_l = b_l (n+l+h)$, and $S=h v_h + l v_l$. 
In this equilibrium, $h$ high types vote $-b_h$ on $h-1$ other high types and vote $-b_l$ on $l$ low types and low types start sabotaging $h$ high types with $-b_h$.

\begin{align}
				                 & & \frac{v_l} {S-h b_h(h-1) - h b_l l - b_h h} - \frac{v_l} {S-h b_h(h-1)- h b_l l - b_h(h-1)} 	\\[+10pt]
	\Leftrightarrow			& & \frac{v_l b_h} {(S-h b_h(h-1) - h b_l l - b_h h)(S-h b_h(h-1)- h b_l l - b_h(h-1))}  	\nonumber	\\[+10pt]
	\Leftrightarrow			& & \frac{v_l b_h} {(S-h b_h h - h b_l l )(S + b_h - b_h h^2 - h b_l l)} \nonumber
\end{align}

% ############## PROOF: Order of switches ##############
\paragraph{Proof: Order of Bounds.}
To establish the correct order of bounds, we show that maximum marginal gain for low types to sabotage high types is higher than the maximum marginal gain for high types to sabotage low types.

\begin{align}
				& & \frac{v_h b_l}{(S-h b_h(h-1)-b_l l)(S-h b_h(h-1) - b_l(l-1))} < \frac{v_l b_h} {(S-h b_h h)(S - b_h h^2 + b_h)} 	%\\[+10pt]
%				& & \frac{v_h b_l}{(S-h b_h(h-1)-b_l l)(S-h b_h(h-1) - b_l(l-1))} - \frac{v_k b_h} {(S-h b_h h)(S - b_h h^2 + b_h)} < 0 	\\[+10pt]
%				& & \frac{v_h b_l [(S-h b_h h)(S - b_h h^2 + b_h)] - v_k b_h [(S-h b_h(h-1)-b_l l)(S-h b_h(h-1) - b_l(l-1))]} {[(S-h b_h(h-1)-b_l l)(S-h b_h(h-1) - b_l(l-1))] [(S-h b_h h)(S - b_h h^2 + b_h)]} < 0 	\\[+10pt]
\end{align}

% The real proof is: just plug in numbers and see that it holds. To see this, check $bounds.R$. 
This inequality holds given $v_l < v_h$ and $b_l < b_h$ which are true by construction.\hfill $\blacksquare$
\vspace{0.5cm}

This shows that low types have more to gain from sabotaging high types than high types have to gain from sabotaging low types. Consequently, as the cost for sabotage decrease, low types will switch from not sabotaging anyone to sabotaging high types before high types will switch from sabotaging only other high types to sabotaging all other high types and low types.
As a consequence, the relevant bound for low types switching to sabotaging high types is the one where only high types sabotage high types but no low types (option H2 above). 
Furthermore, the correct bound for high types to sabotage low types is one where everyone already sabotages high types (option L1 above).
Given that we have now established a sequence of bounds, we establish that the set of possible Nash equilibria is NE = \{no sab, h sab h, l sab h, h sab l, l sab l\}

% ####### Nash Equilibria ###########
\subsection{Nash Equilibria}
This contest has many Nash equilibria. Which equilibrium materializes depends on the size of the groups $n$, $l$, and $h$, the costs $c_s$ and $c_p$, and the performance gap $g$ between high and low types. 

\begin{proposition} \label{proposition}
	In contests with a positive performance gap $g \ge 1$ between high and low types, the following seven states are the set of possible Nash equilibria.
\end{proposition}
% \fxnote{Isnt this dependent on measuring the costs on the same dimension? Only then all EQ without selfpromotion but sabotage do not manifest.}

\begin{enumerate}[(NE1)]
	\item $s_i = (0, 0, 0)$ for $i \in H$, and $s_k = (0, 0, 0)$ for $k \in L$. No one self-promotes, and no one sabotages anyone.
	\item $s_i = (0, 0, 0)$ for $i \in H$, and $s_k = (0, 0, 1)$ for $k \in L$. Low types engage in self-promotion, and no one sabotages anyone.
	\item $s_i = (0, 0, 1)$ for $i \in H$, and $s_k = (0, 0, 1)$ for $k \in L$. All agents engage in self-promotion, and no one sabotages anyone.
	\item $s_i = (h-1, 0, 1)$ for $i \in H$, and $s_k = (0, 0, 1)$ for $k \in L$. All agents engage in self-promotion, high types sabotage each other but do not sabotage low types, and low types do not sabotage.
	\item $s_i = (h-1, 0, 1)$ for $i \in H$, and $s_k = (h, 0, 1)$ for $k \in L$. All agents engage in self-promotion, high types sabotage each other but do not sabotage low types, and low types sabotage only high types.
	\item $s_i = (h-1, l, 1)$ for $i \in H$, and $s_k = (h, 0, 1)$ for $k \in L$. All agents engage in self-promotion, high types sabotage all other agents of both types, and low types sabotage only high types.
	\item $s_i = (h-1, l, 1)$ for $i \in H$, and $s_k = (h, l-1, 1)$ for $k \in L$. All agents engage in self-promotion, and all agents sabotage all other agents of both types.
\end{enumerate}

\paragraph{Proof.} This follows directly from Lemma \ref{lemma1} (low types will engage in self-promotion before high types do), Lemmas \ref{selfOverSabotage}-\ref{sabHighOverLow} which give an order of the bounds, and the additional proof that the benefit for low types sabotaging high types is larger than the benefit of high types sabotaging low types.\hfill $\blacksquare$
\vspace{0.5cm}

% ##### Revisit Performance Gap g #######
\subsection{Revisiting the Performance Gap $g$} \label{sec:PerformanceGap}
It is now interesting to revisit the condition of $g \ge 1$ for self-promotion. It is clear that the closer $b_h$ is to $1$, the less high types have to gain from self-promotion. Consequently, they will only engage in self-promotion if the costs $c_s$ for doing so are increasingly smaller. In the equilibrium condition NE4, for example, in which high types sabotage each other and low types do not sabotage anyone, then $g \ge 1$ is violated if $(n+h+l)b_l > (n+l+1)b_h$ and the resulting gain from self-promotion would be lower for low types than high types. This would then lead to the interesting case where low types do not self-promote while high types do. Considering both self-promotion and sabotage, it is interesting to note the following opposing effect of the performance gap: The larger the performance gap, the more attractive \textit{self-promotion} becomes for \textit{low types} and the more attractive \textit{sabotage} becomes for \textit{high types}. As a result, in contests with a large performance gap, strategic behavior levels the playing field, thus making the contest more attractive to low types.

From the argument above, we can draw another observation: Self-promotion acts as an equalizing mechanism that reduces the performance difference between high and low-skill agents and thus reduces incentives for high-skill agents to sabotage (this can be seen by calculating and comparing Equation \ref{eq:boundHighSabHigh} in our model with and without self-promotion).

% ####### Example ###########
% Remember: Yes, there is just one boundary when agents start to sabotage high types (i.e., it's not a band), but there is a second boundary when agents start to also sabotage low types. So if the purpose of this plot is to show the equilibrium of interest, then yes, it is going to be a band where if costs are too low, we call out of the equilibrium.
\subsection{Example and Additional Predictions} 
%\fxnote{Ben Greiner advised to justify numbers in example e.g. from empirical setting. Furthermore, additional predictions is not intuitive because we havent mentioned predictions before explicitly.}
% TG: Note: We could modify this example to reflect a different empirical context e.g. with numbers on grant funding from NSF, topcoder or www.challenges.gov.
% CR: I think the example is the best we can do. If we make it more realistic, the cost get really low / unrealistic.
% Justifing values is dificult since they are not particularly realistic. 

\begin{figure}[tbh!]
	\begin{center}
		\includegraphics[width=0.80\linewidth]{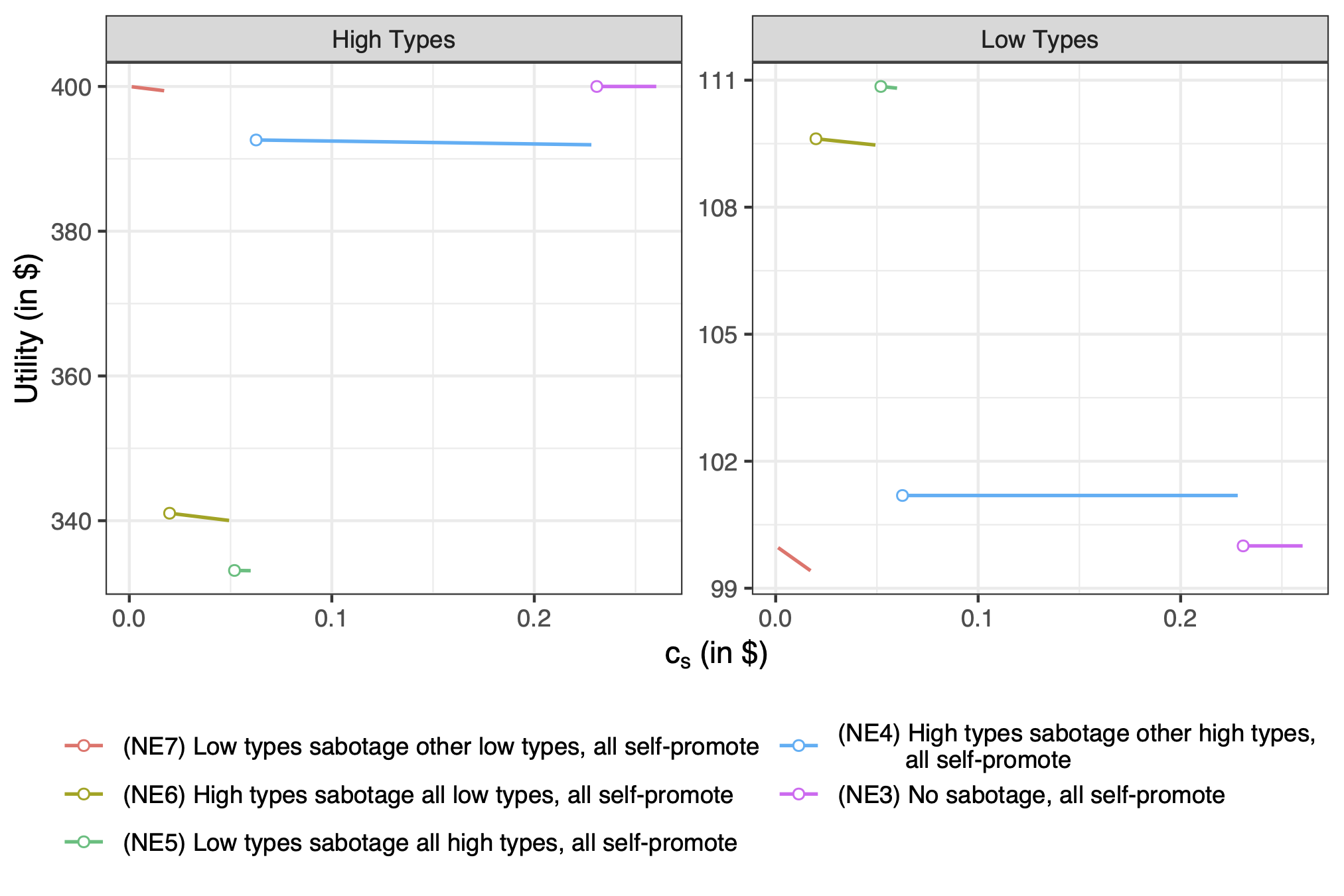}
	\end{center}
	\caption{Equilibria and Utilities.}
	\begin{flushleft}
		\begingroup
		\leftskip4em
		\rightskip\leftskip
		\small{\textit{Note: } According to Lemma 4, all equilibria with some form of sabotage also include self-promotion from all agents.}
		\par
		\endgroup
	\end{flushleft}
	\label{fig:predictionsNEs}
\end{figure}

Let the prize be $M=\$5,000$ and let $h=10$, $l=30$, and $n=100$. Further, let $b_h=0.8$ and $b_l=0.2$ (this corresponds to a common case of a finite rating scale from 0-stars to 5-stars where low type agents produce contest entries of 1-star quality and high type agents produce contest entries of 4-star quality). The resulting performance gap between low and high types is then $g= \frac{b_h}{b_l} \frac{n+1}{n+l+h} =\frac{0.8}{0.2}\frac{100+1}{100+30+10} > 1$ which means that high types always have higher values than low types, even when they are sabotaged. Figure \ref{fig:predictionsNEs} shows agent utility ($Mp$) as a function of cost of sabotage $c_s$. The figure illustrates the transitions between equilibria and their relative sizes in terms of the range of $c_s$ values spanned. The figure suggests that there is one equilibrium in particular that has a large basin of attraction that spans a wide range of $c_s$ values that seem plausible in an online contest (\$0.06 to \$0.23): all agents self-promote and high types sabotage other high types while low types do not sabotage anyone (NE4).

% ###### "Old" version to  make predictions about contest size based on example

% For OrgSci R4 we are taking out contest size
\begin{mycomment}
From the equilibrium predictions above, we develop additional predictions regarding the relationship between strategic behavior (self-promotion and sabotage) and the number of competitors (contest size). The panel in Figure \ref{fig:predictionsCosts} shows some key predictions for changing contest sizes. Regarding self-promotion, comparative statics suggest decreasing marginal utility for this form of strategic behavior. We predict that the response to increased competition is a reduction in self-promotion for both high- and low-skill agents. That is, we predict self-promotion to be less prevalent in more competitive contests.\label{Prediction: SelfPromotionContestSize} 

For sabotage, we predict a broad band of $c_s$ values in which high-skill agents sabotage other high-skill agents. For this prediction, we relax the empirically unlikely assumption that agents face homogeneous costs. Unless the actual cost of sabotage for an agent is exactly at the upper bound, that agent will continue to sabotage all other agents even if one additional high type agent were to join the contest. That is, only if the cost faced by an agent is exactly at the upper limit, will that agent move from sabotaging all other agents to not sabotaging anyone.
Thus, increasing the contest by one additional high type will lead to $2h$ additional acts of sabotage (each of the $h$ agents in the contest sabotages the additional high type, and the additional high type sabotages all $h$ already in the contest). We predict that the response to increased competition is an increase in sabotage by all agents.\label{Prediction: SabotageContestSize}

\begin{figure}[tbh!]
	\begin{center}
		\begin{subfigure}{0.40\linewidth}
			\includegraphics[width=\linewidth]{../../../FIGURES/Rplot-Promotion-ContestSize.png}
			\caption{\textsc{Self-Promotion}}
		\end{subfigure}
		\begin{subfigure}{0.40\linewidth}
			\includegraphics[width=\linewidth]{../../../FIGURES/Rplot-Sabotage-ContestSize.png}
			\caption{\textsc{Sabotage}}
		\end{subfigure}
	\end{center}
	\caption{Equilibrium Strategies for High-Types.}
	\begin{flushleft}
		\begingroup
		\leftskip4em
		\rightskip\leftskip
		\small{\textit{Note: }(a) Marginal utility of self-promotion in contest of changing size. 
			(b) Upper and lower bound of $c_s$ in contests of changing size. If cost $c_s$ fall below this lower bound, low types will also sabotage high types.}
		\par
		\endgroup
	\end{flushleft}
	\label{fig:predictionsCosts}
\end{figure}
\end{mycomment}
%######################################### end contest size

% ####################################
% ########   Empirical Analyses   ##########
% ####################################
\clearpage
\section{Additional Empirical Analyses and Robustness Tests}

\subsection{Distribution of Skill and Contest Size} \label{sec:ExtendedDescriptives}
\begin{figure}[tbh!]
	\begin{center}
		\begin{subfigure}{0.40\linewidth}
			\includegraphics[width=\linewidth]{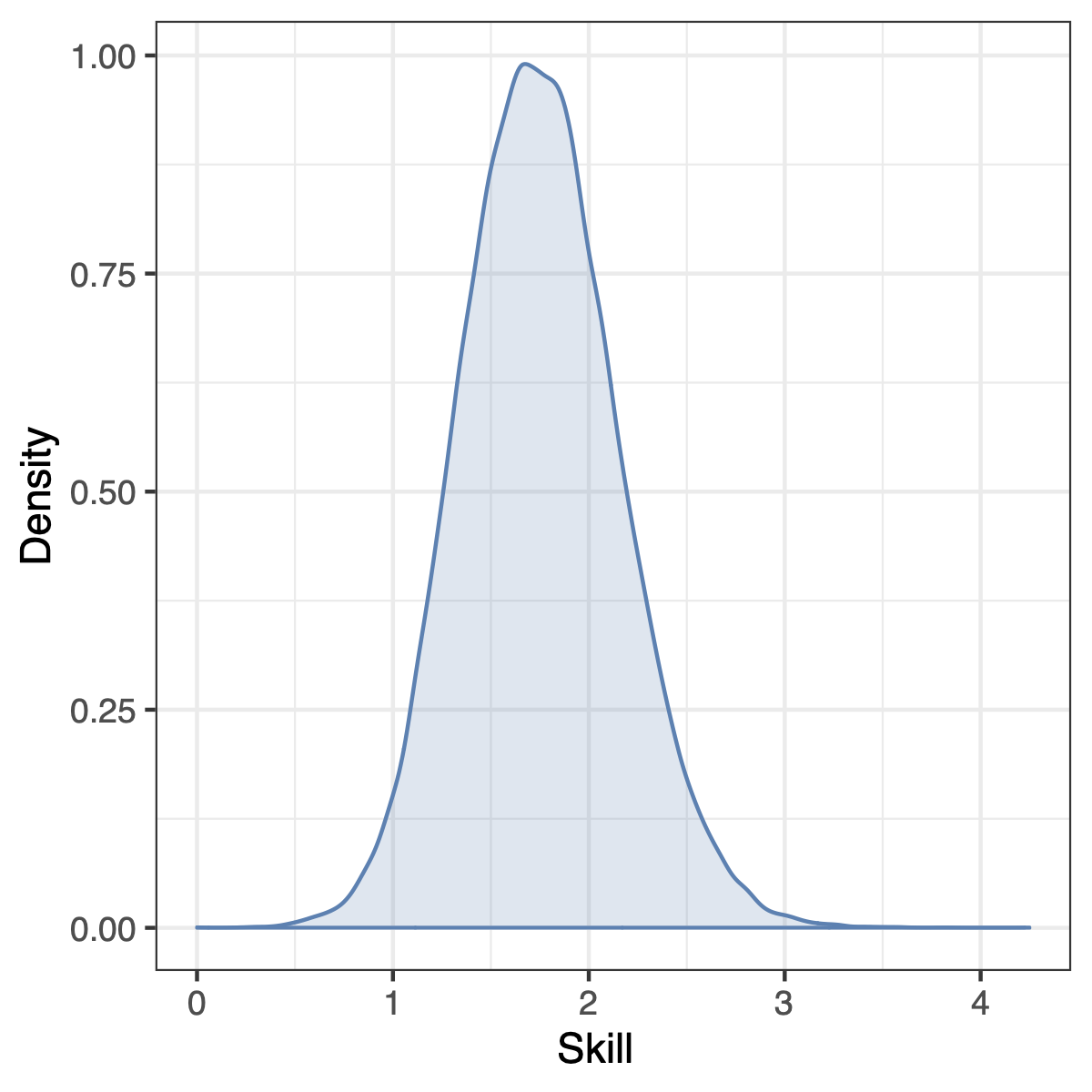}
			\caption{\textsc{Distribution of Skill}}
		\end{subfigure}
		\begin{subfigure}{0.40\linewidth}
			\includegraphics[width=\linewidth]{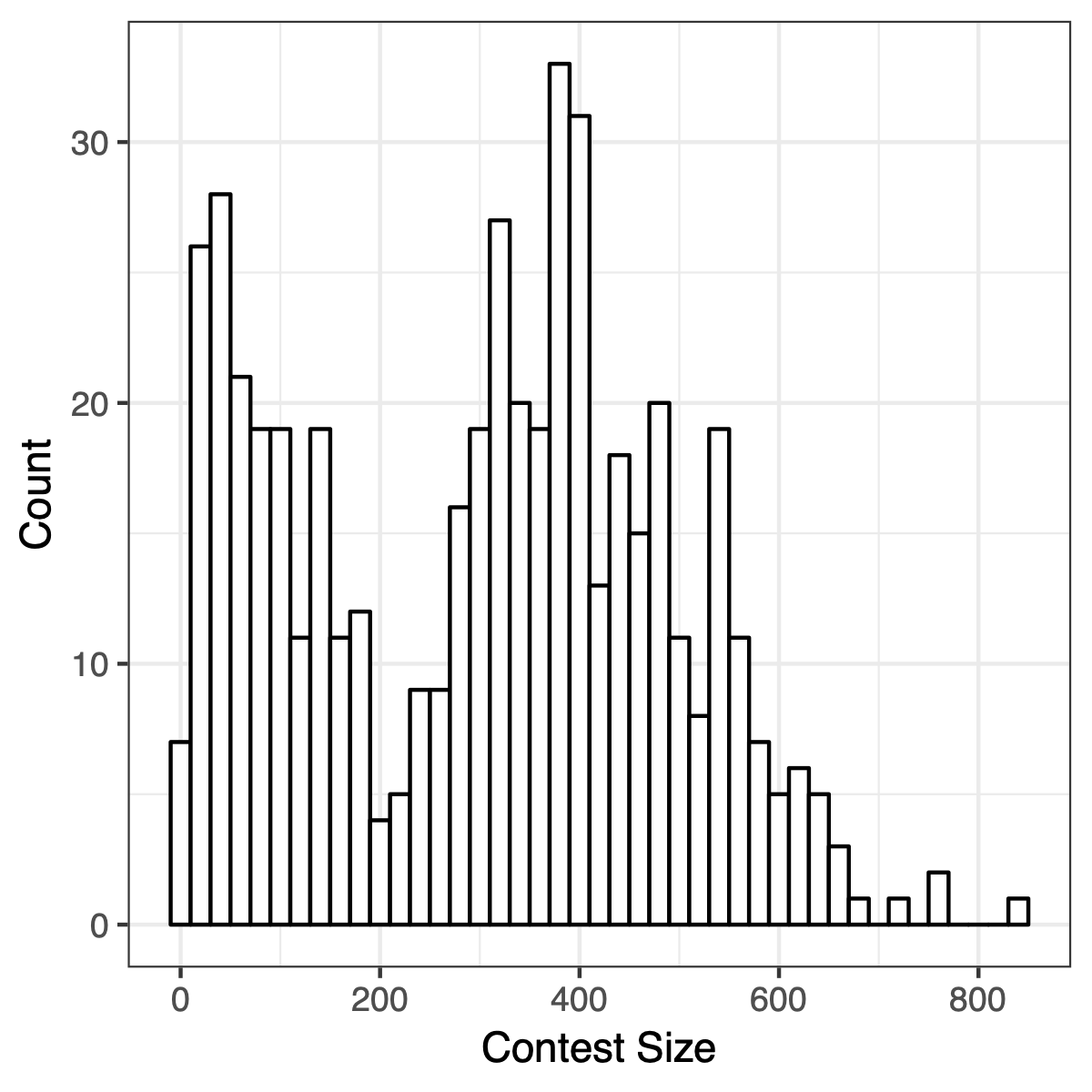}
			\caption{\textsc{Distribution of Contest Size}}
		\end{subfigure}
	\end{center}
	\caption{Distributions of Skill and Contest Size.}
	\begin{flushleft}
		\begingroup
		\leftskip4em
		\rightskip\leftskip
		\small{\textit{Note: }(a) We compute skill as the average rating of all previous submissions by a designer. The skill distribution is right skewed (skewness: 0.21). % 0.21 mean; 0.37 max
			(b) Distribution of contest sizes (i.e., number of contest entries; mean: 302).}
		\par
		\endgroup
	\end{flushleft}
	\label{fig:ContestSize}
\end{figure}

\subsection{Career Length and Participation} \label{sec:ExtendedDescriptives}
On average, competitors have a career length from their first participation to their last participation that spans 42 contests. During their active tenure, they participate in some form in 9.1 contests, submit their own design in 1.5 contests and rate in 9.0. Submitting to a contest tends to be spaced out between periods of just rating. On average, competitors have 1.4 ``flips'' in which they go from rating as outsiders to rating as competitors, or from rating as competitors to rating as outsiders. These changes are the basis of our identification: we contrast rating behavior from weeks in which a participant rates (9 contests on average) and weeks in which participants submit their own design while also rating (on average 1.4 contests). 

\clearpage

\subsection{Rating Distribution of Contest Winners}
\begin{figure}[tbh!]
	\begin{center}
	\includegraphics[width=0.60\linewidth]{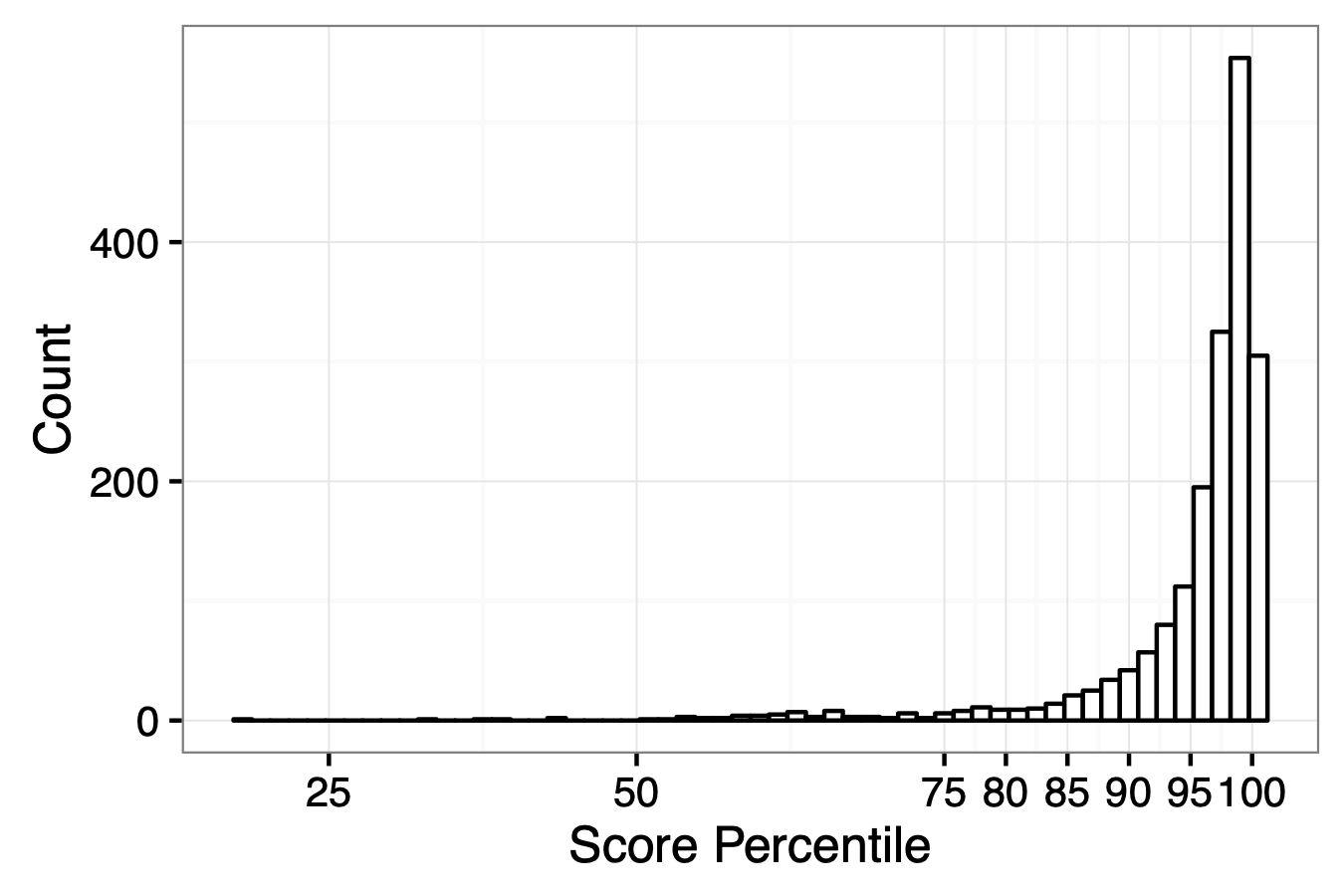}
	\end{center}
	\vspace{-5mm}
	\caption{Percentile score of contest winners.}
	\begin{flushleft}
		\begingroup
		\leftskip4em
		\rightskip\leftskip
	 \small{\textit{Note: }Contest winners scored in the highest percentiles among all submissions in a given week. The median percentile score was 98th percentile.}
	\par
	\endgroup
	\end{flushleft}
	\label{fig:appendix:ratingDistribution}
\end{figure}

We show the percentile scores of contest winners in Figure \ref{fig:appendix:ratingDistribution}.
% \clearpage

% #################################################
% ##                              Heterogeneity                                     ##
% #################################################
\subsection{Heterogeneous Contestant Skill} \label{appendix:Het}
\paragraph{Subsample.} As we move from our overall analysis (Table \ref{table:mainRegression}) to our examination of heterogeneity by skill (Table \ref{tab:HetEffect}), we have to restrict our analysis to a subsample of the data: observations for which skill measures for both the source (i.e., competitor casting the rating) and the target (i.e., competitor who submitted the contest entry being rated) are available. This means we have to exclude ratings by competitors who have never before submitted (and hence source skill is unknown) and ratings of submissions by first-time submitters (for whom target skill is unknown). In the interpretation of the main results, one may consequently worry about sampling of subsets of the data. Table \ref{table:robustnessSubSample} shows the main results across subsamples. We repeat the analysis of the full data (Model 1), and the subset for which both skill measures are missing (Model 4; i.e., the data used in Table \ref{tab:HetEffect}). 

Subsamples do not appear to drive our effects. The coefficients for sabotage remain stable across the subsamples ranging from $-0.007$ to $-0.011$. Never-before submitters are harsh raters and excluding these data increases the positivity bias (Model 2). Competitors greet first-timers with more positive ratings and the positivity bias decreases when we exclude those data (Model 3).
%\fxnote{We go from a large positivity bias (large negative coefficient) to a smaller positivity bias (small negative coefficient). so the data we exclude must have been harsh} The fact that positivity bias increases slightly in the subsample, suggests that the findings on heterogeneity are not driven by the use of a subsample but may be even more pronounced in the full sample.

% neg sign: positive bias -- more positive ratings -- lower likelihood to rate zery
% larger neg: even lower likelihood to rate zero -- even more positive -- larger positivity bias

\begin{table}[h!]
\begin{center}
\begin{scriptsize}
\begin{tabular}{@{\extracolsep{5pt}}l D{.}{.}{3.3} D{.}{.}{3.3} D{.}{.}{3.3} D{.}{.}{3.3} D{.}{.}{3.3} }
\toprule
 & \multicolumn{4}{c}{Linear Probability} \\
 \cline{2-5} \\[-5pt]
 Dependent Variable: & \multicolumn{4}{c}{\textbf{Sabotage:} 0-Star Rating} \\
 \cline{2-5} \\[-5pt]
 & \multicolumn{1}{c}{} 		& \multicolumn{1}{c}{Exclude ratings by  } 		& \multicolumn{1}{c}{Exclude ratings of} 			& \multicolumn{1}{c}{} \\
 & \multicolumn{1}{c}{All data} 	& \multicolumn{1}{c}{never before submitters} 	& \multicolumn{1}{c}{submissions by first-timers} 	& \multicolumn{1}{c}{Exclude both} \\
 \cline{2-2}  \cline{3-3} \cline{4-4} \cline{5-5} \\[-5pt]
 & \multicolumn{1}{c}{(1)} & \multicolumn{1}{c}{(2)} & \multicolumn{1}{c}{(3)} & \multicolumn{1}{c}{(4)} \\
\midrule
Submitted to same contest: Yes & -0.008^{***} & -0.011^{***} & -0.007^{***} & -0.011^{***} \\
                               & (0.000)      & (0.000)      & (0.000)      & (0.000)      \\
Rate own submission: Yes       & -0.195^{***} & -0.195^{***} & -0.186^{***} & -0.186^{***} \\
                               & (0.001)      & (0.001)      & (0.001)      & (0.001)      \\
\textit{Individual} 		& \multicolumn{1}{c}{\textit{Fixed}} & \multicolumn{1}{c}{\textit{Fixed}}& \multicolumn{1}{c}{\textit{Fixed}}& \multicolumn{1}{c}{\textit{Fixed}}  \\
\textit{Submission}		& \multicolumn{1}{c}{\textit{Fixed}} & \multicolumn{1}{c}{\textit{Fixed}}& \multicolumn{1}{c}{\textit{Fixed}}& \multicolumn{1}{c}{\textit{Fixed}}  \\
\midrule
%R$^2$ (full model)             & 0.375        & 0.395        & 0.372        & 0.391        \\
%R$^2$ (proj model)             & 0.001        & 0.002        & 0.001        & 0.002        \\
Adj. R$^2$        & 0.372        & 0.391        & 0.367        & 0.386        \\
%Adj. R$^2$ (proj model)        & -0.005       & -0.006       & -0.006       & -0.006       \\
Num. obs.                      & \multicolumn{1}{c}{38,102,880}     & \multicolumn{1}{c}{27,188,751}     & \multicolumn{1}{c}{26,317,775}     & \multicolumn{1}{c}{18,787,584}     \\
\bottomrule
\multicolumn{5}{l}{\tiny{$^{***}p<0.001$, $^{**}p<0.01$, $^*p<0.05$}}
\end{tabular}
\end{scriptsize}
\caption{Analysis of subsamples.}
\label{table:robustnessSubSample}
\end{center}
\end{table}

% #################################################
% ##                         Cluster Level of SE                                  ##
% #################################################
\subsection{Robustness of Cluster Level of Standard Errors}
In the main text, we cluster standard errors to account for possible dependence of ratings of the same submissions and in the same contest. One may alternatively be concerned about capturing autocorrelation at the rater level. In table \ref{tab:HetEffectMeanIndivLevel} we examine the robustness of our specification and cluster on the individual level instead. While we do see some changes in the standard errors, $p$-values are generally much smaller than $0.001$ (i.e., $1e-7$) so that our substantial conclusions are not affected by the level of clustering of standard errors.

\begin{table}[h!]
\begin{center}
\begin{scriptsize}
\begin{tabular}{@{\extracolsep{5pt}}l D{.}{.}{7.4} D{.}{.}{7.4}}
\toprule
 & \multicolumn{2}{c}{Linear Probability} \\
 \cline{2-3}\\[-5pt]
 Dependent Variable: & \multicolumn{1}{c}{\bf Sabotage} 	 & \multicolumn{1}{c}{\bf Self-Promotion} \\
				 & \multicolumn{1}{c}{0-Star Rating} & \multicolumn{1}{c}{5-Star Rating} \\
 \cline{2-2} \cline{3-3}\\[-5pt]
 & \multicolumn{1}{c}{(1)} & \multicolumn{1}{c}{(2)} \\
 \midrule
Submitted to same contest: Yes                         & -0.057^{***} & -0.000       \\
                                                       & (0.008)      & (0.001)      \\
Rate own submission: Yes                               & -0.179^{***} & 0.996^{***}  \\
                                                       & (0.003)      & (0.016)      \\
Target skill                                         & -0.112^{***} & 0.008        \\
                                                       & (0.011)      & (0.012)      \\
Source skill                                         & -0.049^{**}  & -0.105^{***} \\
                                                       & (0.017)      & (0.015)      \\
Target skill $\times$ Source skill                 & 0.007        & 0.042^{***}  \\
                                                       & (0.006)      & (0.006)      \\
Submitted to same contest: Yes \\
\quad $\times$ Target skill & 0.015^{***}  &              \\
                                                       & (0.003)      &              \\
\quad $\times$ Source skill & 0.010^{***}  &              \\
                                                       & (0.003)      &              \\
Rate own submission: Yes $\times$ Source skill       &              & -0.080^{***} \\
                                                       &              & (0.009)      \\
\textit{Individual} 		& \multicolumn{1}{c}{\textit{Fixed}} & \multicolumn{1}{c}{\textit{Fixed}} \\
\textit{Contest}		& \multicolumn{1}{c}{\textit{Fixed}} & \multicolumn{1}{c}{\textit{Fixed}} \\
\midrule
%R$^2$ (full model)                                     & 0.361        & 0.210        \\
%R$^2$ (proj model)                                     & 0.014        & 0.058        \\
Adj. R$^2$                                & 0.360        & 0.208        \\
%Adj. R$^2$ (proj model)                                & 0.012        & 0.056        \\
Num.\ obs.                                             & \multicolumn{2}{c}{18,787,584} \\
\bottomrule
\multicolumn{3}{l}{\tiny{$^{***}p<0.001$, $^{**}p<0.01$, $^*p<0.05$}}
\end{tabular}
\end{scriptsize}
\caption{Robustness test of heterogeneous effects by skill level with errors clustered on the individual-level.}
\vspace{6pt}
\begin{flushleft}
	\begingroup
	\leftskip4em
	\rightskip\leftskip
\small{\textit{Note:} Skill is measured as the average quality of all past submissions. Contest-level fixed effects are used instead of submission-level as the ability of agents is time invariant at the submission level. Standard errors are in parentheses, clustered at the individual level.}
\par
\endgroup
\end{flushleft}
\label{tab:HetEffectMeanIndivLevel}
\end{center}
\end{table}

% #################################################
% ##                         Measure of Skill                                     ##
% #################################################
\clearpage
\subsection{Skill Measure}
\begin{figure}[ht]
	\begin{center}
		\begin{subfigure}{0.49\linewidth}
			\includegraphics[width=\linewidth]{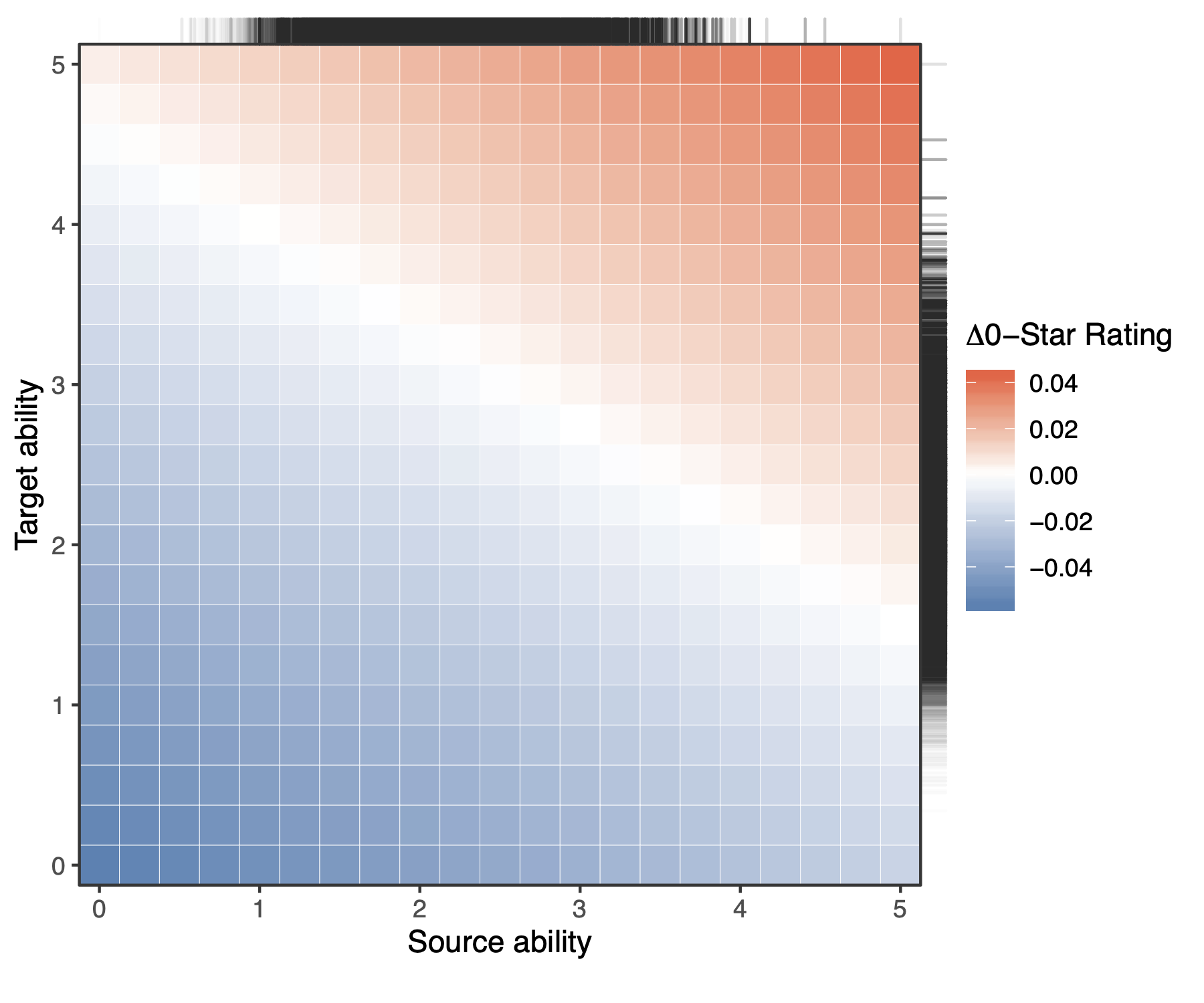}
			\caption{\textsc{Sabotage}}
		\end{subfigure}
		\begin{subfigure}{0.40\linewidth}
			\includegraphics[width=\linewidth]{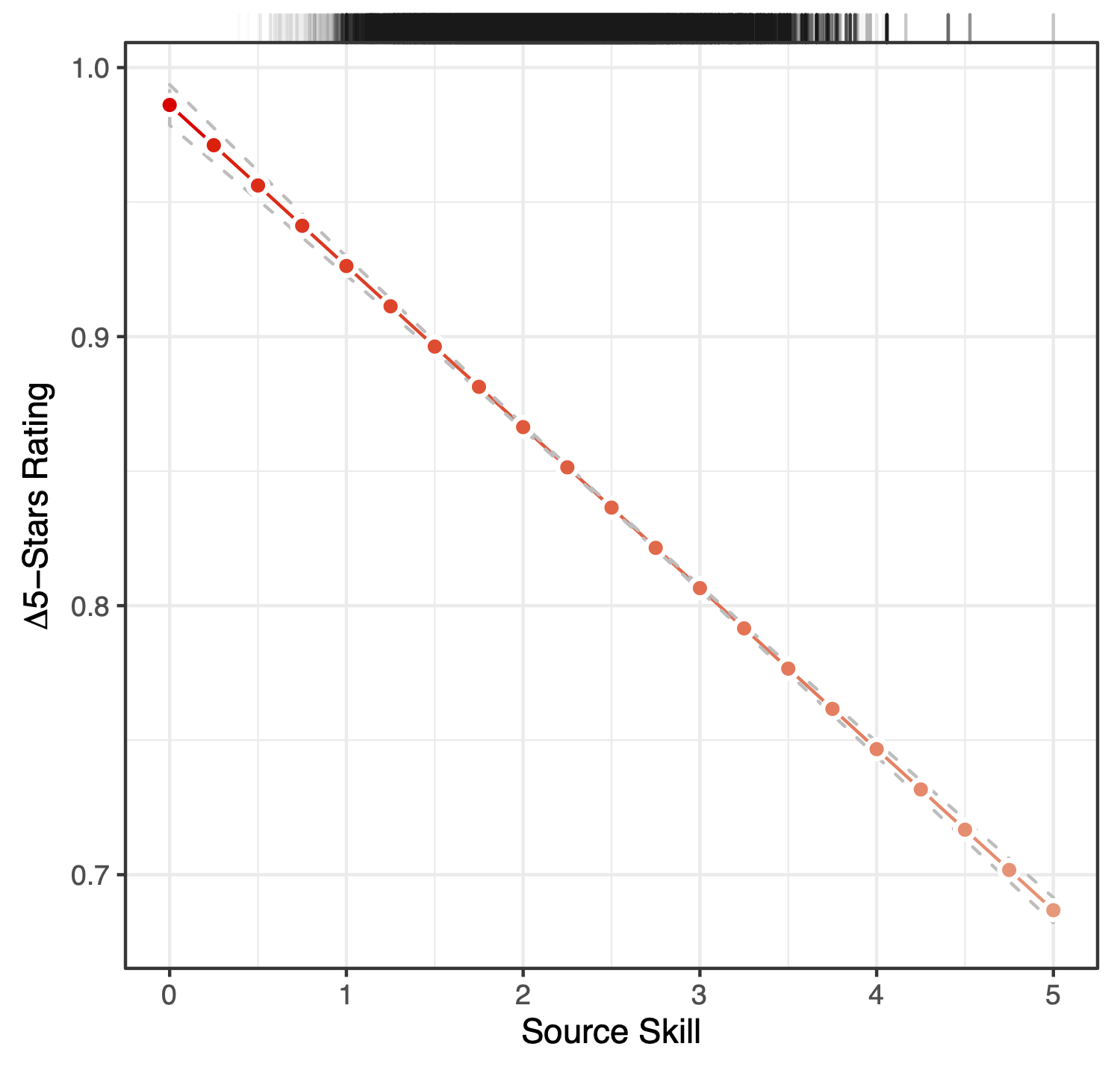}
			\caption{\textsc{Self-Promotion}}
		\end{subfigure}
		\caption{Strategic behavior by competitors of heterogeneous skill (max skill).}
		\begin{flushleft}
			\begingroup
			\leftskip4em
			\rightskip\leftskip
			\small{\textit{Note: }\textbf{Panel A (Sabotage).} Relative change in probability of rating 0-star when competing compared to not competing across skill levels. Outer margins show the distribution of data. There are over 15,000 (10,000) observations for source (target) skill greater than 4.5.
				\textbf{Panel B (Self-promotion).} Relative change in probability of rating 5-star when rating own submission compared to submissions by others of the same skill (error band is 95\% confidence interval).} 
			\par
			\endgroup	
		\end{flushleft}	% CONFIRMED: Check with code, it is 95% CI not SE
		
		\label{fig:HetMax}
	\end{center}
\end{figure}

The analyses presented in the main body measure the idea generation skill of individuals as the average quality of all past submissions. 
We repeat that analysis using the quality of the best design they submitted in the past (Table \ref{tab:HetEffectMax} and Figure \ref{fig:HetMax}). We find very similar results.

\begin{table}[h!]
\begin{center}
\begin{scriptsize}
\begin{tabular}{@{\extracolsep{5pt}}l D{.}{.}{7.4} D{.}{.}{7.4}}
\toprule
 & \multicolumn{2}{c}{Linear Probability} \\
 \cline{2-3}\\[-5pt]
 Dependent Variable: & \multicolumn{1}{c}{0-Star Rating} & \multicolumn{1}{c}{5-Star Rating} \\
 \cline{2-2} \cline{3-3}\\[-5pt]
 & \multicolumn{1}{c}{(1)} & \multicolumn{1}{c}{(2)} \\
 \midrule
Submitted to same contest: Yes                         & -0.055^{***} & -0.000       \\
                                                       			& (0.004)      & (0.001)      \\
Rate own submission: Yes                               	& -0.179^{***} & 0.986^{***}  \\
                                                       			& (0.004)      & (0.007)      \\
Target max skill                                         		& -0.066^{***} & -0.001       \\
                                                       			& (0.002)      & (0.001)      \\
Source max skill                                         		& -0.019^{***} & -0.084^{***} \\
                                                       			& (0.004)      & (0.003)      \\
Target max skill $\times$ Source max skill                 	& 0.002        & 0.027^{***}  \\
                                                       			& (0.001)      & (0.001)      \\
Submitted to same contest: Yes $\times$ Target max skill & 0.012^{***}  &              \\
                                                       			& (0.001)      &              \\
Submitted to same contest: Yes $\times$ Source max skill & 0.008^{***}  &              \\
                                                       			& (0.001)      &              \\
Rate own submission: Yes $\times$ Source ability       &              & -0.060^{***} \\
                                                       			&              & (0.003)      \\
\textit{Individual} 		& \multicolumn{1}{c}{\textit{Fixed}} & \multicolumn{1}{c}{\textit{Fixed}} \\
\textit{Contest}		& \multicolumn{1}{c}{\textit{Fixed}} & \multicolumn{1}{c}{\textit{Fixed}} \\
\midrule
% NOTE: DEFINITELY ContestID fe, NOT submissionID
% Cluster ContestID SE
%R$^2$                                                  & 0.361        & 0.213        \\
Adj.\ R$^2$                                            & 0.360        & 0.211        \\
Num.\ obs.                                             & \multicolumn{2}{c}{18,787,584} \\
\bottomrule
\multicolumn{3}{l}{\tiny{$^{***}p<0.001$, $^{**}p<0.01$, $^*p<0.05$}}
\end{tabular}
\end{scriptsize}
\caption{Estimates of heterogeneous effects using quality of the best prior design as skill measure.}
\begin{flushleft}
	\begingroup
	\leftskip4em
	\rightskip\leftskip
	\small{\textit{Note:} Contest-level fixed effects are used instead of submission-level because the skill of agents is time invariant at the submission level. Standard errors are in parentheses, clustered at the contest level.}
	\par
	\endgroup
\end{flushleft}
\label{tab:HetEffectMax}
\end{center}
\end{table}

% #############################################
% ##              Sabotage - Sources of Sabotage                 ##
% #############################################
\clearpage
\subsection{Sources of Sabotage}
%\textit{Note: There are actually several different mechanisms the reviewers are hinting at: (1) Why you sabotage (at all); (2) why not sabotage more (the cost side of it); (3) changes over time (dynamics); and (4) heterogeneity (why some people might do it more than others). I was first thinking about structuring this section in a way that tackles the different types of mechanisms but I realized that coming up with a good structure is difficult and that giving it structure creates the impression of a level of comprehensiveness that this section maybe doesn't have. I now attempt to keep it more conversational and less formal theory testing ``we test if it moral costs are driving it''.}

In this section we explore tenure, experience, and being a past winner as different mechanism behind the decision to sabotage others (Table \ref{tab:Mechanisms}). Similar to the analysis in the main text, we introduce interaction terms between the \textit{Submitted to same contest} dummy with different time-varying measures of individual attributes. All models control for source and target skill and thus need to be interpreted as drivers above and beyond the finding that sabotage is predominantly used among high-skill competitors targeting other high-skill competitors.

% \input{../../01-Analysis-Writeup-LaTeX/TABLE_MECH0_ALL_clusterSE_MANUAL}
% used in OrgSci R1 - \input{../../01-Analysis-Writeup-LaTeX/TABLE_MECH0_ALL_clusterSE_MANUAL_SIMPLE}
% used in OrgSci R2 
\begin{table}[h!]
\begin{center}
\begin{scriptsize}
\begin{tabular}{@{\extracolsep{5pt}}l D{.}{.}{3.3} D{.}{.}{3.3} D{.}{.}{3.3}   }
\toprule
 & \multicolumn{3}{c}{Linear Probability} \\
 \cline{2-4}\\[-5pt]
 Dependent Variable: & \multicolumn{3}{c}{\textbf{Sabotage:} 0-Star Rating} \\
				\cline{2-4} \\[-5pt]
				& \multicolumn{1}{c}{Tenure} 	& \multicolumn{1}{c}{Experience} & \multicolumn{1}{c}{Past Success} \\
				\cline{2-2} \cline{3-3} \cline{4-4}  \\[-5pt]
				& \multicolumn{1}{c}{(1)} & \multicolumn{1}{c}{(2)} & \multicolumn{1}{c}{(3)}  \\
\midrule
Submitted to same contest: Yes                                & -0.068^{***} & -0.049^{***} & -0.058^{***} \\
                                                              & (0.001)      & (0.001)      & (0.001)      \\
Rate on own submission: Yes                                   & -0.192^{***} & -0.196^{***} & -0.194^{***} \\
                                                              & (0.001)      & (0.001)      & (0.001)      \\
Tenure (in weeks)                                             & 0.017^{***}  &              &              \\
                                                              & (0.000)      &              &              \\
Rating experience (log)                                       &              & 0.031^{***}  &              \\
                                                              &              & (0.000)      &              \\
Submission experience (log)                                   &              & -0.010^{***} &              \\
                                                              &              & (0.000)      &              \\
Source is past winner                                         &              &              & -0.026^{***} \\
                                                              &              &              & (0.000)      \\
Target is past winner                                         &              &              & 0.012^{***}  \\
                                                              &              &              & (0.000)      \\
Submitted to same contest: Yes \\
\quad $\times$ Tenure                & 0.000        &              &              \\
                                                              & (0.000)      &              &              \\
\quad  $\times$ Rating experience     &              & -0.010^{***} &              \\
                                                              &              & (0.000)      &              \\
\quad  $\times$ Submission experience &              & 0.011^{***}  &              \\
                                                              &              & (0.000)      &              \\
\quad  $\times$ Source is past winner &              &              & 0.021^{***}  \\
                                                              &              &              & (0.004)      \\
\quad  $\times$ Target is past winner &              &              & 0.013^{***}  \\
                                                              &              &              & (0.000)      \\
\textit{Skill controls}		& \multicolumn{1}{c}{\textit{Yes}} & \multicolumn{1}{c}{\textit{Yes}} & \multicolumn{1}{c}{\textit{Yes}} \\
\textit{Individual} 		& \multicolumn{1}{c}{\textit{Fixed}} & \multicolumn{1}{c}{\textit{Fixed}} & \multicolumn{1}{c}{\textit{Fixed}} \\
\textit{Submission}		& \multicolumn{1}{c}{\textit{Fixed}} & \multicolumn{1}{c}{\textit{Fixed}} & \multicolumn{1}{c}{\textit{Fixed}} \\
% Target skill                                                  & 0.021^{***}  & 0.021^{***}  & 0.017^{***}  \\
%                                                              & (0.003)      & (0.003)      & (0.003)      \\
% Source skill                                                  & -0.021^{***} & -0.032^{***} & -0.016^{***} \\
%                                                               & (0.001)      & (0.001)      & (0.001)      \\
% Target skill $\times$ Source skill                            & -0.006^{***} & -0.006^{***} & -0.005^{***} \\
%                                                               & (0.000)      & (0.000)      & (0.000)      \\
% Submitted to same contest: Yes $\times$ Target skill          & 0.019^{***}  & 0.018^{***}  & 0.015^{***}  \\
%                                                               & (0.000)      & (0.000)      & (0.000)      \\
% Submitted to same contest: Yes $\times$ Source skill          & 0.011^{***}  & 0.010^{***}  & 0.008^{***}  \\
%                                                               & (0.000)      & (0.000)      & (0.000)      \\
\midrule
% R$^2$                                                         & 0.396        & 0.396        & 0.396        \\
Adj.\ R$^2$                                                   & 0.391        & 0.391        & 0.391        \\
Num.\ obs.                                                    & \multicolumn{3}{c}{27,188,751} \\
\bottomrule
\multicolumn{4}{l}{\tiny{$^{***}p<0.001$; $^{**}p<0.01$; $^{*}p<0.05$}}
\end{tabular}
\end{scriptsize}
\caption{Regression of different mechanisms underlying strategic behavior.}
\label{tab:Mechanisms}
\end{center}
\end{table}

% used in OrgSci R3 \input{../../01-Analysis-Writeup-LaTeX/TABLE_MECH0_ALL_clusterSE_SubFE_MANUAL}
% \input{../../01-Analysis-Writeup-LaTeX/TABLE_MECH0_ALL_clusterSE_SubFE_Round4_MANUAL}

First, we explore tenure on the platform (measured in months between the date they signed up and the date a rating was submitted). We find no significant effect (Model 1). We also tested the interaction with a quadratic term (not shown) but find no effect either. Second, we find mixed results with regard to submission and rating experience (Model 2). While community members with higher rating experience appear to be less likely to sabotage, those with higher submission experience appear to be more likely. This is consistent with the interpretation that community members with a strong competitive motivation (making many submissions) are more likely to act strategically compared to community members who may be more socially motivated (evaluating many submissions by others).
Third, we find past winners are significantly more likely to sabotage and are more likely to be the targets of sabotage.

% #####################################################################
%          				   LENIENCY -   Standard Deviation
% #####################################################################
\clearpage
\section{Leniency as Variation within Contests}
Table \ref{tab:Leniency} shows regression on the contest-level using standard deviation of ratings a user casts as dependent variable. We find that when individuals have submitted to the same contest, they make better use of the full rating spectrum and submit ratings with a higher standard deviation (Model 1: $\beta = 0.037; p < 0.001$). The effect is amplified by skill so that higher-skilled individuals submit ratings with an even higher standard deviation (positive interaction between skill and having submitted to the same contest in Model 2: $\beta = 0.017; p < 0.001$).

\begin{table}[h!]
\begin{center}
\begin{scriptsize}
\begin{tabular}{@{\extracolsep{5pt}}l D{.}{.}{4.4} D{.}{.}{4.4} }
\toprule
 & \multicolumn{2}{c}{OLS} \\
 \cline{2-3} \\[-5pt]
 Dependent Variable: & \multicolumn{2}{c}{\bf Std.Dev. of Ratings} \\
 \cline{2-3} \\[-5pt]
 & \multicolumn{1}{c}{(1)} & \multicolumn{1}{c}{(2)} \\
\midrule
Submitted to same contest: Yes                         & 0.037^{***} & 0.004        \\
                                                       & (0.002)     & (0.009)      \\
Source ability                                         &             & -0.031^{***} \\
                                                       &             & (0.007)      \\
Submitted to same contest: Yes  &             & 0.017^{***}  \\
\quad $\times$ Source ability                                                       &             & (0.005)      \\
\textit{Individual} 		& \multicolumn{1}{c}{\textit{Fixed}} & \multicolumn{1}{c}{\textit{Fixed}}  \\
\textit{Contest}		& \multicolumn{1}{c}{\textit{Fixed}} & \multicolumn{1}{c}{\textit{Fixed}}  \\
\midrule
% R$^2$ (full model)                                     & 0.371       & 0.371        \\
% R$^2$ (proj model)                                     & 0.001       & 0.001        \\
Adj. R$^2$                                & 0.299       & 0.299        \\
% Adj. R$^2$ (proj model)                                & -0.112      & -0.112       \\
% Num. groups: UserID                                    & 44278       & 44278        \\
% Num. groups: ContestID                                 & 511         & 511          \\
Num.~obs.                                              & 439,882      & 439,882       \\
\bottomrule
\multicolumn{3}{l}{\tiny{$^{***}p<0.001$; $^{**}p<0.01$; $^{*}p<0.05$}}
\end{tabular}
\end{scriptsize}
\caption{User-contest level analysis of standard deviations of ratings cast.}
\label{tab:Leniency}
\end{center}
\end{table}

% #############################
% ##              Substitution                 ##
% #############################
\clearpage
\subsection{Substitution of sabotage and self-promotion}

% For OrgSci R2: taking histogram out
% \begin{figure}[htbp]
%	\begin{center}
%		\includegraphics[width=.5\linewidth]{../../../FIGURES/Rplot-Substituion}
%		\caption{{Rating distribution of (a) all competitors and (b) competitors who rated others but not themselves.}}
%		\vspace{12pt}
%		\label{fig:substituion}
%	\end{center}
%\end{figure}

\begin{table}[h!]
\begin{center}
\begin{scriptsize}
\begin{tabular}{@{\extracolsep{5pt}}l D{.}{.}{3.4} D{.}{.}{3.4} }
\toprule
 & \multicolumn{2}{c}{Linear Probability} \\
 \cline{2-3}\\[-5pt]
 Dependent Variable: & \multicolumn{2}{c}{\textbf{Sabotage:} 0-Star Rating} \\
 \cline{2-3} \\[-5pt]
 & \multicolumn{1}{c}{(1)}  & \multicolumn{1}{c}{(2)}  \\
 \midrule
Submitted to same contest: Yes & -0.006^{***} & -0.057^{***} \\
                               & (0.000)      & (0.000)      \\
Rate own submission: Yes       & -0.194^{***} & -0.205^{***} \\
                               & (0.001)      & (0.000)      \\
Self-promoted in this contest  & -0.002^{**}  & 0.031^{***}  \\
                               & (0.001)      & (0.001)      \\
\textit{Individual} 		& \multicolumn{1}{c}{\textit{Fixed}} & \multicolumn{1}{c}{\textit{Fixed}}  \\
\textit{Submission}		& \multicolumn{1}{c}{\textit{Fixed}} & \multicolumn{1}{c}{\textit{Fixed}} \\
\midrule
% R$^2$                          & 0.375        & 0.068        \\
Adj.\ R$^2$                    & 0.372        & 0.064        \\
Num.\ obs.                                       & \multicolumn{2}{c}{38,102,880} \\
\bottomrule
\multicolumn{3}{l}{\tiny{$^{***}p<0.001$; $^{**}p<0.01$; $^{*}p<0.05$}}
\end{tabular}
\end{scriptsize}
\caption{Regression models investigating whether individuals use sabotage and self-promotion together or substitute one for the other.}
\label{tab:Substitution}
\end{center}
\end{table}

% #############################################
% ##              Robustness Tests                ##
% #############################################

\subsection{Robustness: Dyad Fixed Effects}\label{sec:dyadFE}
As a robustness test, we investigate dyadic patterns underlying our findings of sabotage. Research on dyadic rivalry may be an alternative explanation \citep{kilduff2016whatever,grad2022rivalry}. For example, individuals may ``sabotage'' others with whom they have competed in the past and against whom they have lost. One may hence wonder if there are (or emerge over time) dyads in which individuals systematically sabotage others. Some dyads may rate each other with 0-stars whether or not they are competitors in the current contest and rivalry could be an alternative explanation to the strategic motivation (i.e., the behavior is founded on the subjective feeling of rivalry instead of the strategic motivation to increase ones chance of winning the contest prize). If so, that may lead to an overestimation of our effects in the sense that we attribute it to being ``strategic'' rather than ``rivalrous''. To investigate this alternative explanation, we estimate a variation of our main model with dyad-level fixed effects (Table \ref{table:DyadFE}). Since rivalry is generally assumed to require some time to form \citep{kilduff2016whatever,grad2022rivalry}, we restrict our analysis to dyads with ten or more encounters (e.g., $A$ rating five of $B$'s submissions and $B$ rating five of $A$'s submissions). Note that the effect we report in the table is driven by many fewer observations: only those dyads who rate each other at least once in the same contest and once outside the same contest contribute to the parameter estimate. Controlling for unobserved dyadic rivalries through these dyad-level fixed effects (instead of our usual individual- and submission-level fixed effects) we find that likelihood to sabotage others when competing (the main effect) \textit{increases} by more than half ($-0.070$ vs.~$-0.043$). If the ``sabotage'' effect we report in our estimates were driven by unobserved rivalries, then we would expect the effect of rating ones competitors to \textit{decrease}. Furthermore, we find significant positive effects for sabotage committed by high-skilled targeting other high-skilled just as in the analysis reported in the main text (interaction term coefficients $\beta = 0.015; p < 0.001$ for sources and $\beta = 0.002; p = 0.053$ for targets). This suggests that sabotage happens outside, and on top of, any pre-existing dyadic rivalries. That is, even within the same dyad, we still find a difference in the rating behavior between competing for the same prize vs.~not competing. In summary, dyadic rivalry patterns are not sufficient to explain the strategically motivated sabotage.

\begin{table}[h!]
\begin{center}
\begin{scriptsize}
\begin{tabular}{@{\extracolsep{5pt}}l D{.}{.}{3.4} }
\toprule
 & \multicolumn{1}{c}{Linear Probability} \\
 \cline{2-2}\\[-5pt]
Dependent Variable: 	& \multicolumn{1}{c}{\textbf{Sabotage:} 0-Star Rating} \\
 \cline{2-2} \\[-5pt]
 & \multicolumn{1}{c}{10+ Encounters}   \\
 \cline{2-2} \\[-5pt]
 & \multicolumn{1}{c}{(1)}  \\
\midrule
Submitted to same contest: Yes                       & -0.043^{***}   \\
                                                     & (0.004)        \\
% Rate own submission: Yes                             &                \\
%                                                      &                \\
Target skill                                         & -0.029^{***}   \\
                                                     & (0.008)        \\
Source skill                                         & -0.025^{**}    \\
                                                     & (0.009)        \\
Target skill $\times$ Source skill                   & -0.006^{\cdot} \\
                                                     & (0.004)        \\
Submitted to same contest: Yes \\
\quad $\times$ Target skill & 0.002^{\dagger}  \\
                                                     & (0.001)        \\
\quad $\times$ Source skill & 0.015^{***}    \\
                                                     & (0.001)        \\
\textit{Source-Target Dyad} 		& \multicolumn{1}{c}{\textit{Fixed}} \\
\midrule
% R$^2$ (full model)                                   & 0.482          \\
% R$^2$ (proj model)                                   & 0.001          \\
Adj. R$^2$                               & 0.448          \\
% Adj. R$^2$ (proj model)                              & -0.065         \\
Num.~obs.                                            & \multicolumn{1}{c}{36,94,941}        \\
Num.~dyads                               & \multicolumn{1}{c}{22,9404}         \\
\bottomrule
\multicolumn{2}{l}{\tiny{$^{***}p<0.001$; $^{**}p<0.01$; $^{*}p<0.05$; $^{\cdot}p<0.1$}}
\end{tabular}
\end{scriptsize}
\caption{Robustness tests using dyad fixed effects.}
\label{table:DyadFE}
\end{center}
\end{table}

% #############################################
% ##              Natural Experiment 1: Prize Increase                 ##
% #############################################
\subsection{Natural Experiment 1: Prize Increase}

\begin{figure}[htbp]
	\begin{center}
		\includegraphics[width=0.6\linewidth,valign=c]{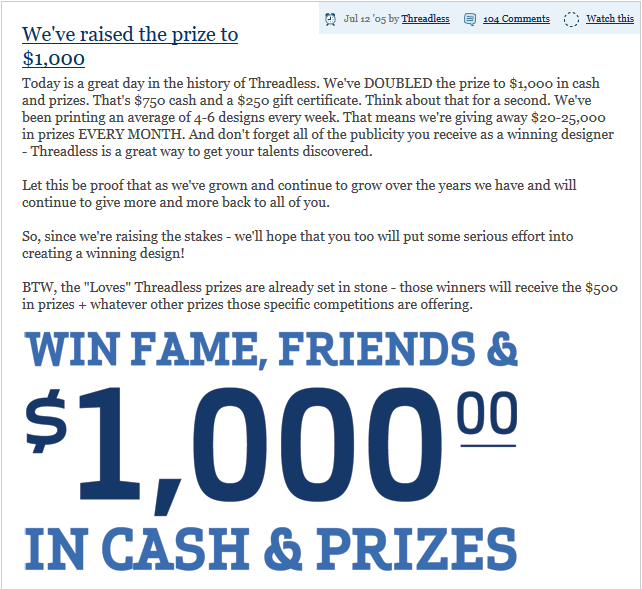}
		\caption{Screenshot natural experiment sabotage.}
		\begin{flushleft}
			\begingroup
			\leftskip4em
			\rightskip\leftskip
			\small{\textit{Note: }Screenshot of the Threadless website showing the announcement of the prize increase posted on July 12, 2005.}
			\par
			\endgroup
		\end{flushleft}
		\label{fig:BlogPrize}
	\end{center}
\end{figure}
\clearpage

% #############################################
% ##              Natural Experiment 2: Scoring Rule Change                 ##
% #############################################
\subsection{Natural Experiment 2: Scoring Rule Change}

\begin{figure}[htbp]
	\begin{center}
		\includegraphics[width=0.90\linewidth,valign=c]{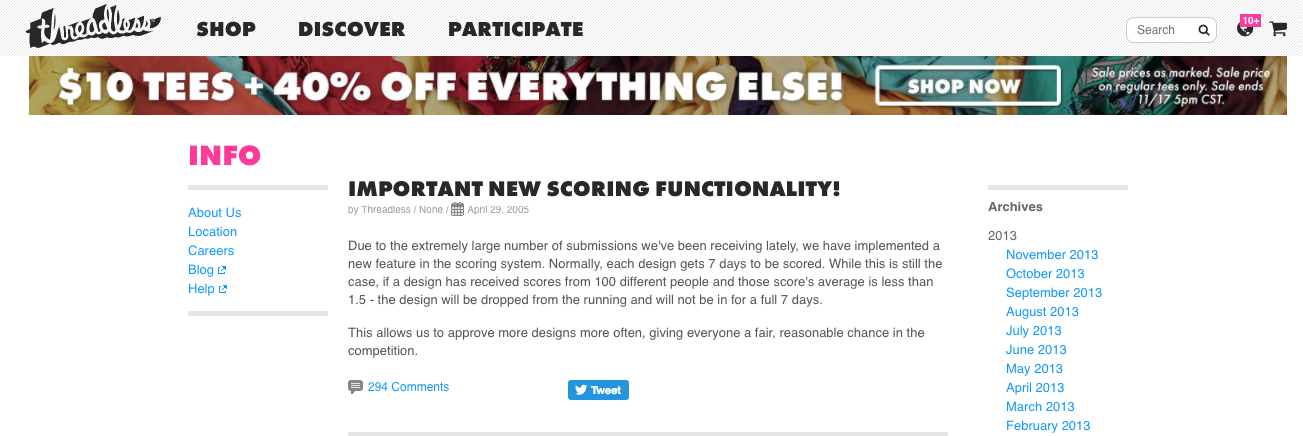}
		\caption{Screenshot natural experiment self-promotion}
		\begin{flushleft}
			\begingroup
			\leftskip4em
			\rightskip\leftskip
			\small{\textit{Note: }Screenshot of the Threadless website showing the announcement of the scoring rule change posted on April 29, 2005.}
			\par
			\endgroup
		\end{flushleft}
		\label{fig:BlogScoring}
	\end{center}
\end{figure}

% #############################################
% #############################################
% ##             Organization - Level Outcomes                  ##
% #############################################
% #############################################

% ##################################
% ##         Long-Term Participation             ##
% ##################################
\subsection{Long-Term Participation}
\begin{figure}[htbp]
	\begin{center}
	\includegraphics[width=.6\linewidth]{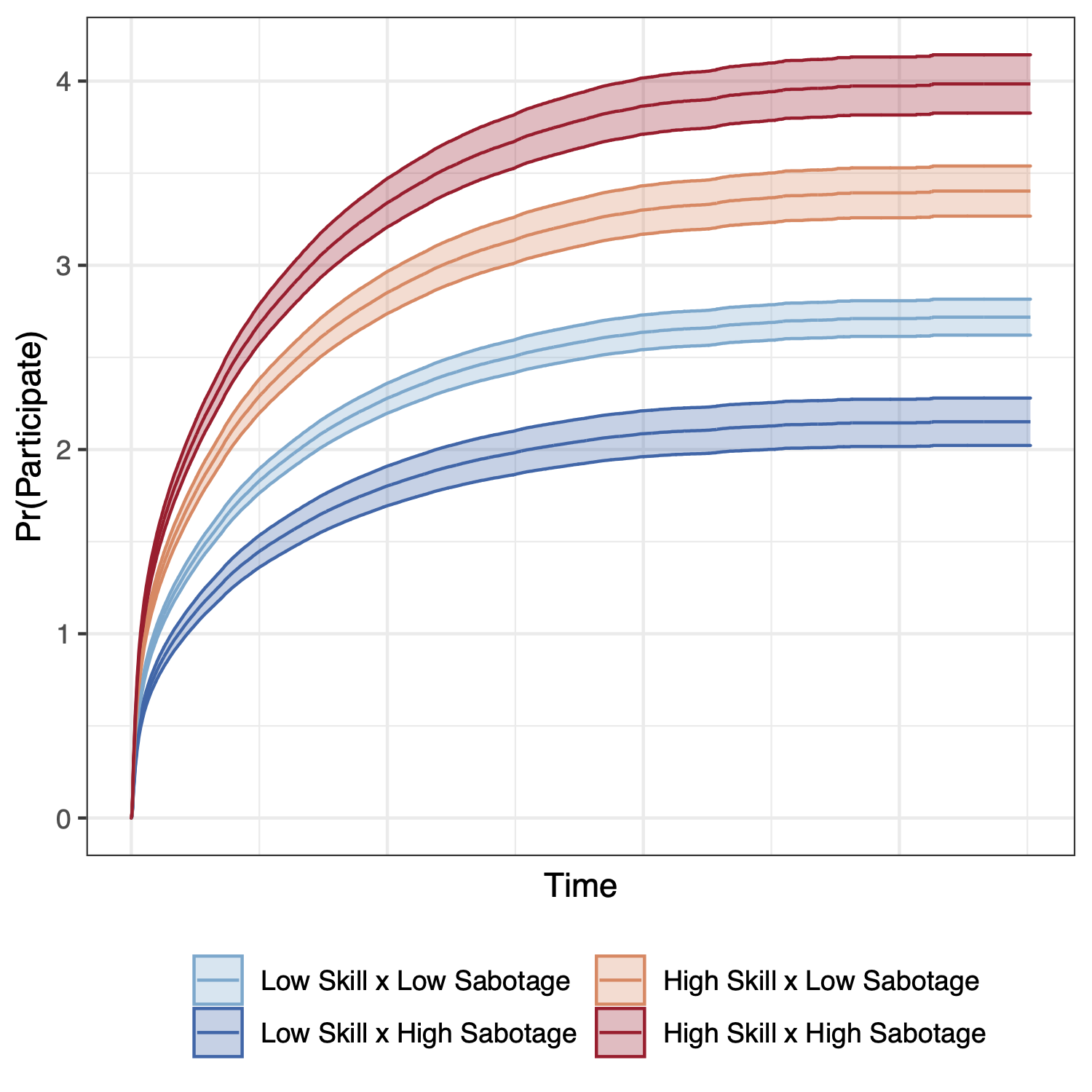}
	\caption{Likelihood of participating in a future contest.}
		\begin{flushleft}
			\begingroup
			\leftskip4em
			\rightskip\leftskip
			\small{\textit{Note: }	Likelihood for high skilled (90th percentile) and low skilled (10th percentile) individuals based on experience in the current contest for individuals who received high (90th percentile) and low (10th percentile) levels of sabotage (cumulative hazard; based on Model 1 from Table \ref{table:longTerm}; controls include rating received, number of ratings received, amount of competition, quality of competition)..}
			\par
			\endgroup
		\end{flushleft}

	\label{fig:HazardRate}
	\end{center}
\end{figure}

Table \ref{table:longTerm} in the main text shows coefficient estimates for the hazard rate of participation in the next round. The hazard rate can also be shown visually (Figure \ref{fig:HazardRate}). The figure contrasts how high- versus low-skilled competitors (10th vs. 90th percentile, respectively) react to receiving high versus low levels of sabotage (also 10th vs. 90th percentile, respectively).

% #############################################
% ##.                 WELFARE ANALYSIS - Ratings             ##
% #############################################
\section{Effect of Strategic Rating on Selection of Contest Winners}

Does strategic rating behavior affect the rank-order of contest submissions, especially given the generally large contests we observe? To assess the impact of \textit{self-promotion}, we exclude all ratings from designers who rated their own designs, recalculate average ratings for submissions, and then re-rank all submissions based on the new averages. We find that, keeping everything else constant, in seven out of 511 contests (1.4\%), the winner of the contest changes; and in 28 of 511 contests (5.5\%), there is a change in at least one of the top three ranks. 
To analyze the effect of \textit{sabotage}, we exclude all ratings from competitors as they might be strategically motivated. Note that this also includes votes from designers on their own designs, i.e., it contains the previous reported effects. We find that in 12\% of the contests, the winner of the contest changes; and that in 48\% of the contests, there is a change in the top three ranks. The effect is especially pronounced in close contests where the contest winner would change in 25\% of cases and 65\% would see a change in the top three. This can be considered an upper boundary for the effect of strategic behavior via the rating mechanism in our setting.
% Excluding the 1.4\% of cases that changed due to self-promotion (5.5\% for a change in the top three) that can be attributed to self-promotion, that leaves changes of 10.6\% (42.5\% for top three) that can be attributed to sabotage.

This change in rankings, however, may be simply due to the fact that removing \textit{any} ratings changes the resulting ranking. So to complement this first test, we compare the effect of removing the ratings of potentially strategic raters, to the effect of removing the ratings of an equal amount of randomly selected raters. Removing random raters, contest winner changes on average 10.7\% of cases across our bootstrap simulation, while contest winners change 12\% of the time when removing ratings by potentially strategic raters.
In close contests, contest winner changes 13.8\% in the null model vs.~25\% winner change when removing potentially strategic raters. Removing strategic ratings affect outcomes more than just removing random raters which lends further support to our claim of the important role of strategic rating. Overall, these two analyses suggest that strategic behavior of idea generators during idea selection, does affect Threadless' ability to identify the most promising ideas. The effect is strongest in close competition where up to 25\% of contest winners may change and around two-thirds of contests would have changes in the top-three.

% #############################################
% ##              Removing Random Ratings                 ##
% #############################################
% \subsection{Removing Random Ratings} \label{sec:RemoveRandom}
To further quantify the effect of strategic rating, we compute a baseline of how much a contests rank-order would change if the ratings of some randomly chosen individuals were removed. For each contest, we compute the number of competitors who did cast potentially strategic ratings, and select the same amount of raters at random. To remove the same number of ratings, we need to remove about 20\% more raters since competitors who submitted to the same contest rate more submissions that those who did not. 
We remove their ratings and recompute the ranking to see how much has changed. We repeat this 500 times for all contests in our data in order to get a robust null model of how much rankings would change after removing random raters. 

% Overall, given that Threadless has paid approximately \$775,000 in prize money over the observation period, we find that up to \$93,000 may have been payed to the ``wrong'' contestant due to strategic behavior.

\end{document}